\documentclass[a4paper,11pt]{article}%
\usepackage{amssymb}
\usepackage{graphicx}
\usepackage{amsbsy}
\usepackage{amsmath}
\usepackage{cite}
\usepackage{slashed}
\usepackage{amsfonts}
\usepackage{hyperref}%
\hypersetup{
colorlinks   = true,
linkcolor    = blue,
citecolor    = blue,
}

\newcommand{\BN}{\mathcal{B}}
\newcommand{\LN}{\mathcal{L}}

\newcommand{\be}{\begin{equation}}
\newcommand{\ee}{\end{equation}}
\newcommand{\ba} {\begin{equation}\begin{aligned}}
\newcommand{\ea} {\end{aligned}\end{equation}}
\newcommand{\bg} {\begin{equation}\begin{gathered}}
\newcommand{\eg} {\end{gathered}\end{equation}}

\newcommand{\TeV}{\ \text{TeV}}

\newcommand\nnfootnote[1]{%
  \begin{NoHyper}
  \renewcommand\thefootnote{}\footnote{#1}%
  \addtocounter{footnote}{-1}%
  \end{NoHyper}}

\ifx\pdfoutput\relax\let\pdfoutput=\undefined\fi
\newcount\msipdfoutput
\ifx\pdfoutput\undefined\else
\ifcase\pdfoutput\else
\ifx\paperwidth\undefined\else
\ifdim\paperheight=0pt\relax\else\pdfpageheight\paperheight\fi
\ifdim\paperwidth=0pt\relax\else\pdfpagewidth\paperwidth\fi
\fi\fi\fi

\begin{document}
\thispagestyle{empty} \setcounter{page}{0} \begin{flushright} June 2022\\
\end{flushright}

\vskip          3 true cm

\begin{center}
{\huge Leptoquarks, axions and the unification of}\\[0.5cm]
{\huge B, L, and Peccei-Quinn symmetries$^{1}$ } \\[1.9cm]

\textsc{Fernando Arias-Arag\'{o}n}$^{2}$, \textsc{Christopher Smith}$^{3}%
$\vspace{0.5cm}\\[9pt]\smallskip{\small \textsl{\textit{Laboratoire de
Physique Subatomique et de Cosmologie, }}}\linebreak%
{\small \textsl{\textit{Universit\'{e} Grenoble-Alpes, CNRS/IN2P3, Grenoble
INP, 38000 Grenoble, France}.}} \\[1.9cm]\textbf{Abstract}\smallskip
\end{center}

\begin{quote}
\noindent In this paper, axion models supplemented by leptoquarks and diquarks are systematically analyzed. Turning on some couplings to and among these latter states permits to unify the Peccei-Quinn symmetry with baryon ($\BN$) and lepton ($\LN$) numbers, such that the axion becomes associated to the spontaneous breaking of the three $U(1)$ symmetries. All possible four- and six-fermion patterns of $\BN$ and $\LN$ violation are discussed, including those inducing proton decay, with $\Delta \BN = 1$ and $\Delta \LN = \pm 1, \pm 3$, neutron-antineutron oscillations with $\Delta \BN = 2$, and Majorana neutrino masses with $\Delta \LN = 2$. Scenarios in which one or two axion fields necessarily appear in any $\BN$ and/or $\LN$ violating operators are also constructed. Nucleon decays would then necessarily involve an axion in the final state, while neutron-antineutron oscillations would only happen in an axionic background. This could have implications for the neutron lifetime puzzle, and more generally, opens the door to new phenomenological and cosmological applications.

\nnfootnote{$^{1}\;$Title changed from \textit{One $U(1)$ to Rule them All: In the Realm of Leptoquarks, the Axion Shines}}
\nnfootnote{$^{2}\;$arias@lpsc.in2p3.fr}
\nnfootnote{$^{3}\;$chsmith@lpsc.in2p3.fr}
\end{quote}

\newpage%

\hrule
\vskip          1 true cm

\setcounter{tocdepth}{2}
\tableofcontents

\vskip          1 true cm
\hrule

\section{Introduction}

Nowadays, the axion mechanism represents our best solution to the longstanding strong CP puzzle, that is, the non-observation of CP violation in the strong interactions that should have manifested itself as an electric dipole moment for the neutron~\cite{Abel:2020pzs}.

The axion mechanism relies on the spontaneous breaking of a new symmetry, the Peccei-Quinn (PQ) symmetry~\cite{PQ}, and on the subsequent realignment of the associated Goldstone boson, the axion~\cite{Weinberg:1977ma,Wilczek:1977pj}, by strong interaction effects that kills off any CP violation in the QCD Lagrangian. This solution is thus tailored to the problem it is intended to solve and, as such, may appear a bit ad-hoc. In addition, unsuccessful experimental searches for the axion have ruled out its simplest incarnation, leaving us with essentially two classes of scenarios in which the axion is extremely light (well below the eV scale) and very weakly coupled to normal matter: the Kim, Shifman, Vainshtein, Zakharov (KSVZ)~\cite{KSVZ} framework in which new very heavy colored fermions are introduced, and the Dine, Fischler, Srednicki, Zhitnitsky (DFSZ)~\cite{DFSZ} scenario in which at least two Higgs doublets are required. Though the strong CP puzzle is extremely serious, additional motivations appear desirable to justify such departures from the Standard Model (SM) matter content. To that avail, knowing that the axion could also make up for the observed dark matter (DM) offers a strong incentive to pursue this route~\cite{DMaxion}. 

Yet, current axion models cannot explain why the DM relic density is so close to that of baryonic matter. Though this may be totally coincidental, it nevertheless suggests a link between DM and baryogenesis~\cite{Kaplan:1991ah}, another prominent cosmological enigma. Actually, it suggests DM is not foreign to baryon $\mathcal{B}$ or lepton $\mathcal{L}$ number (see Ref.~\cite{Alonso-Alvarez:2021oaj} and references therein for a recent analysis), or that DM is somehow related to $\mathcal{B}$ being spontaneously broken~\cite{Dulaney:2010dj}. In parallel, there have been many attempts at involving axions in the baryogenesis mechanism, see e.g. Refs.~\cite{Craig:2010au,Servant:2014bla,Jeong:2018ucz,Co:2019wyp,Krauss:2022usd,Domcke:2020kcp}, though in general still relying on the SM anomalous $\mathcal{B}+\mathcal{L}$ effects. 

With this motivation in mind, our goal here is to design models in which the PQ symmetry is, at the fundamental level, entangled with $\mathcal{B}$ and $\mathcal{L}$. As a matter of principle, accidental symmetries are not particularly attractive, but while we can live with the PQ symmetry, assuming some dynamics hide behind it, $\mathcal{B}$ and $\mathcal{L}$ cannot be viable since, as said before, the electroweak non-perturbative dynamics break them, and baryogenesis asks for their violation. By unifying the PQ symmetry with $\mathcal{B}$ and $\mathcal{L}$, all three are broken spontaneously, but a single Goldstone field remains, the axion (for some recent works along this line, see Refs.~\cite{Reig:2018yfd,Ohata:2021rkh}). In this way, the complex scalar field whose pseudoscalar component is the axion becomes charged under $\mathcal{B}$ and $\mathcal{L}$ and, at the high scale, protects the model from additional $\mathcal{B}$ and/or $\mathcal{L}$ violation. At the same time, though the axion has no charge, it inherits a $\mathcal{B}$ and/or $\mathcal{L}$ violating phenomenology. Whether this is sufficient to relate the DM and baryonic relic densities remains to be seen, and is beyond the scope of the present paper, but we think these constructions may direct us in the right direction.

In this paper, we will use scalar and vector leptoquarks and diquarks to entangle the PQ, $\mathcal{B}$, and $\mathcal{L}$ symmetries. Such states are well motivated in various theoretical settings (see Ref.~\cite{Dorsner:2016wpm} for a review) and, furthermore, supported by a number of anomalies like the $W$ boson mass~\cite{Crivellin:2020ukd}, B decays~\cite{LQreview} or $(g-2)_{\mu}$~\cite{ColuccioLeskow:2016dox,Dorsner:2019itg,Crivellin:2020tsz}, or even combinations of them~\cite{Crivellin:2019dwb,Athron:2022qpo,Bhaskar:2022vgk}. Our goal is to systematically analyze the $\mathcal{B}$ and/or $\mathcal{L}$ symmetry breaking patterns that can arise combining the DFSZ and KSVZ scenarios with leptoquarks and diquarks and, in each case, to analyze the impact on the axion phenomenology.

The paper is organized as follows. In section~\ref{Sec2a}, we briefly introduce the KSVZ and DFSZ axion models and, in Section~\ref{Sec2b}, discuss in some details the ambiguities arising from the $\mathcal{B}$ and $\mathcal{L}$ fermionic currents~\cite{Quevillon:2020hmx,Quevillon:2020aij}. Then in section~\ref{Sec2c}, we set up the leptoquark and diquark sector, describing all the possible $\mathcal{B}$ and $\mathcal{L}$ explicit breaking patterns achievable with these states. This forms the basis for combining the axion and leptoquark/diquark sectors in Sec.~\ref{SecAxLQDQ}. We analyze first the KSVZ setting in Sec.~\ref{SecKSVZ} and describe the $(\Delta\mathcal{B},\Delta\mathcal{L})=(1,\pm1), (2,0), (1,\pm3)$ spontaneous breaking patterns, further adding to them a spontaneously generated $(\Delta\mathcal{B},\Delta\mathcal{L})=(0,2)$ seesaw mechanism for neutrino masses. These scenarios are then trivially adapted to the DFSZ setting in Sec.~\ref{SecDFSZ}. In the final Sec.~\ref{SecSpont}, we show how to force $(\Delta\mathcal{B},\Delta\mathcal{L})$ effects to involve one or more axion fields. The phenomenology is then quite different, and we briefly describe some possible consequences for the neutron lifetime anomaly or neutron-antineutron oscillation experiments. Finally, our results are summarized in Sec.~\ref{Ccl}.

\section{Axion and leptoquark models}

In this section, the KSVZ~\cite{KSVZ} and DFSZ~\cite{DFSZ} axion models are introduced, and their connection to baryon and lepton numbers, $\mathcal{B}$ and $\mathcal{L}$, are detailed. Then, we introduce separately the leptoquarks and diquarks that can be coupled to SM fermions, and discuss how their couplings drive specific $\mathcal{B}$ and $\mathcal{L}$ violating patterns. This sets the stage for the next section, where both axion models and leptoquarks/diquarks will be put together.

\subsection{Introducing the KSVZ and DFSZ models\label{Sec2a}}

In both the KSVZ and DFSZ constructions, the axion emerges as the pseudoscalar component of a complex scalar field. This state is neutral under all the SM gauge interactions, $\phi=(\mathbf{1},\mathbf{1},0)$ under $SU(3)_{C}\otimes SU(2)_{L}\otimes U(1)_{Y}$, but its kinetic term is invariant under the rephasing $\phi\rightarrow e^{i\alpha}\phi$. This invariance is promoted to a
spontaneously broken symmetry $U(1)_{\phi}$ by postulating a rephasing invariant scalar potential with the usual Mexican hat shape, $V(\phi^{\dagger}\phi)=\mu^{2}\phi^{\dagger}\phi+\lambda(\phi^{\dagger}\phi)^{4}$, $\mu^{2}<0$ and $\lambda>0$. In that case, the components of $\phi$ can be written
\begin{equation}
\phi=\frac{1}{\sqrt{2}}(v_{\phi}+\rho)\exp(i\eta_{\phi}/v_{\phi})\ ,
\label{PhiPolar}
\end{equation}
with $\eta_{\phi}$ the associated Goldstone boson and $v_{\phi}^{2}=-\mu^{2}/\lambda$ the vacuum expectation value (VEV). As the breaking scale $v_{\phi}$ naturally tunes all the $\eta_{\phi}$ couplings, it is assumed much higher than the electroweak scale to avoid exclusion bounds.

To solve the strong CP puzzle, $\eta_{\phi}$ must interact with SM particles~\cite{Weinberg:1977ma,Wilczek:1977pj}, in particular with gluons via a $\eta_{\phi}G^{\alpha,\mu\nu}\tilde{G}_{\mu\nu}^{a}$ coupling~\cite{PQ}. What differentiates the KSVZ and DFSZ models is how these couplings are introduced. The former~\cite{KSVZ} adds a vector-like colored fermion $\Psi_{L,R}\sim(\mathbf{R},\mathbf{T},Y)$ for some complex representation $\mathbf{R}$ of $SU(3)_{C}$, but otherwise arbitrary weak representation $\mathbf{T}$ and hypercharge $Y$, and postulates the Lagrangian (the rest of the SM couplings are understood)
\begin{align}
\mathcal{L}_{\mathrm{KSVZ}}  &  =\partial_{\mu}\phi^{\dagger}\partial^{\mu
}\phi-V(\phi)+\bar{\Psi}_{L,R}(i\slashed D)\Psi_{L,R}+(y\phi\bar{\Psi}_{L}\Psi_{R}+h.c.)\nonumber\\
&  -\bar{u}_{R}\mathbf{Y}_{u}q_{L}H-\bar{d}_{R}\mathbf{Y}_{d}q_{L}H^{\dagger
}-\bar{e}_{R}\mathbf{Y}_{e}\ell_{L}H^{\dagger}-\bar{\nu}_{R}\mathbf{Y}_{\nu
}\ell_{L}H+h.c.\ . \label{KSVZ0}%
\end{align}
The covariant derivative acting on $\Psi_{L,R}$ is as appropriate to its chosen gauge quantum numbers. What characterizes this model is first that the Goldstone boson of the PQ symmetry does not mix with that of the $SU(2)_{L}\otimes U(1)_{Y}$ breaking (the phase of the Higgs doublet $H$). Thus, the axion is simply $a^{0}=\eta_{\phi}$, and it has no direct coupling to any of the SM particles. It only couples to $\Psi_{L}$ and $\Psi_{R}$, which necessarily have different charges under $U(1)_{\phi}$. Then, axion to SM gauge boson couplings first arise at one-loop, via anomalous $\Psi_{L,R}$ triangle loops, while those to SM fermions require a further gauge boson loop. Since $\Psi_{L,R}$ can be massive in the electroweak unbroken phase, its loops do not break $SU(2)_{L}\otimes U(1)_{Y}$ and the couplings to gauge bosons have the $SU(2)_{L}\otimes U(1)_{Y}$ invariant form~\cite{Georgi:1986df}
\begin{equation}
\mathcal{L}_{\mathrm{KSVZ}}^{eff}=-\frac{1}{16\pi^{2}v_{\phi}}a^{0}(g_{s}%
^{2}d_{L}C_{C}G_{\mu\nu}^{a}\tilde{G}^{a,\mu\nu}+g^{2}d_{C}C_{L}W_{\mu\nu}%
^{i}\tilde{W}^{i,\mu\nu}+g^{\prime2}d_{L}d_{C}C_{Y}B_{\mu\nu}\tilde{B}^{\mu
\nu})\ , \label{GaugeCouplAno}
\end{equation}
with the quadratic invariants and dimensions of the $\mathbf{R}$ and $\mathbf{T}$ representations denoted $C_{C,L}$ and $d_{C,L}$, and $C_{Y}=Y^{2}/4$.

The DFSZ model~\cite{DFSZ} does not introduce new fermions, but requires two Higgs doublets. The important couplings are%
\begin{align}
\mathcal{L}_{\mathrm{DFSZ}}  &  =\partial_{\mu}\phi^{\dagger}\partial^{\mu
}\phi-V(\phi^{\dagger}\phi)+\phi^{2}H_{u}^{\dagger}H_{d}+V(H_{u}^{\dagger
}H_{u},H_{d}^{\dagger}H_{d})\nonumber\\
&  -\bar{u}_{R}\mathbf{Y}_{u}q_{L}H_{u}-\bar{d}_{R}\mathbf{Y}_{d}q_{L}%
H_{d}^{\dagger}-\bar{e}_{R}\mathbf{Y}_{e}\ell_{L}H_{d}^{\dagger}-\bar{\nu}%
_{R}\mathbf{Y}_{\nu}\ell_{L}H_{u}+h.c.\ . \label{DFSZ0}%
\end{align}
The potentials and Yukawa couplings are invariant under three independent $U(1)$s, corresponding to the rephasing of $\phi$, $H_{u}$, and $H_{d}$. A combination of these is explicitly removed by the mixing term $\phi^{2}H_{u}^{\dagger}H_{d}$ (we could equally take $\phi H_{u}^{\dagger}H_{d}$, but at the cost of introducing a new mass scale), so that only two Goldstone bosons arise. Explicitly, if we adopt for $H_{u,d}$ a polar representation similar as in Eq.~(\ref{PhiPolar}), with their pseudoscalar components denoted as $\eta_{u,d}$ and their VEVs as $v_{u,d}$, the $\phi^{2}H_{u}^{\dagger}H_{d}$ coupling translates as a mass term for the combination $\pi^{0}\sim2\eta_{\phi}/v_{\phi}-\eta_{u}/v_{u}+\eta_{d}/v_{d}$. One of the two remaining Goldstone bosons is eaten by the $Z$ boson. Since $H_{u,d}$ have the same hypercharge, the would-be Goldstone state $G^{0}$ must be $G^{0}\sim v_{u}\eta_{u}+v_{d}\eta_{d}$. The last remaining Goldstone mode, orthogonal to both $\pi^{0}$ and $G^{0}$, stays massless and is the axion:
\begin{equation}
a^{0}\sim\eta_{\phi}+\frac{v_{EW}}{v_{\phi}}\sin2\beta(\cos\beta\eta_{u}
-\sin\beta\eta_{d})+\mathcal{O}(v_{EW}^{2}/v_{\phi s}^{2})\ , \label{DFSZA0}%
\end{equation}
with $\tan\beta=v_{u}/v_{d}$ and $v_{EW}^{2}=v_{u}^{2}+v_{d}^{2}\approx (246\,$GeV)$^{2}$. The net result of all this is that the axion components in $H_{u,d}$ are suppressed by $v_{u,d}/v_{\phi}$. The leading couplings of the axion to SM particles come from the Yukawa couplings, with
\begin{equation}
\mathcal{L}_{\mathrm{DFSZ}}^{eff}=-i\frac{v_{EW}}{v_{\phi}}\sin2\beta
\sum_{f=u,d,e}\frac{m_{f}}{v_{EW}}\chi_{P}^{f}\,a^{0}\bar{\psi}_{f}\gamma
_{5}\psi_{f}\ ,\ \ \chi_{P}^{u}=\frac{1}{\tan\beta}\ ,\ \chi_{P}^{d}=\chi
_{P}^{e}=\tan\beta\ . \label{PseudoCoupl}%
\end{equation}
To reach this form, the mass terms are identified as $\sin\beta v_{EW}\mathbf{Y}_{u}\equiv\sqrt{2}\mathbf{m}_{u}$ and $\cos\beta v_{EW}\mathbf{Y}_{d,e}\equiv\sqrt{2}\mathbf{m}_{d,e}$ and the fermions are rotated to their mass basis. In the DFSZ setting, the axion couplings to gauge bosons only arise through SM fermion loops. As shown in Ref.~\cite{Quevillon:2019zrd} (see also Refs.~\cite{Bonnefoy:2020gyh,Quevillon:2021sfz}), starting from the pseudoscalar couplings in Eq.~(\ref{PseudoCoupl}), the final couplings to gauge boson do not have the form shown in Eq.~(\ref{GaugeCouplAno}), but instead explicitly break $SU(2)_{L}\otimes U(1)_{Y}$ invariance. Naively, this is easily understood since SM fermions only acquire masses after the $SU(2)_{L}\otimes U(1)_{Y}$ breaking.

\subsection{Introducing baryon and lepton numbers\label{Sec2b}}

In the following, when introducing leptoquark states, baryon and lepton numbers $\mathcal{B}$ and $\mathcal{L}$ will play a central role. The purpose in this section is to gather a few important facts about the interplay of these global symmetries with the PQ symmetry. Additional information on this topic can be found in Ref.~\cite{Quevillon:2020hmx}.

By definition, the $U(1)$ symmetry associated to the axion state is called the PQ symmetry. Given the scalar couplings described in the previous section, the PQ charges of all the scalar states are well-defined in the KSVZ and DFSZ models. Explicitly, we have in the KSVZ setting
\begin{equation}
\begin{tabular}[c]{ccc}\hline
KSVZ & $\phi$ & $H$\\\hline
$U(1)_{\phi}$ & $1$ & $0$\\
$U(1)_{H}$ & $0$ & $1$\\\hline
\end{tabular}
\ \ \ \ \ \Longrightarrow%
\begin{tabular}[c]{ccc}\hline
KSVZ & $\phi$ & $H$\\\hline
$U(1)_{PQ}$ & $1$ & $0$\\
$U(1)_{Y}$ & $0$ & $1$\\\hline
\end{tabular}
\end{equation}
and in the DFSZ, choosing the two independent $U(1)$ symmetries as those associated to Higgs doublet rephasings\footnote{The PQ charges of $\phi$, $H_{u}$ and $H_{d}$ are simply the coefficients of $\eta_{\phi,u,d}$ in Eq.~(\ref{DFSZA0}), up to a choice of normalization.},
\begin{equation}
\begin{tabular}[c]{cccc}\hline
DFSZ & $\phi$ & $H_{u}$ & $H_{d}$\\\hline
$U(1)_{Hu}$ & $1/2$ & $1$ & $0$\\
$U(1)_{Hd}$ & $-1/2$ & $0$ & $1$\\\hline
\end{tabular}
\ \ \ \ \ \Longrightarrow
\begin{tabular}[c]{cccc}\hline
DFSZ & $\phi$ & $H_{u}$ & $H_{d}$\\\hline
$U(1)_{PQ}$ & $(x+1/x)/2$ & $x$ & $-1/x$\\
$U(1)_{Y}$ & $0$ & $1$ & $1$\\\hline
\end{tabular}
\ \ \label{DFSZScalars}
\end{equation}
with the conventional notation $\tan\beta\equiv1/x$. Note that the $U(1)_{Y}$ and $U(1)_{PQ}$ charges of the two Higgs doublets are not `orthogonal', reflecting the fact that the original $U(1)_{Hu}$ and $U(1)_{Hd}$ charges for the three states $(\phi,H_{u},H_{d})$ were not. Also, it is important to keep in mind that though well-defined, these PQ charges are only defined in the electroweak broken phase, since they are function of $x\equiv v_{d}/v_{u}$.

For fermions, identifying the PQ charge is less trivial because the Yukawa couplings allow for two global symmetries, $\mathcal{B}$ and $\mathcal{L}$ (no particular structure is assumed for $\mathbf{Y}_{u,d,e,\nu}$, so individual flavors are not conserved a priori). Looking at the Lagrangian, the KSVZ model prescribes
\begin{equation}
\begin{tabular}[c]{ccccccccc}\hline
KSVZ & $\Psi_{L}$ & $\Psi_{R}$ & $q_{L}$ & $u_{R}$ & $d_{R}$ & $\ell_{L}$ &
$e_{R}$ & $\nu_{R}$\\\hline
$U(1)_{PQ}$ & $\alpha$ & $\alpha-1$ & $\beta$ & $\beta$ & $\beta$ & $\gamma$ &
$\gamma$ & $\gamma$\\
$U(1)_{Y}$ & $Y$ & $Y$ & $1/3$ & $4/3$ & $-2/3$ & $-1$ & $-2$ & $0$\\\hline
\end{tabular}
\label{KSVZfermions}
\end{equation}
where $\alpha$, $\beta$, and $\gamma$ are arbitrary, and correspond to conserved $\Psi$ number, baryon number, and lepton number, respectively. Similarly, for the DFSZ model,
\begin{equation}
\begin{tabular}[c]{ccccccc}\hline
DFSZ & $q_{L}$ & $u_{R}$ & $d_{R}$ & $\ell_{L}$ & $e_{R}$ & $\nu_{R}$\\\hline
$U(1)_{PQ}$ & $\beta$ & $\beta+x$ & $\beta-1/x$ & $\gamma$ & $\gamma-1/x$ &
$\gamma+x$\\
$U(1)_{Y}$ & $1/3$ & $4/3$ & $-2/3$ & $-1$ & $-2$ & $0$\\\hline
\end{tabular}
\label{DFSZfermions}
\end{equation}
Since $\beta$ and $\gamma$ are aligned with baryon and lepton numbers, it is tempting to set $\beta=\gamma=0$. This is not acceptable. For the DFSZ scenario, all the SM fermions do couple to the axion, but these couplings are not $SU(2)_{L}\otimes U(1)_{Y}$ invariant. Looking at Eq.~(\ref{PseudoCoupl}), no value of $\beta$ or $\gamma$ makes perfect sense since the PQ charge of the Dirac $u$ and $d$ states are different, so that of $q_{L}$ cannot be defined. The situation appears simpler in the KSVZ case, where it seems rather natural to set $\beta=\gamma=0$ since the SM fermions are not directly coupled to the scalar field $\phi$. Yet, even that is not tenable.

To see this, let us set off a seesaw mechanism~\cite{TypeI}. Given the quantum numbers of the $\nu_{R}$ field, we can either allow for a Majorana mass term $M_{R}\bar{\nu}_{R}^{\mathrm{C}}\nu_{R}$, a coupling $\phi\bar{\nu}_{R}^{\mathrm{C}}\nu_{R}$, or a coupling $\phi^{\dagger}\bar{\nu}_{R}^{\mathrm{C}}\nu_{R}$. These three cases are mutually exclusive since they impose different PQ charges to $\nu_{R}$. Let us consider the $\phi^{\dagger}\bar{\nu}_{R}^{\mathrm{C}}\nu_{R}$ case, which in effect identifies the PQ symmetry with lepton number symmetry, and the axion with the Majoron~\cite{Langacker:1986rj,Shin:1987xc,Clarke:2015bea} (see also \cite{Heeck:2019guh}). It imposes non-zero values for $\gamma$~\cite{Quevillon:2020hmx}
\begin{align}
\text{KSVZ}  &  :\phi^{\dagger}\bar{\nu}_{R}^{\mathrm{C}}\nu_{R}\rightarrow\gamma=\frac{1}{2}\ ,\\
\text{DFSZ}  &  :\phi^{\dagger}\bar{\nu}_{R}^{\mathrm{C}}\nu_{R}\rightarrow\gamma=\frac{1-3x^{2}}{4x}\ .
\end{align}
In both cases, the PQ current acquires a component aligned with the lepton number current, $J_{\mathcal{L}}^{\mu}=\bar{\ell}_{L}\gamma^{\mu}\ell_{L}+\bar{e}_{R}\gamma^{\mu}e_{R}+\bar{\nu}_{R}\gamma^{\mu}\nu_{R}$. In other words, $\ell_{L}$ and/or $e_{R}$ do end up PQ charged also. Yet, in the KSVZ case, a look at the Lagrangian shows that neither are directly coupled to $\phi$. Because of $\phi^{\dagger}\bar{\nu}_{R}^{\mathrm{C}}\nu_{R}$, the axion does end up coupled to right-handed neutrinos, with a $a^{0}\rightarrow\nu_{R}\nu_{R}$ vertex, but no such $\Delta\mathcal{L}=2$ coupling exists with the other leptons since it is forbidden by hypercharge. Only at the cost of extra Higgs doublet insertions could a $a^{0}\rightarrow\nu_{L}\nu_{L}$ exist, as arising from an effective PQ- and hypercharge-neutral operator $\phi^{\dagger}H\ell_{L}H\ell_{L}$ (or $\phi^{\dagger}H_{u}\ell_{L}H_{u}\ell_{L}$ in the DFSZ model), while obviously, any $\Delta\mathcal{L}=2$ coupling to charged lepton would require either extra gauge fields, or charged Higgs bosons.

The ambiguous nature of the PQ charges of fermions is not purely academic. In most phenomenological studies of the axion, the starting point is the effective Lagrangian that is obtained by reparametrizing fermion fields to make them PQ neutral (even if that is usually not explicitly stated):
\begin{equation}
\psi\rightarrow\exp(-iPQ(\psi)a^{0}/v_{\phi})\psi\ , \label{ReparamG}%
\end{equation}
where $\psi$ denotes generically the PQ-charged fermions. Since the underlying physics is PQ neutral, this looks innocuous. Yet, it modifies the Lagrangian of the model in two important ways. First, it removes the axion field from Yukawa interactions (both for the SM and heavy fermions, if present), and replaces them by shift-symmetric derivative couplings of the axion to the
fermionic PQ current, as adequate for a Goldstone boson
\begin{equation}
\delta\mathcal{L}_{\text{\textrm{Der}}}=\frac{\partial_{\mu}a^{0}}{v_{\phi}%
}J_{PQ}^{\mu}\ ,\ J_{PQ}^{\mu}=\sum_{\psi}PQ(\psi)\bar{\psi}\gamma^{\mu}\psi\ .
\end{equation}
Second, the PQ symmetry being anomalous, the fermion reparametrizations in Eq.~(\ref{ReparamG}) change the fermionic measure. To account for this, one must introduce anomalous couplings to the gauge bosons,
\begin{equation}
\delta\mathcal{L}_{\text{\textrm{Jac}}}=\frac{a^{0}}{16\pi^{2}v_{\phi}}\left(
g_{s}^{2}\mathcal{N}_{C}G_{\mu\nu}^{a}\tilde{G}^{a,\mu\nu}+g^{2}%
\mathcal{N}_{L}W_{\mu\nu}^{i}\tilde{W}^{i,\mu\nu}+g^{\prime2}\mathcal{N}%
_{Y}B_{\mu\nu}\tilde{B}^{\mu\nu}\frac{{}}{{}}\right)  \;,
\end{equation}
where the coefficients $\mathcal{N}_{C,L,Y}$ are functions of the PQ charges of all the fermions, and generically given by
\begin{equation}
\mathcal{N}_{X}=\sum_{\psi}PQ(\psi)C_{X}(\psi)\ ,
\end{equation}
with $C_{C,L,Y}(\psi)$ the quadratic invariant of the field $\psi$ under $SU(3)_{C}$, $SU(2)_{L}$ or $U(1)_{Y}$. The effective Lagrangian
\begin{equation}
\mathcal{L}_{\text{\textrm{Eff}}}=\delta\mathcal{L}_{\text{\textrm{Jac}}%
}+\delta\mathcal{L}_{\text{\textrm{Der}}}\ , \label{AxionEL}%
\end{equation}
is in general the basis in which the axion phenomenology is studied, with the common further assumption that $\delta\mathcal{L}_{\text{\textrm{Der}}}$ is model-dependent and subleading compared to the model independent $\delta\mathcal{L}_{\text{\textrm{Jac}}}$. Yet, since the PQ charge of the fermions are ambiguous, both $\delta\mathcal{L}_{\text{\textrm{Der}}}$ and $\delta\mathcal{L}_{\text{\textrm{Jac}}}$ are also ambiguous. This is most striking in the DFSZ case, where $\mathcal{N}_{L}\sim3\beta+\gamma$. This conundrum was analyzed in Ref.~~\cite{Quevillon:2019zrd}, where in particular it was shown that $\delta\mathcal{L}_{\text{\textrm{Der}}}$ and $\delta\mathcal{L}_{\text{\textrm{Jac}}}$ do in fact contribute at the same order to physical observables, and that this ensures all the ambiguities in $\delta\mathcal{L}_{\text{\textrm{Der}}}$ and $\delta\mathcal{L}_{\text{\textrm{Jac}}}$ cancel each other systematically. This means that the couplings to (chiral) gauge bosons cannot be read off $\delta\mathcal{L}_{\text{\textrm{Jac}}}$, and that $\delta\mathcal{L}_{\text{\textrm{Der}}}$ cannot be neglected.

For our purpose, it is important to emphasize how this translates for the baryon and lepton numbers. Thus, consider the KSVZ scenario with the fermion charges in Eq.~(\ref{KSVZfermions}), keeping $\alpha$, $\beta$, and $\gamma$ arbitrary, and let us perform the reparametrization of Eq.~(\ref{ReparamG}) for all the fermions. The PQ current is then identified as
\begin{equation}
J_{PQ}^{\mu}=\bar{\Psi}_{R}\gamma^{\mu}\Psi_{R}+\alpha J_{\Psi}^{\mu}+3\beta
J_{\mathcal{B}}^{\mu}+\gamma J_{\mathcal{L}}^{\mu}\ ,
\end{equation}
where%
\begin{align}
J_{\Psi}^{\mu}  &  =\bar{\Psi}_{L}\gamma^{\mu}\Psi_{L}+\bar{\Psi}_{R}%
\gamma^{\mu}\Psi_{R}=\bar{\Psi}\gamma^{\mu}\Psi\ ,\\
J_{\mathcal{B}}^{\mu}  &  =\frac{1}{3}\bar{q}_{L}\gamma^{\mu}q_{L}+\frac{1}%
{3}\bar{u}_{R}\gamma^{\mu}u_{R}+\frac{1}{3}\bar{d}_{R}\gamma^{\mu}d_{R}%
=\frac{1}{3}\bar{u}\gamma^{\mu}u+\frac{1}{3}\bar{d}\gamma^{\mu}d\ ,\ \\
J_{\mathcal{L}}^{\mu}  &  =\bar{\ell}_{L}\gamma^{\mu}\ell_{L}+\bar{e}%
_{R}\gamma^{\mu}e_{R}+\bar{\nu}_{R}\gamma^{\mu}\nu_{R}=\bar{e}\gamma^{\mu
}e+\bar{\nu}\gamma^{\mu}\nu\ .
\end{align}
At first sight, one may think to discard the vector currents $J_{\Psi}^{\mu}$, $J_{\mathcal{B}}^{\mu}$, and $J_{\mathcal{L}}^{\mu}$ from the derivative interactions since upon integration by part, $\partial_{\mu}a^{0}\bar{\psi}\gamma^{\mu}\psi=-a^{0}\partial_{\mu}\bar{\psi}\gamma^{\mu}\psi=-a^{0}\bar{\psi}(m-m)\psi=0$. This is incorrect though. The vector Ward identity does not survive to the presence of chiral gauge interactions. While $J_{\Psi}^{\mu}$ can indeed safely be discarded since $\Psi$ is vector-like, the baryon and lepton currents are anomalous in the presence of chiral gauge fields:
\begin{equation}
\partial_{\mu}J_{\mathcal{B}}^{\mu}=\partial_{\mu}J_{\mathcal{L}}^{\mu}%
=-\frac{N_{f}}{16\pi^{2}}\left(  \frac{1}{2}g^{2}W_{\mu\nu}^{i}\tilde
{W}^{i,\mu\nu}-\frac{1}{2}g^{\prime2}B_{\mu\nu}\tilde{B}^{\mu\nu}\right)  \ .
\end{equation}
Obviously, these contributions trivially cancel the $\beta$ and $\gamma$-dependent Jacobian terms generated by the fermion reparametrization, which have precisely the same form and origin. Thus, in the KSVZ setting, it seems
that the sole role of the SM fermions derivative interactions aligned with the $\mathcal{B}$ and $\mathcal{L}$ current is to kill the correspondingly spurious anomalous gauge interactions.

There is a problem in this reasoning though. This cancellation occurs whether a $\phi^{\dagger}\bar{\nu}_{R}^{\mathrm{C}}\nu_{R}$ coupling is assumed initially present or not, since the value of $\gamma$ is irrelevant. This is puzzling since in the presence of $\phi^{\dagger}\bar{\nu}_{R}^{\mathrm{C}}\nu_{R}$, the axion should retain some couplings to $\nu_{R}$. In the above argument, the step at which we lost the $a\nu_{R}\nu_{R}$ coupling is in the Ward identity. After the spontaneous symmetry breaking (SSB), $\mathcal{L}$, as part of the PQ symmetry, is no longer conserved and the equation of motion (EoM) of $\nu_{R}$ breaks explicitly the anomalous vector Ward identity. In practice, $(\partial_{\mu}a^{0}/v)\bar{\nu}_{R}\gamma^{\mu}\nu_{R}$ does generate the $(M_{R}/v)a^{0}\nu_{R}\nu_{R}$ coupling. This means that whether the axion is coupled to $\nu_{R}$ or not is not apparent at the level of the effective axion Lagrangian, but hides in the EoM of $\nu_{R}$.\ Further, these EoM spoil the $1/v_{\phi}$ scaling of the effective Lagrangian operators, since they contain terms of $\mathcal{O}(v_{\phi})$. Phenomenologically, this failure of the effective interactions to manifestly exhibit all the possible axion interactions is clearly an important point to keep in mind.

To conclude, let us stress again:

\begin{itemize}
\item The PQ symmetry has some room for $\mathcal{B}$ and/or $\mathcal{L}$ violating effects. In the presence of such violation, the PQ symmetry eats part of the $\mathcal{B}$ and $\mathcal{L}$ global $U(1)$s, and the PQ current inherits some $J_{\mathcal{B}}^{\mu}$ and/or $J_{\mathcal{L}}^{\mu}$ components.

\item Incorporating a $\mathcal{B}$ and/or $\mathcal{L}$ component in the PQ current does not modify the leading order axion to gauge boson couplings.

\item The $\mathcal{B}$ and/or $\mathcal{L}$ components of PQ current do not tell us much about the couplings of the axion to SM fermions. Most of the $\partial_{\mu}a^{0}J_{\mathcal{B}}^{\mu}$ and $\partial_{\mu}a^{0}J_{\mathcal{L}}^{\mu}$ couplings are just there to cancel spurious local anomalous terms.

\item Any $\mathcal{B}$ and/or $\mathcal{L}$ violating couplings must break explicitly the (already anomalous) $\mathcal{B}$ and/or $\mathcal{L}$ vector Ward identities. In their presence, the EoM of the SM fermions will ensure the derivative interactions $\partial_{\mu}a^{0}J_{\mathcal{B}}^{\mu}$ and $\partial_{\mu}a^{0}J_{\mathcal{L}}^{\mu}$ do include the expected $\Delta\mathcal{B}$ and/or $\Delta\mathcal{L}$ couplings of the axion.
\end{itemize}

As we will see in the following, introducing leptoquark states often forces us to entangle $\mathcal{B}$ and/or $\mathcal{L}$ with the PQ symmetry. These points are thus crucial to understand the phenomenological consequences.

\subsection{Introducing leptoquarks and diquarks\label{Sec2c}}

Leptoquarks (LQ) are scalars or vectors that couple simultaneously to a quark-lepton pair, while diquarks (DQ) couple to quark pairs (for a review, see e.g. Ref.~\cite{Dorsner:2016wpm}). Given the quantum numbers of the SM fermions, only a finite number of LQ and DQ can couple to normal matter, and only a few of them can have both LQ and DQ couplings. Though the full list of possible LQ and DQ states is well-known, let us nevertheless go through this construction as it will play an important role in the following, and permits to conveniently introduce our notations.

All the LQ are color triplets, while DQ are triplets (using $1\supset 3\otimes3\otimes3$) or sexplets (using $1\supset3\otimes3\otimes\bar{6}$). From the point of view of $SU(2)_{L}$, these states can be either triplet, doublets, or singlets, depending on the involved SM fermions. Once $SU(2)_{L}\otimes SU(3)_{C}$ contractions are set, the hypercharge is then fixed to accommodate specific couplings to SM fermions. In this regard, one should remember that scalars couple to $\bar{\psi}_{L}\psi_{R}$ or $\bar{\psi}_{R}\psi_{L}$, vectors to $\bar{\psi}_{R}\gamma_\mu\psi_{R}$ or $\bar{\psi}_{L}\gamma_\mu\psi_{L}$, and that charge conjugation $\mathrm{C}$ flips the chirality. This means that a scalar can couple to $\bar{\psi}_{R}^{\mathrm{C}}\psi_{R}$ for example. By constructing all possible pairs of SM leptons, including conjugate fields, the standard list of possible states are recovered, with the scalars LQ states
\begin{align}
(\mathbf{3},\mathbf{2},+1/3)  &  :S_{2}^{1/3}\times(\bar{d}_{R}\ell_{L}\ ,\ \bar{q}_{L}\nu_{R})\ ,\nonumber\\
(\mathbf{3},\mathbf{2},+7/3)  &  :S_{2}^{7/3}\times(\bar{u}_{R}\ell_{L}\ ,\ \bar{q}_{L}e_{R})\ ,\nonumber\\
(\mathbf{3},\mathbf{1},-2/3)  &  :S_{1}^{2/3}\times(\bar{d}_{R}\nu_{R}^{\mathrm{C}}\ \ ,\ \bar{u}_{R}e_{R}^{\mathrm{C}}\ ,\ \bar{q}_{L}\ell_{L}^{\mathrm{C}})\ ,\ \ (\mathbf{3},\mathbf{3},-2/3):S_{3}^{2/3}\times\bar
{q}_{L}\ell_{L}^{\mathrm{C}}\ ,\label{LQ1}\\
(\mathbf{3},\mathbf{1},+4/3)  &  :S_{1}^{4/3}\times\bar{u}_{R}\nu_{R}^{\mathrm{C}}\ ,\nonumber\\
(\mathbf{3},\mathbf{1},-8/3)  &  :S_{1}^{8/3}\times\bar{d}_{R}e_{R}^{\mathrm{C}}\ ,\nonumber
\end{align}
and the vector LQ states$\ $%
\begin{align}
(\mathbf{3},\mathbf{2},+1/3)  &  :V_{2,\mu}^{1/3}\times(\bar{u}_{R}\gamma^{\mu}\ell_{L}^{\mathrm{C}}\ ,\ \bar{q}_{L}\gamma^{\mu}\nu_{R}^{\mathrm{C}})\ ,\nonumber\\
(\mathbf{3},\mathbf{2},-5/3)  &  :V_{2,\mu}^{5/3}\times(\bar{d}_{R}\gamma^{\mu}\ell_{L}^{\mathrm{C}}\ ,\ \bar{q}_{L}\gamma^{\mu}e_{R}^{\mathrm{C}})\ ,\nonumber\\
(\mathbf{3},\mathbf{1},+4/3)  &  :V_{1,\mu}^{4/3}\times(\bar{u}_{R}\gamma^{\mu}\nu_{R}\ ,\ \bar{d}_{R}\gamma^{\mu}e_{R}\ ,\ \bar{q}_{L}\gamma^{\mu}\ell_{L})\ ,\ \ (\mathbf{3},\mathbf{3},+4/3):V_{3,\mu}^{4/3}\times\bar{q}%
_{L}\gamma^{\mu}\ell_{L}\ ,\label{LQ2}\\
(\mathbf{3},\mathbf{1},10/3)  &  :V_{1,\mu}^{10/3}\times\bar{u}_{R}\gamma^{\mu}e_{R}\ ,\nonumber\\
(\mathbf{3},\mathbf{1},-2/3)  &  :V_{1,\mu}^{2/3}\times\bar{d}_{R}\gamma^{\mu}\nu_{R}\ .\nonumber
\end{align}
Many notations exist for these states, in particular $S_{i}$, $\tilde{S}_{i}$, $\bar{S}_{i}$ when several states occur with the same $SU(3)_{C}\otimes SU(2)_{L}$ quantum numbers~\cite{Dorsner:2016wpm}. Here, we denote all states as color triplets $S_{t}^{y}$ or $V_{t}^{y}$, with $t$ the $SU(2)_{L}$ dimensionality and $y$ the absolute value of the $U(1)_{Y}$ hypercharge. Note also that $V_{1,\mu}^{2/3}$ and $S_{1}^{4/3}$ exist only in the presence of $\nu_{R}$, and are thus often discarded. Concerning diquarks, there are only six possible combinations of quark fields, leading to%
\begin{align}
(\mathbf{3},\mathbf{2},+1/3)  &  :V_{2,\mu}^{1/3}\times\bar{d}_{R}^{\mathrm{C}}\gamma^{\mu}q_{L}\ ,\nonumber\\
(\mathbf{3},\mathbf{2},-5/3)  &  :V_{2,\mu}^{5/3}\times\bar{u}_{R}^{\mathrm{C}}\gamma^{\mu}q_{L}\ ,\nonumber\\
(\mathbf{3},\mathbf{1},-2/3)  &  :S_{1}^{2/3}\times(\bar{q}_{L}^{\mathrm{C}}q_{L}\ ,\ \bar{d}_{R}^{\mathrm{C}}u_{R})\ ,\ (\mathbf{3},\mathbf{3},-2/3):S_{3}^{2/3}\times\bar{q}_{L}^{\mathrm{C}}q_{L}\ ,\label{LQ3}\\
(\mathbf{3},\mathbf{1},+4/3)  &  :S_{1}^{4/3}\times\bar{d}_{R}^{\mathrm{C}}d_{R}\ ,\nonumber\\
(\mathbf{3},\mathbf{1},-8/3)  &  :S_{1}^{8/3}\times\bar{u}_{R}^{\mathrm{C}}u_{R}\ .\nonumber
\end{align}
All these states are already present in the LQ list. Note that each of the above quark state can also couple to a DQ transforming like $\mathbf{\bar{6}}$ under $SU(3)_{C}$, with the same $SU(2)_{L}\otimes U(1)_{Y}$ quantum numbers. In that case, they do not have LQ couplings. We will adopt the same notation for these states, relying on the context to make clear whether they transform as $\mathbf{3}$ or $\mathbf{\bar{6}}$.

\begin{table}[t] \centering
\begin{tabular}[c]{l}
\begin{tabular}[c]{lllllll}\hline
$\Delta\mathcal{B}$ & $\Delta\mathcal{L}$ & Dim. &
\multicolumn{2}{l}{Operators (no $\nu_{R}$)} &  & \\\hline
$+0$ & $+2$ & $5$ & $H^{\dagger2}\ell_{L}^{2}$ &  &  & \\
$+1$ & $+1$ & $6$ & $q_{L}^{3}\ell_{L}$ & $u_{R}^{2}d_{R}e_{R}$ & $q_{L}%
u_{R}d_{R}\ell_{L}$ & $q_{L}^{2}u_{R}e_{R}$\\
$+1$ & $-1$ & $7$ & $H^{\dagger}d_{R}^{3}\ell_{L}^{\mathrm{C}}$ & $Hd_{R}%
^{2}q_{L}e_{R}^{\mathrm{C}}$ & $Hd_{R}^{2}u_{R}\ell_{L}^{\mathrm{C}}$ &
$Hq_{L}^{2}d_{R}\ell_{L}^{\mathrm{C}}$\\
$+2$ & $+0$ & $9$ & $d_{R}^{4}u_{R}$ & $d_{R}^{3}u_{R}q_{L}^{2}$ & $d_{R}%
^{2}q_{L}^{4}$ & \\
$+1$ & $+3$ & $9$ & $u_{R}^{2}q_{L}\ell_{L}^{3}$ & $u_{R}^{3}\ell_{L}^{2}%
e_{R}$ &  & \\
$+1$ & $-3$ & $10$ & $Hd_{R}^{3}\ell_{L}^{\mathrm{C},3}$ &  &  &
\end{tabular}
\\
\begin{tabular}[c]{llllllll}\hline
$\Delta\mathcal{B}$ & $\Delta\mathcal{L}$ & Dim. &
\multicolumn{2}{l}{Operators (one $\nu_{R}$)} &  &  & \\\hline
$+0$ & $+2$ & $5$ & $H^{\dagger2}e_{R}\nu_{R}$ &  &  &  & \\
$+1$ & $+1$ & $6$ & $q_{L}^{2}d_{R}\nu_{R}$ & $d_{R}^{2}u_{R}\nu_{R}$ &  &  &
\\
$+1$ & $-1$ & $7$ & $H^{\dagger}d_{R}^{2}q_{L}\nu_{R}^{\mathrm{C}}$ &
$Hd_{R}q_{L}u_{R}\nu_{R}^{\mathrm{C}}$ & $Hq_{L}^{3}\nu_{R}^{\mathrm{C}}$ &  &
\\
$+2$ & $+0$ & $9$ & -- &  &  &  & \\
$+1$ & $+3$ & $9$ & $d_{R}u_{R}^{2}\ell_{L}^{2}\nu_{R}$ & $d_{R}q_{L}u_{R}%
\ell_{L}^{2}\nu_{R}$ & $u_{R}^{3}e_{R}^{2}\nu_{R}$ & $u_{R}^{2}q_{L}\ell
_{L}e_{R}\nu_{R}$ & $q_{L}^{2}u_{R}\ell_{L}^{2}\nu_{R}$\\
$+1$ & $-3$ & $10$ & $Hd_{R}^{3}\ell_{L}^{\mathrm{C}}e_{R}^{\mathrm{C}}\nu
_{R}^{\mathrm{C}}$ & $Hd_{R}^{2}q_{L}\ell_{L}^{\mathrm{C},2}\nu_{R}%
^{\mathrm{C}}$ &  &  & \\\hline
\end{tabular}
$\ $
\end{tabular}
$\ $
\caption{Leading effective operators with non-trivial $(\Delta\mathcal{B},\Delta\mathcal{L})$ charges in the SM, involving no or one $\nu_R$ field.
We do not include redundant patterns, e.g. all the $(\Delta\mathcal{B},\Delta\mathcal{L})=n\times(0,2),n\times
(1,1),...$ with $n=2,3,...$ operators, or operators of higher dimensions within each $(\Delta\mathcal{B},\Delta\mathcal
{L})$ class. With even more fields, the next unique patterns involve eight fermions, and induce $(\Delta\mathcal{B},\Delta\mathcal{L})=(1,5)$ transitions at dimension 12, and $(\Delta\mathcal{B},\Delta\mathcal{L})=(1,-5)$ transitions at dimension 13 (with an extra Higgs field). All these processes involve at least one $\nu
_{R}$ field at these orders. Still higher in dimensionality, $(\Delta\mathcal{B},\Delta\mathcal{L})=(3,1)$ and $(\Delta\mathcal{B},\Delta\mathcal{L})=(1,7)$ come at the ten-fermion level, via dimension-15 operators. Only the SM Higgs doublet $H$ is used in the Table together with SM fermions, but the extension to the two-Higgs-doublet model is trivial.}%
\label{TableLQBL}
\end{table}

\begin{figure}[ptb]
\begin{center}
\includegraphics[height=2.3134in,width=5.3195in]{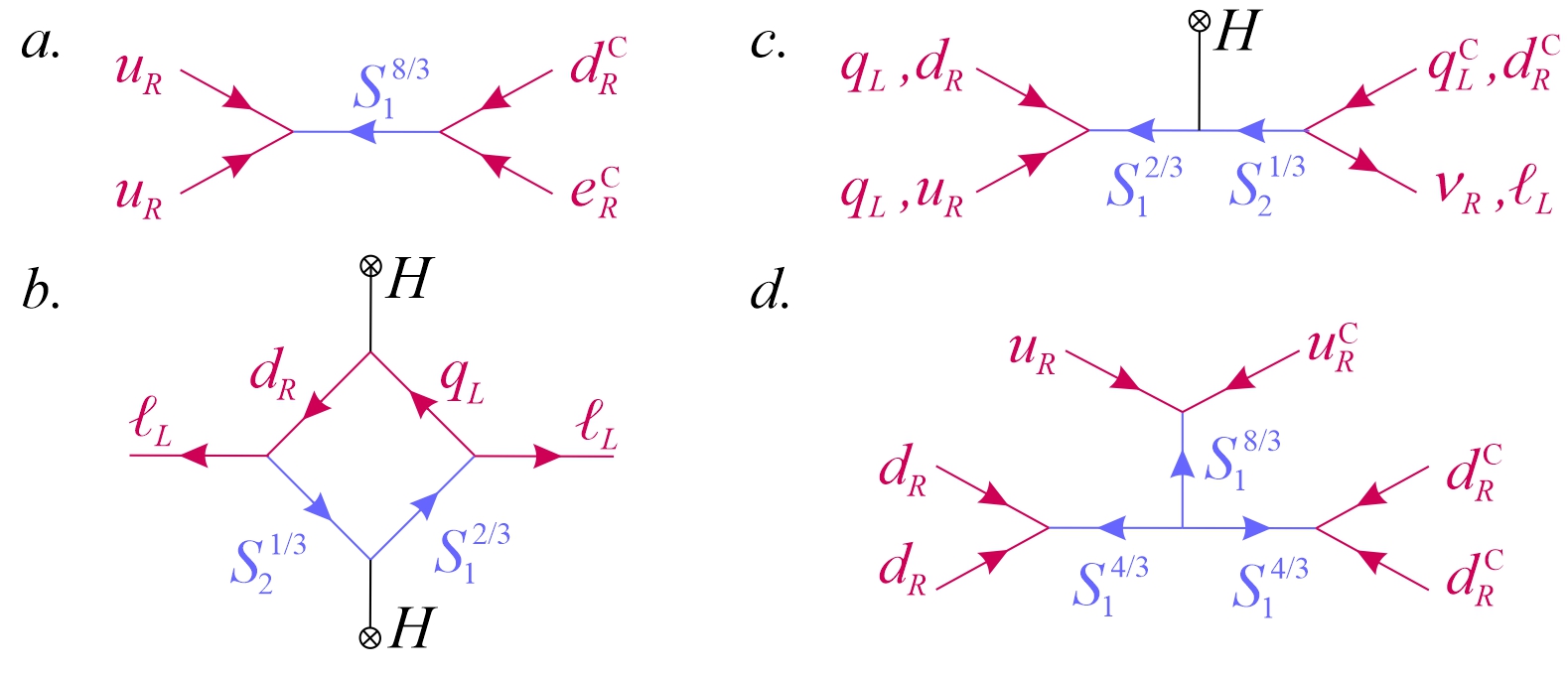}
\caption{LQ/DQ processes inducing proton decay via $(\Delta\mathcal{B},\Delta\mathcal{L})=(1,1)$ operators ($a.$), a neutrino Majorana mass term ($b.$), proton decay via $(\Delta\mathcal{B},\Delta\mathcal{L})=(1,-1)$ operators ($c.$), and neutron-antineutron oscillations via a $(\Delta\mathcal{B},\Delta\mathcal{L})=(2,0)$ operator ($d.$).}%
\label{Fig1}
\end{center}
\end{figure}

Introducing scalar or vector states that couple to quarks and leptons can impact the global $\mathcal{B}$ and $\mathcal{L}$ symmetries (for a recent review, see e.g. Ref.~\cite{Assad:2017iib}). Depending on which states are introduced and, if several of them are present, depending also on how they are coupled, the symmetry pattern can be quite different. Actually, these symmetry patterns are reminiscent of those of the possible effective operators involving SM fields but carrying non-trivial $\mathcal{B}$ and/or $\mathcal{L}$ charges~\cite{Weinberg,WeinbergPRD22,Weldon:1980gi}. Those are listed in Table~\ref{TableLQBL}. This connection is easily understood from tree diagrams with the external fermions linked together by virtual LQ/DQ exchanges\footnote{The notation LQ/DQ generically refers to any of the pure LQ, pure DQ, or mixed LQ/DQ state introduced in Eqs.~(\ref{LQ1}),~(\ref{LQ2}), and~(\ref{LQ3}).}. Obviously, these external fermion states must be $SU(3)_{C}\otimes SU(2)_{L}\otimes U(1)_{Y}$ invariant since the LQ/DQ are. Further, operators with six or less fermions are the most relevant when only renormalizable interactions among the LQ/DQ are present. Being colored, these states can at most have quadratic or cubic interactions, hence induce four or six fermion interactions. More complicated fermion interactions can arise, but they would require multiple cubic interactions, and would not open additional phenomenologically interesting channels. Indeed, the above set contains already the $(\Delta\mathcal{B},\Delta\mathcal{L})=(0,2)$ operators for neutrino masses, $(\Delta\mathcal{B},\Delta\mathcal{L})=(2,0)$ operators for neutron-antineutron oscillations, and all the others for proton decay. Note, finally, that one can understand why some states have both LQ and DQ couplings while others do not from the fact that dimension-six operators are necessarily $(\Delta\mathcal{B},\Delta\mathcal{L})=(1,1)$, see Table~\ref{TableLQBL}. As tree-level exchanges of states with both LQ and DQ couplings (Fig.~\ref{Fig1}$a$) must match onto these operators, only $V_{2}^{y}$ and $S_{1}^{y}$ can occur since they couple to a quark-lepton (or antiquark-antilepton) pair\footnote{This condition is sometimes quantified using $\mathcal{F}=3\mathcal{B}+\mathcal{L}$ as a quantum numbers~\cite{Dorsner:2016wpm}, so that those states with both LQ and DQ couplings have $\mathcal{F}=\pm2$, and the others $\mathcal{F}=0$. We prefer here to use $\mathcal{B}\pm\mathcal{L}$.}.

With the above picture in mind, let us see in more details how the various
$(\Delta\mathcal{B},\Delta\mathcal{L})$ patterns of Table~\ref{TableLQBL} can arise. Note that most of the following mechanisms have already been described elsewhere, see for instance Refs.~\cite{Klapdor-Kleingrothaus:2002rvk,Kovalenko:2002eh,Arnold:2012sd,FileviezPerez:2015mlm,Dorsner:2022twk}, but this is repeated here in some details as it constitutes the basis for the discussions in the next sections, where these patterns will be induced spontaneously.

\begin{enumerate}
\item[A.] Exact $U(1)_{\mathcal{B}}\otimes U(1)_{\mathcal{L}}:$ Whenever a given $S$ or $V$ state with only LQ or DQ coupling is present, $\mathcal{B}$ and $\mathcal{L}$ can still be unambiguously defined. The LQ or DQ state simply carries some specific $\mathcal{B}$ and $\mathcal{L}$ quantum numbers, but overall, $U(1)_{\mathcal{B}}\otimes U(1)_{\mathcal{L}}$ is still exact. This remains true even in the presence of several different states, so long as they do not couple together.

\item[B.] Exact $U(1)_{\mathcal{B}-\mathcal{L}}:$ When a state with both LQ and DQ couplings is present, the symmetry gets reduced to $U(1)_{\mathcal{B}-\mathcal{L}}$, with the $\mathcal{B}-\mathcal{L}$ quantum numbers $-2/3$ for $S_{1}^{y}$ and $V_{2}^{y}$, $+1/3$ and $-1$ for quarks and leptons, respectively. This remains true if more than one DQ/LQ state is present provided any couplings among them is compatible with these charge assignments, which further requires the $\mathcal{B}-\mathcal{L}$ quantum numbers of $S_{2}^{y}$ and $V_{1}^{y}$ to be $+4/3$. For example, a scenario with $S_{1}^{2/3}$ and $S_{1}^{4/3}$ but without an $S_{1}^{2/3}S_{1}^{2/3}S_{1}^{4/3}$ interaction, or with $S_{2}^{7/3}$, $S_{1}^{2/3}$ and a coupling $H^{\dagger}S_{2}^{7/3}S_{1}^{2/3}S_{1}^{2/3}$, or with $S_{2}^{1/3}$, $S_{1}^{2/3}$ and a coupling $HS_{2}^{1/3}S_{1}^{2/3}S_{1}^{2/3}$ all preserve $U(1)_{\mathcal{B}-\mathcal{L}}$ (note that the antisymmetric color contraction requires at least two different $S_{1}^{2/3}$). For all these scenarios, the $S$ and/or $V$ mass has to be pushed at the GUT scale since $(\Delta\mathcal{B},\Delta\mathcal{L})=(1,1)$ operators induce proton decay (Fig.~\ref{Fig1}$a$).

\item[C.] No exact $U(1):$ In the presence of two states having different $\mathcal{B}-\mathcal{L}$ quantum numbers, there is no remaining global symmetry whenever those states have all their gauge-allowed couplings to SM fermions turned on, and when they are coupled together. For example, introducing both $S_{2}^{1/3}$ and $S_{1}^{2/3}$ with a $\mu HS_{2}^{1/3\dagger}S_{1}^{2/3}$ coupling, $U(1)_{\mathcal{B}}$ and $U(1)_{\mathcal{L}}$ are entirely broken. As seen earlier, $(\Delta\mathcal{B},\Delta\mathcal{L})=(1,1)$ proton decay is induced by $S_{1}^{2/3}$, pushing its mass to the GUT range. But the total absence of global $U(1)$s means the other classes of $(\Delta\mathcal{B},\Delta\mathcal{L})$ operators are also generated. The simplest is the $(\Delta\mathcal{B},\Delta\mathcal{L})=(0,2)$ operator, generating neutrino masses via the diagram of Fig.~\ref{Fig1}$b$.

\item[D.] Exact $U(1)_{\mathcal{B}}:$ Adding to the scenarios A a seesaw mechanism for neutrino masses, i.e., a $\bar{\nu}_{R}^{\mathrm{C}}\nu_{R}$ term, then $U(1)_{\mathcal{L}}$ is explicitly broken but $U(1)_{\mathcal{B}}$ remains exact, preventing proton decay. The same pattern can be obtained using mixing terms among some carefully chosen LQ/DQ states, such that an effective neutrino mass term is generated but proton decay cannot occur. For example, introducing $S_{2}^{1/3}$, $S_{1}^{2/3}$, the mixing term $\mu HS_{2}^{1/3\dagger}S_{1}^{2/3}$ but turning off the DQ couplings of $S_{1}^{2/3}$ (or alternatively, with the mixing term $\mu S_{2}^{1/3}S_{2}^{1/3}S_{1}^{2/3}$ but turning off the LQ couplings of $S_{1}^{2/3}$), the dimension-five $(\Delta\mathcal{B},\Delta\mathcal{L})=(0,2)$ operator arises, see Fig.~\ref{Fig1}$b$. In these scenarios, $S_{2}^{1/3}$, $S_{1}^{2/3}$ acquire well defined $\mathcal{B}$ numbers, $U(1)_{\mathcal{B}}$ is conserved, and proton decay is forbidden.

\item[E.] Exact $U(1)_{\mathcal{B}+\mathcal{L}}:$ Another possible symmetry pattern corresponds to taking again $S_{2}^{1/3}$, $S_{1}^{2/3}$, and the $\mu HS_{2}^{1/3\dagger}S_{1}^{2/3}$ coupling but turning off the LQ couplings of $S_{1}^{2/3}$ (or with $\mu S_{2}^{1/3}S_{2}^{1/3}S_{1}^{2/3}$ but turning off the DQ couplings of $S_{1}^{2/3}$). In this case, no neutrino masses can be generated, but proton decay is back. Yet, the proton decay channels do not match those induced by the dimension-six Weinberg operators. With the $\mu HS_{2}^{1/3\dagger}S_{1}^{2/3}$ coupling, the simplest processes lead to the dimension-seven $(\Delta\mathcal{B},\Delta\mathcal{L})=(1,-1)$ effective operators, see Fig.~\ref{Fig1}$c$, while the $\mu S_{2}^{1/3}S_{2}^{1/3}S_{1}^{2/3}$ coupling generates $(\Delta\mathcal{B},\Delta\mathcal{L})=(1,-1)$ transitions but with an extra lepton-antilepton pair.

\item[F.] Exact $U(1)_{\mathcal{L}}:$ Another pattern is obtained by introducing several states but now allowing only for DQ couplings, and turning on some mixing terms (this kind of construction was considered recently e.g. in Refs.~\cite{Arnold:2012sd,FileviezPerez:2015mlm}). These latter mixings are necessary since otherwise, $U(1)_{\mathcal{B}}\otimes U(1)_{\mathcal{L}}$ remains exact. The simplest scenarios are those with $S_{1}^{2/3}$, $S_{1}^{4/3}$, and the cubic coupling $\mu S_{1}^{2/3}S_{1}^{2/3}S_{1}^{4/3}$, or $S_{1}^{4/3}$, $S_{1}^{8/3}$, and the cubic coupling $\mu S_{1}^{4/3}S_{1}^{4/3}S_{1}^{8/3}$. In both cases, only the DQ couplings are allowed, and $S_{1}^{4/3}$ ($S_{1}^{8/3}$) must transform as $\mathbf{\bar{6}}$ in the first (second) case, respectively. As a result, neither neutrino masses nor proton decay are induced, but the dimension-nine $(\Delta\mathcal{B},\Delta\mathcal{L})=(2,0)$ operators do arise, and contribute to neutron-antineutron oscillations, see Fig.~\ref{Fig1}$d$.

\item[G.] Exact $U(1)_{3\mathcal{B}-\mathcal{L}}:$ As for the $(\Delta\mathcal{B},\Delta\mathcal{L})=(2,0)$ case, dimension-nine $(\Delta\mathcal{B},\Delta\mathcal{L})=(1,3)$ operators are attainable by taking $S_{1}^{2/3}$, $S_{1}^{4/3}$, and the cubic coupling $\mu S_{1}^{2/3}S_{1}^{2/3}S_{1}^{4/3}$, or $S_{1}^{4/3}$, $S_{1}^{8/3}$, and the cubic coupling $\mu S_{1}^{4/3}S_{1}^{4/3}S_{1}^{8/3}$, but turning on only the LQ couplings (since all LQ transform as $\mathbf{3}$, the color contraction requires three different LQ to be present). Yet, only interactions involving $\nu_{R}$ can occur because of the LQ coupling of $S_{1}^{4/3}$ to $\bar{u}_{R}\nu_{R}^{\mathrm{C}}$, so proton decay is suppressed. The dimension-nine $(\Delta\mathcal{B},\Delta\mathcal{L})=(1,3)$ operators not involving $\nu_{R}$ require a combination of scalar and vector LQ, for example $S_{1}^{2/3}V_{2}^{1/3}V_{2}^{1/3}$ can induce both $\bar{q}_{L}\ell_{L}^{\mathrm{C}}\bar{u}_{R}\gamma_{\mu}\ell_{L}^{\mathrm{C}}\bar{u}_{R}\gamma^{\mu}\ell_{L}^{\mathrm{C}}$ and $\bar{u}_{R}e_{R}^{\mathrm{C}}\bar{u}_{R}\gamma_{\mu}\ell_{L}^{\mathrm{C}}\bar{u}_{R}\gamma^{\mu}\ell_{L}^{\mathrm{C}}$.

\item[H.] Exact $U(1)_{3\mathcal{B}+\mathcal{L}}:$ While the previous two patterns rely on cubic interactions among the $S_{1}^{y}$ and $V_{2}^{y}$ states, this pattern rather needs to involve only the $S_{2}^{y}$ and $V_{1}^{y}$ states. Furthermore, since the Higgs field appears in the six-fermion $(\Delta\mathcal{B},\Delta\mathcal{L})=(1,-3)$ proton decay operators of Table~\ref{TableLQBL}, the simplest mechanisms should be based on a quartic coupling H-LQ-LQ-LQ. At first sight, the simplest would be the $H^{\dagger}S_{2}^{1/3}S_{2}^{1/3}S_{2}^{1/3}$ coupling, but the antisymmetric color contraction vanishes identically since $S_{2}^{1/3}$ has only two $SU(2)_L$ degrees of freedom~\cite{Crivellin:2021ejk}. The simplest mechanism then necessarily involves either two different $S_{2}^{1/3}$ states, or both scalar and vector LQ, in which case three different LQ states must be introduced.
\end{enumerate}

This concludes our list of symmetry patterns. It is quite remarkable that a relatively simple scenario exists for all the possible patterns of Table~\ref{TableLQBL}, with in each case the `orthogonal' $(\Delta\mathcal{B},\Delta\mathcal{L})$ pattern remaining as an exact global $U(1)$ symmetry. What this list does not show is that actually, not so many other scenarios do exist to generate most of these symmetry-breaking patterns. Indeed, in most cases, allowing for several LQ/DQ states, both scalar and vector, and some couplings among them, one simply ends up with no global symmetries. The interesting situations in which some global symmetries do remain are quite constrained, and those can be classified once and for all.

First, notice that at the renormalizable level, there are only two classes of couplings among the LQ/DQ: those with bilinear color contractions, typically $\mathbf{3}\otimes\mathbf{\bar{3}}$ or $\mathbf{6}\otimes\mathbf{\bar{6}}$, and those with cubic contractions, typically $\mathbf{3}\otimes\mathbf{3}\otimes\mathbf{3}$ or $\mathbf{3}\otimes\mathbf{3}\otimes\mathbf{\bar{6}}$. For the former, barring partial derivatives acting on the LQ/DQ fields, the only non-trivial LQ/DQ bilinear couplings compatible with the SM gauge symmetries are%
\begin{equation}
HS_{2}^{1/3\dagger}S_{1}^{2/3}\ ,\ HS_{1}^{4/3\dagger}S_{2}^{1/3}%
\ ,\ HS_{2}^{7/3\dagger}S_{1}^{4/3}\ ,\ HV_{2,\mu}^{1/3\dagger}V_{1}^{2/3,\mu
}\ ,\ HV_{1,\mu}^{4/3\dagger}V_{2}^{1/3,\mu}\ ,\ HV_{1,\mu}^{2/3\dagger}%
V_{2}^{5/3,\mu}\ . \label{LQOpsBmL}
\end{equation}
The $HS_{2}^{1/3\dagger}S_{1}^{2/3}$ coupling was used to illustrate the symmetry patterns, but all the others are completely similar: $U(1)_{\mathcal{B}}$ and $U(1)_{\mathcal{L}}$ are entirely broken when all the LQ/DQ couplings are present (case C), $U(1)_{\mathcal{B}}$ stays exact with only LQ couplings (case D), or $U(1)_{\mathcal{B}+\mathcal{L}}$ remains if $S_{1}^{y}$ or $V_{2}^{y}$ have only DQ couplings (case E). This last situation is probably the most interesting phenomenologically since each coupling in Eq.~(\ref{LQOpsBmL}) produces a specific subset of the dimension-seven $(\Delta\mathcal{B},\Delta\mathcal{L})=(1,-1)$ operators in Table~\ref{TableLQBL}.

For cubic interactions, though there are a total of $37$ such couplings, most of them involve LQ/DQ of different $\mathcal{B}-\mathcal{L}$ charges and conserve either $U(1)_{\mathcal{B}-\mathcal{L}}$ (case B) or $U(1)_{\mathcal{B}+\mathcal{L}}$ (case E). Yet, compared to the dimension 6 and 7 operators in Table~\ref{TableLQBL}, they necessarily produce an extra lepton-antilepton pair. The symmetry patterns typical of six-fermion states, i.e., leading to the dimension 9 or 10 operators in Table~\ref{TableLQBL}, are obtained with three LQ/DQ with the same $\mathcal{B}-\mathcal{L}$ charge, and this leaves only eight possibilities:
\begin{align}
&  S_{1}^{2/3}V_{2,\mu}^{1/3}V_{2}^{1/3,\mu}\ ,\ S_{1}^{4/3}V_{2,\mu}%
^{1/3}V_{2}^{5/3,\mu}\ ,\ S_{1}^{2/3}S_{1}^{2/3}S_{1}^{4/3}\ ,\ S_{1}%
^{4/3}S_{1}^{4/3}S_{1}^{8/3}\ ,\label{LQOpsB2}\\
&  H^{\dagger}S_{2}^{\prime1/3}S_{2}^{1/3}S_{2}^{1/3}\ ,\ H^{\dagger}S_{2}%
^{1/3}V_{1,\mu}^{2/3}V_{1}^{4/3,\mu}\ ,\ H^{\dagger}S_{2}^{7/3}V_{1,\mu
}^{\prime2/3}V_{1}^{2/3,\mu}\ ,\ HS_{2}^{1/3}V_{1,\mu}^{\prime2/3}%
V_{1}^{2/3,\mu}\ . \label{LQOpsBL3}%
\end{align}
The scenarios in the first line lead to $(\Delta\mathcal{B},\Delta\mathcal{L})=(2,0)$ or $(\Delta\mathcal{B},\Delta\mathcal{L})=(1,3)$ operators (case F and G), and those in the second line to $(\Delta\mathcal{B},\Delta\mathcal{L})=(1,-3)$ operators (case H). Note that for $(\Delta\mathcal{B},\Delta\mathcal{L})=(1,\pm3)$ transitions, the LQ must transform as $\mathbf{3}$, and the color contraction is necessarily antisymmetric. To get a non-vanishing coupling, one of the three $S_{2}^{1/3}$ is primed in the first operator, while one of the two $V_{1,\mu}^{2/3}$ fields is primed in the last two operators of Eq.~(\ref{LQOpsBL3}). This does not apply to $(\Delta\mathcal{B},\Delta\mathcal{L})=(2,0)$ operators, for which it is always possible to take one of the DQ to transform as a symmetric $\mathbf{\bar{6}}$. As a final remark, it should be noted that scalar or vector color-singlet dileptons could also be introduced, opening the door to quartic couplings among the new states, and correspondingly, to eight-fermion $(\Delta\mathcal{B},\Delta\mathcal{L})=(2,\pm2)$ operators~\cite{Helset:2021plg} (see also Ref.~\cite{He:2021sbl} where similar operators are obtained by imposing an additional discrete symmetry on the LQ couplings). This will not be considered here.

Throughout this paper, when estimating bounds on LQ/DQ masses from proton decay or neutron-antineutron oscillations, the LQ/DQ couplings to SM fermions is assumed flavor universal, or at the very least non-hierarchical in flavor space. As was shown in Ref.~\cite{MFVBandL}, this is a strong assumption for $\mathcal{B}$ and/or $\mathcal{L}$ violating operators. The $SU(3)^{5}$ flavor
symmetry would ask instead for a strong hierarchy because of the systematic presence of the three quark generations in all the operators in Table~\ref{TableLQBL}. In the present context, such hierarchies would first require LQ/DQ to carry flavor quantum numbers, and then to extend the minimal flavor violating formalism to the LQ/DQ sector~\cite{Davidson:2010uu}. This will not be analyzed here, but such kind of flavor suppression should be kept in mind, especially given the context in B physics. There, a number of puzzles in leptonic and semileptonic decays can be explained by introducing new LQ states with particular flavor hierarchies (for a recent review, see e.g. Ref.~\cite{LQreview}). Typically, the favored LQ is $V_{1,\mu}^{4/3}\sim(\mathbf{3},\mathbf{1},+4/3)$ thanks to its $q_{L}\gamma^{\mu}\ell_{L}$ couplings, but other states could also occur in principle. The connection of some of these models with axions has been investigated e.g. in Ref.~\cite{Fuentes-Martin:2019bue} (for some considerations of axions in the context of the B physics anomalies see e.g. \cite{Baek:2020ovw}, whereas axions in a more broad flavor context have also been studied in Refs.~\cite{Ema:2016ops,Calibbi:2016hwq,Arias-Aragon:2017eww,Bonnefoy:2020llz}, but to our knowledge, no systematic studies has been performed yet. In the present paper, our goal is mainly to analyze symmetry breaking patterns involving both LQ/DQ and axions, so the LQ/DQ couplings to SM fermions will simply be assumed $\mathcal{O}(1)$ for all flavors whenever deriving bounds on their masses. Turning on non-trivial flavor structures is left for future studies.

\section{Coupling axions to leptoquarks and diquarks\label{SecAxLQDQ}}

In the previous section, we have established the possible global symmetries in the presence of LQ and DQ states. Here, we want to add to these scenarios a KSVZ or DFSZ sector. The consequences are rather different for both models, since the SM fermions can be PQ neutral in the former case, but not in the latter. Yet, so long as the $\phi$ (and the heavy KSVZ fermions $\Psi_{L,R}$) are not directly coupled to the LQ/DQ states, the axion stays rather insensitive to the possible $\mathcal{B}$ and/or $\mathcal{L}$ violation.

To illustrate this, consider the KSVZ scenario. Without direct couplings of $\phi$ or $\Psi_{L,R}$ to the LQ/DQ states, the $U(1)_{\phi}$ symmetry stays separate from $U(1)_{\mathcal{B},\mathcal{L}}$, so the PQ breaking proceeds trivially as
\begin{equation}
U(1)_{\phi}\otimes U(1)_{\mathcal{B}}\otimes U(1)_{\mathcal{L}}\overset
{\text{Explicit}}{\rightarrow}U(1)_{\phi}\otimes U(1)_{X}\simeq U(1)_{PQ}%
\otimes U(1)_{X}\overset{\text{Spontaneous}}{\rightarrow}U(1)_{X}\ ,\ \
\end{equation}
The specific LQ/DQ scenario fixes which global symmetry, 
\begin{equation}
U(1)_{X}=U(1)_{\mathcal{B}}\otimes U(1)_{\mathcal{L}},\ U(1)_{\mathcal{B}%
\pm\mathcal{L}}\ ,U(1)_{\mathcal{B}}\ ,\ U(1)_{\mathcal{L}}%
,\ U(1)_{3\mathcal{B}\pm\mathcal{L}}\ ,...\ , \label{SurvivingU1}%
\end{equation}
survives, by introducing couplings that explicitly break $U(1)_{\mathcal{B}}\otimes U(1)_{\mathcal{L}}\backslash U(1)_{X}$. Yet, the axion does not break $U(1)_{\mathcal{B}}\otimes U(1)_{\mathcal{L}}$ or $U(1)_{X}$, only the dynamics of the SM and LQ/DQ fields does. Of course, the axion being coupled to SM gauge fields and SM fermions, it does end up coupled to leptoquarks and possibly acquires some $\mathcal{B}$ and/or $\mathcal{L}$ violating decay channels, but this is indirect. A good example for this situation is the KSVZ model with a Majorana mass $M_{R}\nu_{R}\nu_{R}$. The Majorana mass term explicitly breaks $U(1)_{\mathcal{L}}$ at all scale, but such that $U(1)_{X}=U(1)_{\mathcal{B}}$ stays exact at all scales. Clearly, the axion dynamics does not break $U(1)_{\mathcal{L}}$, only neutrino masses do. Thus, any $\Delta\mathcal{L}=2$ effect would come indirectly, e.g. as in $a^{0}\rightarrow\bar{\nu}_{R}\nu_{L}\rightarrow\nu_{R}\nu_{L}$. The situation in the DFSZ scenario is similar, though the $U(1)_{PQ}$ arises from a specific combination of $U(1)_{\phi}$ and $U(1)_{Y}$, see Eq.~(\ref{DFSZScalars}). This situation also corresponds to that often found in simple GUT models. For example, in $SU(5)$, gauge interactions break $U(1)_{\mathcal{B}}\otimes U(1)_{\mathcal{L}}$ down to $U(1)_{\mathcal{B}-\mathcal{L}}$ independently of the axion field (for a detailed account of how the PQ, $\mathcal{B}$, and $\mathcal{L}$ symmetries are entangled in the $SU(5)$ setting, see Ref.~\cite{Quevillon:2020aij}).

Our goal is to consider situations in which the symmetry above the PQ scale entangles $U(1)_{\phi}$ within $U(1)_{\mathcal{B}}\otimes U(1)_{\mathcal{L}}$. Breaking $U(1)_{\phi}$ spontaneously then means breaking a linear combination of $\mathcal{B}$ and $\mathcal{L}$ (or both) spontaneously. Taking again the KSVZ scenario for illustration, this is accomplished by introducing some set of couplings that are only invariant under a subgroup of $U(1)_{\phi}\otimes U(1)_{\mathcal{B}}\otimes U(1)_{\mathcal{L}}$. In most cases of interests, $U(1)_{\mathcal{B}}\otimes U(1)_{\mathcal{L}}$ stays active at the high scale, but $\phi$ carries some definite $\mathcal{B}$ and/or $\mathcal{L}$ quantum numbers, so that the breaking chain becomes
\begin{equation}
U(1)_{\phi}\otimes U(1)_{\mathcal{B}}\otimes U(1)_{\mathcal{L}}\overset
{\text{Explicit}}{\rightarrow}U(1)_{\mathcal{B}}\otimes U(1)_{\mathcal{L}%
}\simeq U(1)_{PQ}\otimes U(1)_{X}\overset{\text{Spontaneous}}{\rightarrow
}U(1)_{X}\ .
\end{equation}
The simplest example illustrating this situation is the KSVZ model with the $\phi^{\dagger}\bar{\nu}^\mathrm{C}_{R}\nu_{R}$ couplings, so that $\phi$ becomes a $(\mathcal{B},\mathcal{L})=(0,2)$ state, $U(1)_{PQ}=U(1)_{\mathcal{L}}$ is spontaneously broken, but $U(1)_{X}=U(1)_{\mathcal{B}}$ stays exact. Compared to the previous case, the main difference is that the axion has a $\Delta\mathcal{L}=2$ coupling $a^{0}\rightarrow\nu_{R}\nu_{R}$ of $\mathcal{O}(1)$. Of course, phenomenologically, whether one adds $M_{R}\bar{\nu}^\mathrm{C}_{R}\nu_{R}$ or $\phi^{\dagger}\bar{\nu}^\mathrm{C}_{R}\nu_{R}$ is irrelevant, but this may not be the case for scenarios in which $U(1)_{\mathcal{B}}$ is spontaneously broken. Our goal here is to systematically study these scenarios, taking advantage of the fact that LQ/DQ open many routes to entangle $U(1)_{\phi}$ within $U(1)_{\mathcal{B}}\otimes U(1)_{\mathcal{L}}$ at the renormalizable level (with only SM fields, the $\phi^\dagger\bar{\nu}^\mathrm{C}_{R}\nu_{R}$ coupling is the only possibility). Note, finally, that in the KSVZ context, there is actually an extra global symmetry corresponding to $\Psi$ number, $U(1)_{\Psi}$, that will either survive or be incorporated within $U(1)_{\mathcal{B}}\otimes U(1)_{\mathcal{L}}$ via explicit breaking terms independent of $\phi$. In this way, the final surviving $U(1)_{X}$ is independent of $U(1)_{\Psi}$, and still given by Eq.~(\ref{SurvivingU1}).

In practice, to entangle the $U(1)_{\phi}$ symmetry with the other global symmetries, the strategy is to turn on some direct couplings between $\phi$ and the LQ/DQ, and for the latter, to turn on some or all of their couplings to SM fields such that no direct $\mathcal{B}$ and/or $\mathcal{L}$ violation occurs. It is important to stress that we do not assign $U(1)$ charges to the fields. Instead, we study all possible combinations of global $U(1)$ symmetries that can remain exact and, afterwards, derive the charges of the fields. Indeed, it is well-known that symmetries and charges are entirely fixed given a set of couplings in the Lagrangian, but often one identifies them by inspection, or starts from the charges to infer the allowed couplings. In the present case, as we will see, the set of couplings can be quite large, and the surviving $U(1)$s assign quite intricate charges to the fields. Typically, a naive inspection of the Lagrangian couplings would most likely miss some of the surviving $U(1)$s, or outright fail to identify possible scenarios. In practice, starting from the Lagrangian also provides a very systematic procedure: to find the surviving $U(1)$ symmetries, it suffices to express the charge constraint corresponding to each coupling, and solve this system of equations. When this system is under-determined, each parametric under-determination corresponds to a surviving $U(1)$. The charges of $\phi$ under these $U(1)$ then tell us which combination is spontaneously broken.

\subsection{KSVZ scenarios with leptoquarks and diquarks\label{SecKSVZ}}

Our requirements for the KSVZ scenarios are first that there should be only one Higgs doublet, neutral under the PQ and all global symmetries, and no direct mixing of the heavy fermions $\Psi_{L,R}$ with SM quarks to avoid FCNC or CKM unitarity constraints. Also, our goal is to force proton decay, neutron-antineutron oscillations, or a Majorana mass terms for $\nu_{R}$ (or more generally, neutrino-less double beta decays~\cite{Deppisch:2012nb}) to only arise through the spontaneous symmetry breaking of $U(1)_{\phi}$. Thus, none of these observables should be immediately allowed by LQ/DQ transitions. Typically, the strategy to achieve this is, starting from some Lagrangian with a specific set of couplings among $\phi$, some chosen LQ/DQs, and the SM fermions, to identify the global symmetries, and then make sure these symmetries forbid any other renormalizable Lagrangian couplings. This will be made clear going through specific examples. But, before that, let us describe some generic features of the scenarios and their consequence for the axion effective Lagrangian.

In all scenarios, there will be some $\phi^{2}S_{i}^{\dagger}S_{j}$, $\phi HS_{i}^{\dagger}S_{j}$, and/or $\phi S_{i}S_{j}S_{k}$ couplings. In this representation, the axion ends up coupled to the LQ/DQ, as can be seen plugging in Eq.~(\ref{PhiPolar}) in these couplings (remember $\eta_{\phi}=a^{0}$ in the KSVZ setting). Importantly, these couplings are never suppressed by the PQ breaking scale, since for example
\begin{equation}
\phi S_{i}S_{j}S_{k}\rightarrow\frac{1}{\sqrt{2}}(v_{\phi}+i\eta_{\phi}+...)S_{i}S_{j}S_{k}\ . 
\label{ExampleSSS}
\end{equation}
Though as a matter of principle, the axion $\mathcal{B}$ and/or $\mathcal{L}$ violating couplings are not suppressed by $v_{\phi}$, this scale nevertheless indirectly limits them. Indeed, the leading $v_{\phi}$ term produces a direct coupling among the LQ/DQ such that one falls into any one of the situations described in Sec.~\ref{Sec2c}, with some $U(1)_{X}$ smaller than $U(1)_{\mathcal{B}}\otimes U(1)_{\mathcal{L}}$ remaining exact. At low energy, these LQ/DQ couplings can induce $\mathcal{B}$ and/or $\mathcal{L}$ violating processes, hence set rather strong bounds on the LQ/DQ masses. Now, the largest $v_{\phi}$ is, the tightest these bounds are, so indirectly, the $\mathcal{B}$ and/or $\mathcal{L}$ violating axion couplings to SM fermions decrease for increasing $v_{\phi}$.

Coming back to the point of principle, one may wonder how is it that the axion couplings are not suppressed by $v_{\phi}$ in the effective axion Lagrangian language of Eq.~(\ref{AxionEL}). Indeed, as a result of the $\phi^{2}S_{i}^{\dagger}S_{j}$, $\phi HS_{i}^{\dagger}S_{j}$, and/or $\phi S_{i}S_{j}S_{k}$ couplings, some or all of the LQ/DQ become charged under $U(1)_{PQ}$. This means that if, along with Eq.~(\ref{ReparamG}) for the fermions, we reparametrize them as
\begin{equation}
S_{i}\rightarrow\exp(-iPQ(S_{i})a^{0}/v_{\phi})S_{i}\ , 
\label{ReparamS}
\end{equation}
the axion field is entirely removed from all the Lagrangian couplings. Indeed, the Lagrangian is PQ-symmetric, so the $\exp(ia^{0}/v_{\phi})$ factors always compensate exactly. Their kinetic terms $D_{\mu}S_{i}^{\dagger}D^{\mu}S_{i}$ are not invariant under the reparametrization though, and as for fermions, this is embodied in dimension-five interactions
\begin{equation}
\delta\mathcal{L}_{\text{\textrm{Der}}}=\frac{1}{v_{\phi}}\partial_{\mu}%
a^{0}J_{PQ}^{\mu}\text{ ,\ \ \ }J_{PQ}^{\mu}=\sum_{i}PQ(S_{i})(S_{i}^{\dagger
}(D^{\mu}S_{i})-(D^{\mu}S_{i}^{\dagger})S_{i})+...
\end{equation}
This representation is deceptive because the axion couplings to LQ/DQ appear suppressed by $v_{\phi}$. Yet, they are not suppressed because the EoM of the $S_{i}$ have $\mathcal{O}(v_{\phi})$ terms, like that coming from a $v_{\phi}S_{i}S_{j}S_{k}$ coupling in the example of Eq.~(\ref{ExampleSSS}). The same happens if LQ/DQ are integrated out before the reparametrization Eq.~(\ref{ReparamS}). They then do not occur in $J_{PQ}^{\mu}$, but SM fermions do, and their EoM now have inherited $\mathcal{O}(v_{\phi}/M^{n}$)
terms for some $n$, with $M$ the LQ/DQ mass scale. In all cases, the axion keeps its $\mathcal{O}(v_{\phi}^{0})$ $\mathcal{B}$ and/or $\mathcal{L}$ violating couplings, as it should.

This shows explicitly that the shift-symmetric $\delta\mathcal{L}_{\text{\textrm{Der}}}$ is not well-suited to these scenarios, at least for what concerns couplings to matter fields. For gauge boson, the situation is a bit different. The fermion reparametrization Eq.~(\ref{ReparamG}) generates spurious anomalous interactions to chiral gauge fields that are cancelled by the anomalies in the $J_{PQ}^{\mu}$ current, exactly as before, but the LQ/DQ obviously do not. Thus, for them, the axion effective Lagrangian after the reparametrization of Eq.~(\ref{ReparamS}) correctly captures the fact that triangle graphs with LQ/DQ running in the loop are not anomalous, and vanish at the dimension-five level for a massless axion. Thus, none of the axion to gauge boson couplings is affected by the LQ/DQ at leading order.

\subsubsection{Spontaneous breaking of $\mathcal{B}+\mathcal{L}$}

We have seen that $\mathcal{B}+\mathcal{L}$ is immediately broken whenever a given $S_{i}$ or $V_{i}$ has both LQ and DQ couplings. For example, $S_{1}^{8/3}$ with its couplings to $\bar{d}_{R}e_{R}^{\mathrm{C}}$ and $\bar{u}_{R}^{\mathrm{C}}u_{R}$ can induce $(\Delta\mathcal{B},\Delta\mathcal{L})=(1,1)$ operators and proton decay. A possible strategy to adapt this scenario and force these operators to appear only through the SSB of $\phi$ is to consider two such states, one LQ and one DQ, with a $\phi$-dependent mixing term:
\begin{equation}
\mathcal{L}_{\mathrm{KSVZ+LQ}}=\mathcal{L}_{\mathrm{KSVZ}}+S_{1}^{8/3}\bar
{d}_{R}e_{R}^{\mathrm{C}}+\tilde{S}_{1}^{8/3}\bar{u}_{R}^{\mathrm{C}}%
u_{R}+\phi^{2}S_{1}^{8/3\dagger}\tilde{S}_{1}^{8/3}+h.c.\ , \label{LagrKSVZ1a}%
\end{equation}
with $\mathcal{L}_{\mathrm{KSVZ}}$ given in Eq.~(\ref{KSVZ0}), and LQ/DQ kinetic terms are understood. We also do not write explicitly the LQ/DQ scalar potential terms made of bilinears like $S_{1}^{8/3\dagger}S_{1}^{8/3}$ or $\tilde{S}_{1}^{8/3\dagger}\tilde{S}_{1}^{8/3}$ since those are neutral under any $U(1)$ symmetry. Solving for the $U(1)$ charges of all the fields under the requirement that the Higgs doublet is neutral (to avoid mixing with $U(1)_{Y}$), a triple under-determination remains, which we can identify as\footnote{Evidently, the normalization of each line is free, but that for $U(1)_{\mathcal{B}}$ and $U(1)_{\mathcal{L}}$ is chosen to reproduce conventional quark and lepton $\mathcal{B}$ and $\mathcal{L}$ of $1/3$ and $1$, respectively.}
\begin{equation}
\begin{tabular}[c]{cccccccccccc}\hline
& $\phi$ & $S_{1}^{8/3}$ & $\tilde{S}_{1}^{8/3}$ & $\Psi_{L}$ & $\Psi_{R}$ &
$q_{L}$ & $u_{R}$ & $d_{R}$ & $\ell_{L}$ & $e_{R}$ & $\nu_{R}$\\\hline
$U(1)_{\Psi}$ & $0$ & $0$ & $0$ & $1$ & $1$ & $0$ & $0$ & $0$ & $0$ & $0$ &
$0$\\
$U(1)_{\mathcal{B}}$ & $1/2$ & $1/3$ & $-2/3$ & $-1/2$ & $0$ & $1/3$ & $1/3$ &
$1/3$ & $0$ & $0$ & $0$\\
$U(1)_{\mathcal{L}}$ & $1/2$ & $1$ & $0$ & $-1/2$ & $0$ & $0$ & $0$ & $0$ &
$1$ & $1$ & $1$\\\hline
\end{tabular}
\label{ChargKSVZ1a}
\end{equation}

\begin{figure}[t]
\begin{center}
\includegraphics[height=1.3586in,width=1.5826in]{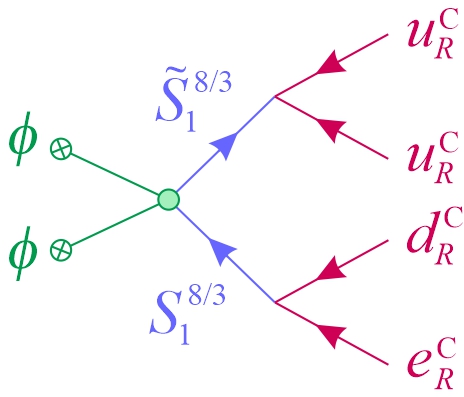}
\caption{Proton decay operator generated by the spontaneous breaking of $U(1)_{\mathcal{B}+\mathcal{L}}$.}%
\label{Fig2}
\end{center}
\end{figure}

What this table shows is that $\phi$ carries a $U(1)_{\mathcal{B}+\mathcal{L}}$ charge, which thus gets spontaneously broken, while $U(1)_{\mathcal{B}-\mathcal{L}}$ stays exact. This model is essentially identical to that introduced long ago in Ref.~\cite{WeinbergPRD22}, except that the Goldstone boson is here identified with the axion. This pattern of symmetry breaking is easily understood from the Lagrangian couplings and the diagram in Fig.~\ref{Fig2}. Plugging in the polar representation of $\phi$, Eq.~(\ref{PhiPolar}), the effective operator at the low-scale is
\begin{equation}
\mathcal{H}_{(\Delta\mathcal{B},\Delta\mathcal{L})=(1,1)}^{eff}=\exp
(2ia^{0}/v_{\phi})\frac{v_{\phi}^{2}}{m_{S}^{2}m_{\tilde{S}}^{2}}\bar{u}%
_{R}^{\mathrm{C}}u_{R}\bar{d}_{R}^{\mathrm{C}}e_{R}+h.c.\ , \label{EffHKSVZ1}%
\end{equation}
where we have identified $\eta_{\phi}$ as the axion $a^{0}$, and denoted the $S_{1}^{8/3}$ and $\tilde{S}_{1}^{8/3}$ masses as $m_{S}$ and $m_{\tilde{S}}$, respectively. Note well that this operator arises entirely through the SSB: the charges in Eq.~(\ref{ChargKSVZ1a}) explicitly prevent a DQ coupling for $S_{1}^{8/3}$, and a LQ coupling for $\tilde{S}_{1}^{8/3}$. Expanding the exponential, the leading term involves only SM particles, and contributes to proton decay. Thus, $m_{S}$ and $m_{\tilde{S}}$ have to be pushed quite high, though a bit lower that in the usual GUT scenarios. For instance, while the scale of the dim-6 operators is typically pushed above $10^{14}$~GeV, we only need $m_{S}\approx m_{\tilde{S}}>10^{11}$~GeV when $v_{\phi}=10^{9}$~GeV. With these parameters, the proton decay modes involving the axion are thus totally negligible. Finally, notice that the axion totally disappears from $\mathcal{H}_{(\Delta\mathcal{B},\Delta\mathcal{L})=(1,1)}^{eff}$ under the reparametrization Eq.~(\ref{ReparamG}), with the PQ charges identified as $(\mathcal{B}+\mathcal{L})/2$. As stated earlier, the $v_{\phi}a^{0}\bar{u}_{R}^{\mathrm{C}}u_{R}\bar{d}_{R}^{\mathrm{C}}e_{R}$ effective coupling would then hide in the $\partial_{\mu}a^{0}J_{PQ}^{\mu}/v_{\phi}$ terms since the quarks and leptons inherit from $v_{\phi}^{2}\bar{u}_{R}^{\mathrm{C}}u_{R}\bar{d}_{R}^{\mathrm{C}}e_{R}$ some $\mathcal{O}(v_{\phi}^{2})$ terms in their EoM.%

As shown in Eq.~(\ref{ChargKSVZ1a}), $U(1)_{\Psi}$ remains as an exact global symmetry, which means that the $\mathcal{B}+\mathcal{L}$ charges of $\Psi_{L,R}$ are not unambiguously defined. To fix them requires $\Psi_{L,R}$
to couple to SM fermions, and this is possible only for some specific gauge quantum numbers. If we further ask that $S_{1}^{8/3}$ ($\tilde{S}_{1}^{8/3}$) should always (never) couple to leptons, the only possibilities are%
\begin{equation}%
\begin{tabular}
[c]{c|cccccc}\hline
$Y,\mathcal{B},\mathcal{L}$ & $S_{1}^{8/3}\bar{\Psi}_{L}\ell_{L}^{\mathrm{C}}$
& $S_{1}^{8/3}\bar{\Psi}_{R}e_{R}^{\mathrm{C}}$ & $S_{1}^{8/3}\bar{\Psi}%
_{R}\nu_{R}^{\mathrm{C}}$ & $\tilde{S}_{1}^{8/3}\bar{\Psi}_{L}^{\mathrm{C}%
}q_{L}$ & $\tilde{S}_{1}^{8/3}\bar{\Psi}_{R}^{\mathrm{C}}u_{R}$ & $\tilde
{S}_{1}^{8/3}\bar{\Psi}_{R}^{\mathrm{C}}d_{R}$\\\hline
$\Psi_{L}$ & $-\dfrac{5}{3},\dfrac{1}{3},0\rule[-0.14in]{0in}{0.36in}$ &
$-\dfrac{2}{3},\dfrac{5}{6},\dfrac{1}{2}$ & $-\dfrac{8}{3},\dfrac{5}{6}%
,\dfrac{1}{2}$ & $\dfrac{7}{3},\dfrac{1}{3},0$ & $\dfrac{4}{3},\dfrac{5}%
{6},\dfrac{1}{2}$ & $\dfrac{10}{3},\dfrac{5}{6},\dfrac{1}{2}$\\
$\Psi_{R}$ & $-\dfrac{5}{3},-\dfrac{1}{6},-\dfrac{1}{2}\rule[-0.14in]%
{0in}{0.36in}$ & $-\dfrac{2}{3},\dfrac{1}{3},0$ & $-\dfrac{8}{3},\dfrac{1}%
{3},0$ & $\dfrac{7}{3},-\dfrac{1}{6},-\dfrac{1}{2}$ & $\dfrac{4}{3},\dfrac
{1}{3},0$ & $\dfrac{10}{3},\dfrac{1}{3},0$\\\hline
\end{tabular}
\end{equation}
These couplings are mutually exclusive since they impose different hypercharges for $\Psi_{L,R}$. Also, $S_{1}^{8/3}\bar{\Psi}_{R}e_{R}^{\mathrm{C}}$ and $\tilde{S}_{1}^{8/3}\bar{\Psi}_{R}^{\mathrm{C}}u_{R}$ would allow for direct $\bar{\Psi}_{R}d_{R}$ and $\bar{\Psi}_{R}u_{R}$ couplings, respectively, hence must be discarded. Note how the peculiar choice of couplings completely twists the $\mathcal{B},\mathcal{L}$ charges, in the sense that they do not correspond to the naive assignments of $\mathcal{B}=1/3$ and $\mathcal{L}=0$ one may have expected for the "heavy quarks" of the KSVZ mechanism. As said before, the charges of the fields have to be deduced from the set of couplings of the Lagrangian, and not the other way around.

Similar scenarios can be constructed using $S_{1}^{2/3}$, $S_{1}^{4/3}$, $V_{2}^{1/3}$ or $V_{2}^{5/3}$. Actually, $S_{1}^{2/3}$ was considered in Ref.~\cite{Reig:2018yfd}, though the model built there is more complicated (here the PQ symmetry is directly identified with $\mathcal{B}+\mathcal{L}$ and only a single Higgs doublet is introduced instead of four). Each time, two such states are taken, with one having LQ couplings, and the other DQ couplings, and a $\phi$-driven mixing term is introduced. The only difference in each case is the specific $(\Delta\mathcal{B},\Delta\mathcal{L})=(1,1)$ operator(s) that can be spontaneously generated, see Table~\ref{TableLQBL}, and thereby, the induced pattern of proton decay modes. In this respect, it is worth to look at the $S_{1}^{4/3}$ scenario, since it has only LQ couplings to $\nu_{R}$:
\begin{equation}
\mathcal{L}_{\mathrm{KSVZ+LQ}}=\mathcal{L}_{\mathrm{KSVZ}}+S_{1}^{4/3}\bar
{u}_{R}\nu_{R}^{\mathrm{C}}+\tilde{S}_{1}^{4/3}\bar{d}_{R}^{\mathrm{C}}%
d_{R}+\phi^{2}S_{1}^{4/3\dagger}\tilde{S}_{1}^{4/3}+\phi\bar{\nu}%
_{R}^{\mathrm{C}}\nu_{R}+h.c.\ . \label{LagrKSVZ1b}%
\end{equation}
Let us also turn on a coupling to $\Psi_{L,R}$, to fix its charges, and for definiteness, let us take $S_{1}^{4/3}\bar{\Psi}_{L}\ell_{L}^{\mathrm{C}}$. Because of the $\phi\bar{\nu}_{R}^{\mathrm{C}}\nu_{R}$ coupling, solving for the $U(1)$ charges of all the fields now leaves a single under-determination:
\begin{equation}%
\begin{tabular}
[c]{cccccccccccc}\hline
& $\phi$ & $S_{1}^{4/3}$ & $\tilde{S}_{1}^{4/3}$ & $\Psi_{L}$ & $\Psi_{R}$ &
$q_{L}$ & $u_{R}$ & $d_{R}$ & $\ell_{L}$ & $e_{R}$ & $\nu_{R}$\\\hline
$U(1)_{PQ}$ & $2$ & $2/3$ & $-10/3$ & $5/3$ & $-1/3$ & $5/3$ & $5/3$ & $5/3$ &
$-1$ & $-1$ & $-1$\\\hline
\end{tabular}
\end{equation}
This time, neither $\mathcal{B}$ nor $\mathcal{L}$ survives. Starting from $U(1)_{\phi}\otimes U(1)_{\mathcal{B}}\otimes U(1)_{\mathcal{L}}$, two $U(1)$s are explicitly broken by $\mathcal{L}_{\mathrm{KSVZ+LQ}}$, while the remaining exact $U(1)$ is identified with $U(1)_{PQ}$ and spontaneously broken by $\phi$. The interest in this scenario is that $S_{1}^{4/3}$ couples only to $\nu_{R}$, whose mass is pushed at the PQ breaking scale by the $\phi\bar{\nu}_{R}^{\mathrm{C}}\nu_{R}$ coupling. At the low-energy scale, the leading proton decay operator will scale as%
\begin{equation}
\mathcal{H}_{(\Delta\mathcal{B},\Delta\mathcal{L})=(1,1)}^{eff}=\exp
(2ia^{0}/v_{\phi})\frac{v_{\phi}^{2}}{m_{S}^{2}m_{\tilde{S}}^{2}}\bar{d}%
_{R}d_{R}^{\mathrm{C}}\bar{u}_{R}\nu_{R}^{\mathrm{C}}+h.c.\rightarrow
\exp(2ia^{0}/v_{\phi})\frac{v_{\phi}v_{EW}}{m_{S}^{2}m_{\tilde{S}}^{2}}\bar{d}%
_{R}d_{R}^{\mathrm{C}}\bar{u}_{R}\nu_{L}^{\mathrm{C}}+h.c.\ .
\label{FinalScale11}%
\end{equation}
Thanks to this extra suppression, the PQ breaking scale, which is also the neutrino seesaw scale, and the LQ/DQ mass scale, can all sit at around $10^{9}$ GeV. They could thus naturally have a common UV origin.

\subsubsection{Spontaneous breaking of $\mathcal{B}-\mathcal{L}$}

\begin{figure}[t]
\begin{center}
\includegraphics[height=1.3984in,width=1.9009in]{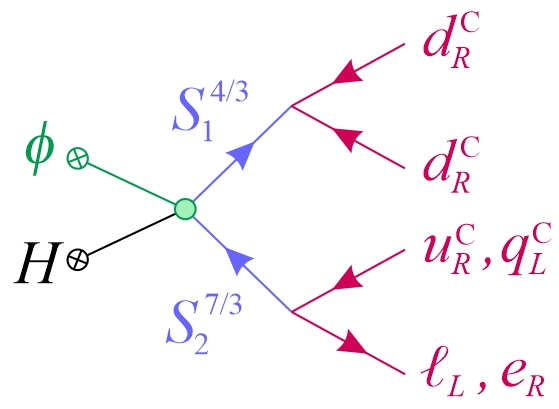}
\caption{Proton decay operator generated by the spontaneous breaking of $U(1)_{\mathcal{B}-\mathcal{L}}$.}
\label{Fig3}
\end{center}
\end{figure}

With only LQ/DQ, scenarios in which $\mathcal{B}-\mathcal{L}$ is explicitly broken typically arise from any one of the $HS_{i}^{\dagger}S_{j}$ or $HV_{i}^{\dagger}V_{j}$ couplings in Eq.~(\ref{LQOpsBmL}). Those couplings always involve a pure LQ state together with a mixed LQ/DQ state. The $(\Delta\mathcal{B},\Delta\mathcal{L})=(1,-1)$ pattern arises when the latter has only DQ couplings. All these scenarios can be adapted to force $\mathcal{B}-\mathcal{L}$ to be broken spontaneously instead of explicitly. Let us take the $HS_{2}^{7/3\dagger}S_{1}^{4/3}$ case as an example, the others being totally similar. To entangle the KSVZ symmetry with $\mathcal{B}-\mathcal{L}$, we start from the Lagrangian
\begin{equation}
\mathcal{L}_{\mathrm{KSVZ+LQ}}=\mathcal{L}_{\mathrm{KSVZ}}+S_{1}^{4/3}\bar
{d}_{R}^{\mathrm{C}}d_{R}+S_{2}^{7/3}(\bar{u}_{R}\ell_{L}+\bar{q}_{L}%
e_{R})+\phi HS_{2}^{7/3\dagger}S_{1}^{4/3}+h.c.\ , \label{LagrKSVZ2}%
\end{equation}
where again kinetic terms and LQ/DQ potential terms are understood. For definiteness, we also include the $S_{1}^{4/3}\bar{\Psi}_{L}\ell_{L}^{\mathrm{C}}$ coupling to get rid of $U(1)_{\Psi}$ and fix the quantum numbers of $\Psi_{L,R}$. Then, there remain only a $U(1)_{\mathcal{B}}\otimes U(1)_{\mathcal{L}}$ symmetry with charges
\begin{equation}
\begin{tabular}[c]{cccccccccccc}\hline
& $\phi$ & $S_{2}^{7/3}$ & $S_{1}^{4/3}$ & $\Psi_{L}$ & $\Psi_{R}$ & $q_{L}$ &
$u_{R}$ & $d_{R}$ & $\ell_{L}$ & $e_{R}$ & $\nu_{R}$\\\hline
$U(1)_{\mathcal{B}}$ & $1$ & $1/3$ & $-2/3$ & $-2/3$ & $-5/3$ & $1/3$ & $1/3$
& $1/3$ & $0$ & $0$ & $0$\\
$U(1)_{\mathcal{L}}$ & $-1$ & $-1$ & $0$ & $-1$ & $0$ & $0$ & $0$ & $0$ & $1$
& $1$ & $1$\\\hline
\end{tabular}
\end{equation}
and $U(1)_{\mathcal{B}-\mathcal{L}}$ is spontaneously broken when $\phi$ acquires its vacuum expectation value. Note that these charges prevent the LQ couplings of $S_{1}^{4/3}$ (taking $\phi HS_{2}^{1/3\dagger}S_{1}^{2/3}$ instead, they would further forbid the $HS_{2}^{1/3}S_{2}^{1/3}S_{2}^{1/3}$ coupling). The final operators are part of the $(\Delta\mathcal{B},\Delta\mathcal{L})=(1,-1)$ dimension-seven ones in Table~\ref{TableLQBL} because of the Higgs doublet appearing in the $\phi HS_{2}^{7/3\dagger}S_{1}^{4/3}$ mixing term (see Fig.~\ref{Fig3}):
\begin{equation}
\mathcal{H}_{(\Delta\mathcal{B},\Delta\mathcal{L})=(1,-1)}^{eff}=\exp
(ia^{0}/v_{\phi})\frac{v_{\phi}}{m_{S}^{2}m_{\tilde{S}}^{2}}H\bar{d}_{R}%
d_{R}^{\mathrm{C}}(\bar{u}_{R}\ell_{L}+\bar{q}_{L}e_{R})+h.c.\ .
\end{equation}
The situation is thus similar to that in Eq.$~$(\ref{FinalScale11}). Further lowering the LQ/DQ scale by about an order of magnitude is possible starting from the $HV_{1,\mu}^{2/3\dagger}V_{2}^{5/3,\mu}$ coupling, as $V_{1,\mu}^{2/3\dagger}$ couples only to $\nu_{R}$.

In this regard, note that all these scenarios are again compatible with a seesaw mechanism. Adding either a $\phi\bar{\nu}_{R}^{\mathrm{C}}\nu_{R}$, $\phi^{\dagger}\bar{\nu}_{R}^{\mathrm{C}}\nu_{R}$, or $M_{R}\bar{\nu}_{R}^{\mathrm{C}}\nu_{R}$ coupling to $\mathcal{L}_{\mathrm{KSVZ+LQ}}$ in Eq.~(\ref{LagrKSVZ2}), a single exact $U(1)$ remains at the PQ scale, with charges
\begin{equation}
\begin{tabular}[c]{ccccccccccccc}\hline
&  & $\phi$ & $S_{2}^{7/3}$ & $\tilde{S}_{1}^{4/3}$ & $\Psi_{L}$ & $\Psi_{R}$
& $q_{L}$ & $u_{R}$ & $d_{R}$ & $\ell_{L}$ & $e_{R}$ & $\nu_{R}$\\\hline
$\phi\bar{\nu}_{R}^{\mathrm{C}}\nu_{R}:$ & $U(1)_{PQ}$ & $2$ & $4/3$ & $-2/3$
& $1/3$ & $-5/3$ & $1/3$ & $1/3$ & $1/3$ & $-1$ & $-1$ & $-1$\\
$\phi^{\dagger}\bar{\nu}_{R}^{\mathrm{C}}\nu_{R}:$ & $U(1)_{PQ}$ & $2/3$ & $0$
& $-2/3$ & $-1$ & $-5/3$ & $1/3$ & $1/3$ & $1/3$ & $1/3$ & $1/3$ & $1/3$\\
$M_{R}\bar{\nu}_{R}^{\mathrm{C}}\nu_{R}:$ & $U(1)_{PQ}$ & $1$ & $1/3$ & $-2/3$
& $1/3$ & $-5/3$ & $1/3$ & $1/3$ & $1/3$ & $0$ & $0$ & $0$\\\hline
\end{tabular}
\end{equation}
For all these cases, the axion still emerges as a massless Goldstone boson, and is associated to both $U(1)_{\mathcal{B}-\mathcal{L}}$ and $U(1)_{\mathcal{L}}$ spontaneous breakings.

\subsubsection{Spontaneous breaking of $\mathcal{B}$}

\begin{figure}[t]
\begin{center}
\includegraphics[height=1.8844in,width=1.7167in]{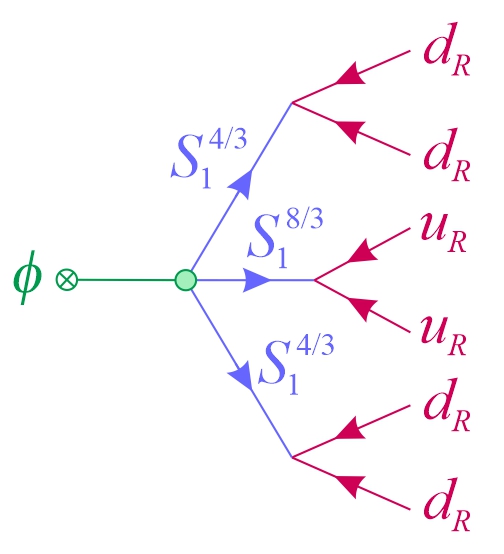}
\caption{Neutron-antineutron oscillation operators generated by the spontaneous breaking of $U(1)_{\mathcal{B}}$.}
\label{Fig4a}
\end{center}
\end{figure}

The spontaneous breaking of $\mathcal{B}$ first arose at the dimension-9 level in Table~\ref{TableLQBL} since it necessarily involves six fermions. As seen in Sec.~\ref{Sec2c}, typical scenarios thus require a cubic coupling among DQ states. Let us start with
\begin{equation}
\mathcal{L}_{\mathrm{KSVZ+LQ}}=\mathcal{L}_{\mathrm{KSVZ}}+S_{1}^{4/3}\bar
{d}_{R}^{\mathrm{C}}d_{R}+S_{1}^{8/3}\bar{u}_{R}^{\mathrm{C}}u_{R}+\phi
S_{1}^{4/3}S_{1}^{4/3}S_{1}^{8/3}+h.c.\ , \label{LagrKSVZ3}%
\end{equation}
where $S_{1}^{4/3}\sim(\mathbf{3},\mathbf{1},+4/3)$ and $S_{1}^{8/3}\sim(\mathbf{\bar{6}},\mathbf{1},-8/3)$. Though not compulsory, we add the coupling $S_{1}^{8/3}\bar{\Psi}_{L}^{\mathrm{C}}q_{L}$ to break $U(1)_{\Psi}$ and fix the charges of $\Psi_{L,R}$. With this Lagrangian, only two $U(1)$s are exact:
\begin{equation}
\begin{tabular}[c]{cccccccccccc}\hline
& $\phi$ & $S_{2}^{8/3}$ & $S_{1}^{4/3}$ & $\Psi_{L}$ & $\Psi_{R}$ & $q_{L}$ &
$u_{R}$ & $d_{R}$ & $\ell_{L}$ & $e_{R}$ & $\nu_{R}$\\\hline
$U(1)_{\mathcal{B}}$ & $2$ & $-2/3$ & $-2/3$ & $1/3$ & $-5/3$ & $1/3$ & $1/3$
& $1/3$ & $0$ & $0$ & $0$\\
$U(1)_{\mathcal{L}}$ & $0$ & $0$ & $0$ & $0$ & $0$ & $0$ & $0$ & $0$ & $1$ &
$1$ & $1$\\\hline
\end{tabular}
\end{equation}
Thus, $U(1)_{PQ}=U(1)_{\mathcal{B}}$ is broken spontaneously by two units, but $U(1)_{\mathcal{L}}$ stays exact. This model is actually very similar to that of Ref.~\cite{Barbieri:1981yr} (see also Ref.~\cite{Berezhiani:2015afa}), except that the Goldstone boson associated to the $U(1)_{\mathcal{B}}$ breaking is identified with the axion. In turn, the axion ends up coupled to neutron pairs, via the diagram shown in Fig.~\ref{Fig4a}. The corresponding operator is
\begin{equation}
\mathcal{H}_{(\Delta\mathcal{B},\Delta\mathcal{L})=(2,0)}^{eff}=\exp
(ia^{0}/v_{\phi})\frac{v_{\phi}}{m_{S^{4/3}}^{4}m_{S^{8/3}}^{2}}\bar{d}%
_{R}^{\mathrm{C}}d_{R}\bar{d}_{R}^{\mathrm{C}}d_{R}\bar{u}_{R}^{\mathrm{C}%
}u_{R}+h.c.\ . \label{ScaleNN}%
\end{equation}
Typical bounds on the scale of the $(\Delta\mathcal{B},\Delta\mathcal{L})=(2,0)$ operators are at around $100$~TeV~\cite{Mohapatra:2009wp,Phillips:2014fgb} if the couplings implicit in Eq.~(\ref{LagrKSVZ3}) are all $\mathcal{O}(1)$. The PQ scale of $10^{9}$~GeV pushes the DQ scale slightly higher than those $100\TeV$, but given that their masses appear to the sixth power, this is marginal (less than an order of magnitude). The presence of the axion also leads to an effective operator%
\begin{equation}
\frac{1}{m_{S^{4/3}}^{4}m_{S^{8/3}}^{2}}a^{0}\bar{d}_{R}^{\mathrm{C}}d_{R}%
\bar{d}_{R}^{\mathrm{C}}d_{R}\bar{u}_{R}^{\mathrm{C}}u_{R}+h.c.\ \rightarrow
\frac{\Lambda_{QCD}^{6}}{m_{S^{4/3}}^{4}m_{S^{8/3}}^{2}}a^{0}\bar
{n}^{\mathrm{C}}\gamma_{5}n+h.c.\ ,
\end{equation}
with the QCD confinement scale $\Lambda_{QCD}$ of the order of $300$~MeV. Because the DQ mass scale is pushed rather high by the dimension-nine operator in Eq.~(\ref{ScaleNN}), this direct coupling is very suppressed. Note, though, that it could have consequences in a cosmological context~\cite{Mohapatra:2009wp}.%

As for the previous two scenarios, a seesaw mechanism can be implemented by adding a $\phi\bar{\nu}_{R}^{\mathrm{C}}\nu_{R}$, $\phi^{\dagger}\bar{\nu}_{R}^{\mathrm{C}}\nu_{R}$, or $M_{R}\bar{\nu}_{R}^{\mathrm{C}}\nu_{R}$ coupling to the Lagrangian in Eq.~(\ref{LagrKSVZ3}). For the former two, this identifies the axion as the Majoron~\cite{Barbieri:1981yr}. The only change is, in some sense, to assign a $\mathcal{B}$ number to $\nu_{R}$, hence by extension, to the leptons:
\begin{equation}
\begin{tabular}[c]{ccccccccccccc}\hline
&  & $\phi$ & $S_{2}^{8/3}$ & $S_{1}^{4/3}$ & $\Psi_{L}$ & $\Psi_{R}$ &
$q_{L}$ & $u_{R}$ & $d_{R}$ & $\ell_{L}$ & $e_{R}$ & $\nu_{R}$\\\hline
$\phi\bar{\nu}_{R}^{\mathrm{C}}\nu_{R}:$ & $U(1)_{PQ}$ & $2$ & $-2/3$ & $-2/3$
& $1/3$ & $-5/3$ & $1/3$ & $1/3$ & $1/3$ & $-1$ & $-1$ & $-1$\\
$\phi^{\dagger}\bar{\nu}_{R}^{\mathrm{C}}\nu_{R}:$ & $U(1)_{PQ}$ & $2$ &
$-2/3$ & $-2/3$ & $1/3$ & $-5/3$ & $1/3$ & $1/3$ & $1/3$ & $1$ & $1$ & $1$\\
$M_{R}\bar{\nu}_{R}^{\mathrm{C}}\nu_{R}:$ & $U(1)_{PQ}$ & $2$ & $-2/3$ &
$-2/3$ & $1/3$ & $-5/3$ & $1/3$ & $1/3$ & $1/3$ & $0$ & $0$ & $0$\\\hline
\end{tabular}
\end{equation}
Note that the charges imposed by the presence of $\phi\bar{\nu}_{R}^{\mathrm{C}}\nu_{R}$ open the door to the $S_{1}^{4/3}\bar{u}_{R}\nu_{R}^{\mathrm{C}}$ coupling also, and thus to direct proton decay via an $S_{1}^{4/3}$ Fermi interaction. For the other two scenarios, proton decay remains forbidden since all its decay modes include an odd number of leptons, but only $\Delta\mathcal{L}=2n$ transitions are made possible by the Lagrangian couplings.

\subsubsection{Spontaneous breaking of $\mathcal{B}\pm3\mathcal{L}$}

\begin{figure}[t]
\begin{center}
\includegraphics[height=2.2329in,width=4.9536in]{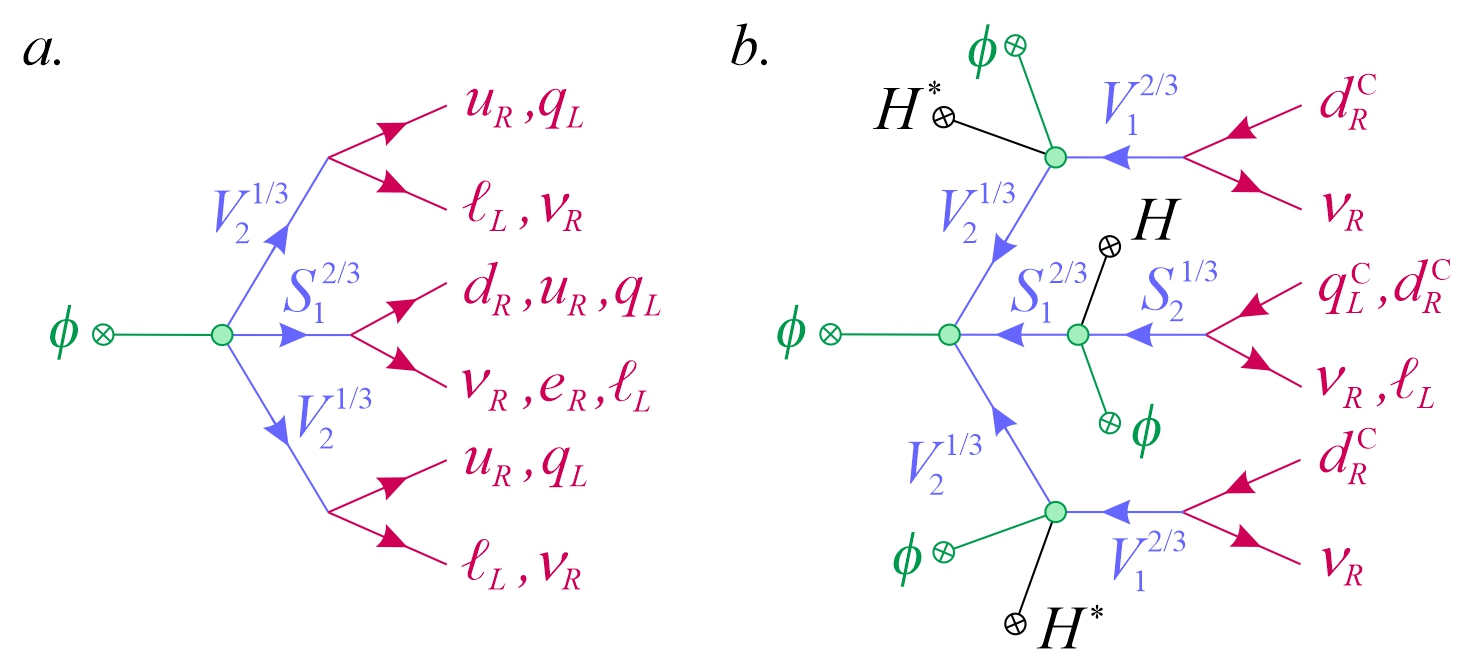}
\caption{Proton decay operators generated by the spontaneous breaking of $U(1)_{\mathcal{B}+3\mathcal{L}}$ ($a.$) and $U(1)_{\mathcal{B}-3\mathcal{L}}$ ($b.$).}
\label{Fig4b}
\end{center}
\end{figure}

From Eq.~(\ref{LQOpsB2}), it is clear that the scenarios leading to $(\Delta\mathcal{B},\Delta\mathcal{L})=(2,0)$ can be adapted to generate $(\Delta\mathcal{B},\Delta\mathcal{L})=(1,3)$ effects. All that is needed is to replace all DQ couplings by LQ couplings. The only difficulty is to account for the antisymmetric color contraction, since LQ necessarily transform as $\mathbf{3}$ under $SU(3)_{C}$. If we insist on introducing at most two different LQ, the only available scenario is%
\begin{equation}
\mathcal{L}_{\mathrm{KSVZ+LQ}}=\mathcal{L}_{\mathrm{KSVZ}}+S_{1}^{2/3}(\bar
{d}_{R}\nu_{R}^{\mathrm{C}}+\bar{u}_{R}e_{R}^{\mathrm{C}}+\bar{q}_{L}\ell
_{L}^{\mathrm{C}})+V_{2,\mu}^{1/3}(\bar{u}_{R}\gamma^{\mu}\ell_{L}%
^{\mathrm{C}}+\bar{q}_{L}\gamma^{\mu}\nu_{R}^{\mathrm{C}})+\phi S_{1}%
^{2/3}V_{2,\mu}^{1/3}V_{2}^{1/3,\mu}+h.c.\ . \label{LagrKSVZ4}%
\end{equation}
As usual, the $U(1)_{\Psi}$ is broken explicitly, this time by adding $V_{2,\mu}^{1/3}\bar{\Psi}_{L}\gamma^{\mu}e_{R}^{\mathrm{C}}$ to force the hypercharge of $\Psi_{L,R}$ to be different from that of SM quarks. If instead of the LQ couplings, DQ couplings were allowed, this scenario produces the $(\Delta\mathcal{B},\Delta\mathcal{L})=(2,0)$ symmetry pattern discussed in the previous section. Now, with these LQ couplings and no DQ couplings, the charges are
\begin{equation}%
\begin{tabular}[c]{cccccccccccc}\hline
& $\phi$ & $S_{1}^{2/3}$ & $V_{2}^{1/3}$ & $\Psi_{L}$ & $\Psi_{R}$ & $q_{L}$ &
$u_{R}$ & $d_{R}$ & $\ell_{L}$ & $e_{R}$ & $\nu_{R}$\\\hline
$U(1)_{\mathcal{B}}$ & $1$ & $1/3$ & $1/3$ & $1/3$ & $-2/3$ & $1/3$ & $1/3$ &
$1/3$ & $0$ & $0$ & $0$\\
$U(1)_{\mathcal{L}}$ & $3$ & $1$ & $1$ & $0$ & $-3$ & $0$ & $0$ & $0$ & $1$ &
$1$ & $1$\\\hline
\end{tabular}
\end{equation}
The PQ symmetry is identified with $U(1)_{\mathcal{B}+3\mathcal{L}}$, and dimension-nine $(\Delta\mathcal{B},\Delta\mathcal{L})=(1,3)$ proton decay operators appear at the low scale, see Fig.~\ref{Fig4b}$a$ (a similar LQ model was proposed in Ref.~\cite{WeinbergPRD22} to break $\mathcal{B}+3\mathcal{L}$ spontaneously). The fact that these operators are dimension-nine allows to lower the LQ scale, but qualitatively, this scenario is not very different from the $\mathcal{B}\pm\mathcal{L}$ ones. Also, a seesaw mechanism can be added with either $\phi\bar{\nu}_{R}^{\mathrm{C}}\nu_{R}$ or $M_{R}\bar{\nu}_{R}^{\mathrm{C}}\nu_{R}$, but not with $\phi^{\dagger}\bar{\nu}_{R}^{\mathrm{C}}\nu_{R}$ as this would allow back the DQ couplings of both $S_{1}^{2/3}$ and $V_{2,\mu}^{1/3}$. It should be noted that these charges allow for the $D^{\mu}HS_{1}^{2/3}V_{2,\mu}^{1/3\dagger}$ and $HD^{\mu}S_{1}^{2/3}V_{2,\mu}^{1/3\dagger}$couplings. If not initially present, they are immediately generated via a fermion loop. Yet, these operators carry $(\Delta\mathcal{B},\Delta\mathcal{L})=(0,0)$ and cannot help create simpler proton decay processes. They could turn on some new four-fermion semileptonic FCNC operators though, but these effects are beyond our scope.%

The final pattern is $(\Delta\mathcal{B},\Delta\mathcal{L})=(1,-3)$, and this one is quite difficult to induce spontaneously. The operators in Eq.~(\ref{LQOpsBL3}) being already of dimension four, we cannot proceed as for the other cases and simply multiply them by $\phi$. One way to proceed is to start with an operator from the $(\Delta\mathcal{B},\Delta\mathcal{L})=(1,3)$
class in Eq.~(\ref{LQOpsB2}), and then switch $\mathcal{L}$ by six units using $\Delta\mathcal{L}=2$ operators of Eq.~(\ref{LQOpsBmL}). For instance, the Lagrangian
\begin{align}
\mathcal{L}_{\mathrm{KSVZ+LQ}}  &  =\mathcal{L}_{\mathrm{KSVZ}}+S_{2}%
^{1/3}(\bar{d}_{R}\ell_{L}+\bar{q}_{L}\nu_{R})+V_{1,\mu}^{2/3}\bar{d}%
_{R}\gamma^{\mu}\nu_{R}\nonumber\\
&  \ \ \ \ +\phi(HS_{2}^{1/3\dagger}S_{1}^{2/3}+H^{\dagger}V_{1}%
^{2/3\dagger,\mu}V_{2,\mu}^{1/3}+S_{1}^{2/3\dagger}V_{2,\mu}^{1/3\dagger}%
V_{2}^{1/3,\mu\dagger})+h.c.\ , \label{LagrKSVZ5}%
\end{align}
does lead to the desired $(\Delta\mathcal{B},\Delta\mathcal{L})=(1,-3)$ pattern, as shown in Fig.~\ref{Fig4b}$b$. With four LQ states, it is certainly more complex than the other scenarios, though it should be noted that there is a certain symmetric flavor to the presence of $S_{2}^{1/3}$, $S_{1}^{2/3}$ and $V_{2,\mu}^{1/3}$, $V_{1,\mu}^{2/3}$. Also, it is not compulsory for $\phi$ to appear in all of the last three couplings, but when it does, only some combinations do give a $\mathcal{B}-3\mathcal{L}$ charge to $\phi$. Further adding $V_{1,\mu}^{2/3}\bar{\Psi}_{R}\gamma^{\mu}e_{R}$ to fix the $\Psi_{L,R}$ quantum numbers, we find%
\begin{equation}
\begin{tabular}[c]{cccccccccccccc}\hline
& $\phi$ & $S_{1}^{2/3}$ & $V_{2}^{1/3}$ & $S_{2}^{1/3}$ & $V_{1}^{2/3}$ &
$\Psi_{L}$ & $\Psi_{R}$ & $q_{L}$ & $u_{R}$ & $d_{R}$ & $\ell_{L}$ & $e_{R}$ &
$\nu_{R}$\\\hline
$U(1)_{\mathcal{B}}$ & $1/4$ & $1/12$ & $1/12$ & $1/3$ & $1/3$ & $7/12$ &
$1/3$ & $1/3$ & $1/3$ & $1/3$ & $0$ & $0$ & $0$\\
$U(1)_{\mathcal{L}}$ & $-3/4$ & $-1/4$ & $-1/4$ & $-1$ & $-1$ & $-3/4$ & $0$ &
$0$ & $0$ & $0$ & $1$ & $1$ & $1$\\\hline
\end{tabular}
\end{equation}
The PQ symmetry is thus indeed $U(1)_{\mathcal{B}-3\mathcal{L}}$. Note that these charges forbid all the SM fermion couplings of $S_{1}^{2/3}$ and $V_{2}^{1/3}$, as well as all other possible cubic interactions among the LQ and DQ states\footnote{Some derivative interactions are possible though, but those necessarily involve the LQs whose SM fermion couplings are forbidden, hence they do not alter the symmetry breaking pattern, and would lead to more suppressed proton decay operators.}. However, given the complicated structure shown in Fig.~\ref{Fig4b}$b$, the final operators are of dimension 16 instead of dimension 10:
\begin{align}
\mathcal{H}_{(\Delta\mathcal{B},\Delta\mathcal{L})=(1,-3)}^{eff}  &
=\frac{\phi^{4}(H^{\dagger}H)}{m_{S1/3}^{2}m_{V2/3}^{4}m_{S2/3}^{2}%
m_{V1/3}^{4}}H^{\dagger}(\bar{d}_{R}\ell_{L}+\bar{q}_{L}\nu_{R})\bar{d}%
_{R}\gamma_{\mu}\nu_{R}\bar{d}_{R}\gamma^{\mu}\nu_{R}\nonumber\\
&  \rightarrow\frac{v_{\phi}^{4}v_{EW}^{3}}{m_{S1/3}^{2}m_{V2/3}^{4}m_{S2/3}%
^{2}m_{V1/3}^{4}}(\bar{d}_{R}\ell_{L}+\bar{q}_{L}\nu_{R})\bar{d}_{R}%
\gamma_{\mu}\nu_{R}\bar{d}_{R}\gamma^{\mu}\nu_{R}\ .
\end{align}
Besides, turning on a seesaw mechanism with $\phi^{\dagger}\bar{\nu}_{R}^{\mathrm{C}}\nu_{R}$ (as $\phi\bar{\nu}_{R}^{\mathrm{C}}\nu_{R}$ would allow back some $S_{1}^{2/3}$ and $V_{2}^{1/3}$ couplings to SM fermions), a further suppression of $(v_{EW}/v_{\phi})^{2}$ to connect two $\nu_{R}$ to light fermions arises. Altogether, assuming a common scale for all the LQs, their mass can be as low as around 100 TeV when $v_{\phi}\approx10^{9}$ GeV. This is much lower than in GUT scenarios, and actually falls within the ballpark of the scale required by neutron-antineutron oscillation from Eq.~(\ref{ScaleNN}).

\subsection{DFSZ scenarios with leptoquarks and diquarks\label{SecDFSZ}}

All the scenarios discussed in the KSVZ case can readily be adapted to the DFSZ model. Basically, one removes the $\Psi_{L,R}$ state but introduces a $\phi^{2}H_{u}^{\dagger}H_{d}$ coupling, while the $\phi$ couples to various combinations of LQ/DQ states exactly as in the KSVZ scenarios. A number of peculiarities are worth mentioning though:

\begin{enumerate}
\item The symmetry patterns are more difficult to analyze in the DFSZ case because the PQ and hypercharge symmetries are entangled, see Eq.~(\ref{DFSZScalars}). Thus, further entangling $U(1)_{\mathcal{B}}$ and $U(1)_{\mathcal{L}}$ with the $U(1)$s associated to $H_{u}$ and $H_{d}$ rephasing blurs the picture completely. In practice, the PQ charges of $\phi$, $H_{u}$, and $H_{d}$ are always fixed to $PQ(H_{u})=x$, $PQ(H_{d})=-1/x$, $PQ(\phi)=\left(  x+1/x\right)  /2$, see Eq.~(\ref{DFSZScalars}), no matter the amount of $U(1)_{\mathcal{B}}$ and $U(1)_{\mathcal{L}}$ that is entangled within $U(1)_{PQ}$.

\item Because $H_{u}$, and $H_{d}$ carry PQ charges, so does the SM fermions, even without the presence of LQ/DQ states. As shown in Eq.~(\ref{DFSZfermions}), these charges have ambiguities reflecting the exact global symmetries. Thus, any entanglement of $U(1)_{\mathcal{B}}$ and $U(1)_{\mathcal{L}}$ with $U(1)_{PQ}$ will be reflected in that arbitrariness. Typically, only one free parameter will remain instead of the $\beta$ and $\gamma$ parameters of Eq.~(\ref{DFSZfermions}). Thus, looking at this remaining arbitrariness permits to identify the combination of $\beta$ and $\gamma$, i.e., $U(1)_{\mathcal{B}}$ and $U(1)_{\mathcal{L}}$, that has been spontaneously broken.

\item Because $U(1)_{PQ}$ has always a component within $U(1)_{Hu}\otimes U(1)_{Hd}$, the PQ charge of the SM fermions are never fully aligned with some combinations of $\mathcal{B}$ and $\mathcal{L}$. As a result, LQ states are often restricted to couple to only a single SM fermion LQ or DQ pair. For example, the gauge symmetries allow both $S_{2}^{1/3}\bar{d}_{R}\ell_{L}$ and $S_{2}^{1/3}\bar{q}_{L}\nu_{R}$, but the PQ charge do not since $PQ(\bar{d}_{R}\ell_{L})=\gamma-\beta+1/x$ and $PQ(\bar{q}_{L}\nu_{R})=\gamma-\beta+x$, and this is true independently of $\beta$ and $\gamma$. In some cases, this actually makes the choice of LQ/DQ couplings more natural than in the KSVZ case, since once some of them are selected, the others are immediately forbidden.

\item With $H_{u,d}$ at hand, many new couplings to LQ/DQ states are a priori possible already in the scalar potential. For instance, replacing $H$ by $H_{u}$ or $H_{d}$ in any of the couplings in Eqs.~(\ref{LQOpsB2}) or~(\ref{LQOpsBL3}) would couple the axion to $\mathcal{B}$ and/or $\mathcal{L}$ violating operators. However, these situations correspond to breaking $U(1)_{\mathcal{B}}$ and/or $U(1)_{\mathcal{L}}$ at the electroweak scale, by entangling them with $U(1)_{Y}$. Indeed, $\mathcal{B}$ and/or $\mathcal{L}$ violating operators would involve $\eta_{u}$ or $\eta_{d}$ (the pseudoscalar components of $H_{u}$ and $H_{d}$), and thus also the Would-be Goldstone associated to $U(1)_{Y}$ since $G^{0}\sim v_{u}\eta_{u}+v_{d}\eta_{d}$. The axion has only tiny $\eta_{u}$ and $\eta_{d}$ components, see Eq.~(\ref{DFSZA0}). Turning on some $H_{u}^{\dagger}H_{d}S_{j}^{\dagger}S_{i}$ couplings would prevent any $G^{0}$ coupling, but would similarly lead to tiny axion couplings via its $\cos\beta\eta_{u}-\sin\beta\eta_{d}$ component. For these reasons, all the scenarios discussed below start from coupling $\phi$ to the LQ/DQ states, so that $\mathcal{B}$ and/or $\mathcal{L}$ are broken at the PQ scale and the axion inherits some large $\mathcal{B}$ and/or $\mathcal{L}$ violating couplings. These scenarios are thus constructed exactly as in the KSVZ case.
\end{enumerate}

After these general comments, let us briefly go through each of the $\mathcal{B}$ and/or $\mathcal{L}$ spontaneous breaking scenarios.

\subsubsection{Spontaneous breaking of $\mathcal{B}+\mathcal{L}$}

By analogy with the KSVZ scenario, Eq.~(\ref{LagrKSVZ1a}), let us take%
\begin{equation}
\mathcal{L}_{\mathrm{DFSZ+LQ}}=\mathcal{L}_{\mathrm{DFSZ}}+S_{1}^{8/3}\bar
{d}_{R}e_{R}^{\mathrm{C}}+\tilde{S}_{1}^{8/3}\bar{u}_{R}^{\mathrm{C}}%
u_{R}+\phi^{2}S_{1}^{8/3\dagger}\tilde{S}_{1}^{8/3}+h.c.\ , \label{LagrDFSZ1}%
\end{equation}
with $\mathcal{L}_{\mathrm{DFSZ}}$ given in Eq.~(\ref{DFSZ0}). Solving for the $U(1)$ charges under the constraint that $PQ(H_{u})=x$, $PQ(H_{d})=-1/x$, which fixes $PQ(\phi)=\left(  x+1/x\right)  /2$, leaves a single under-determination. In this way, we identify the remaining symmetry as $U(1)_{\mathcal{B}-\mathcal{L}}$, with%
\begin{equation}%
\begin{tabular}[c]{ccccccccc}\hline
& $S_{1}^{8/3}$ & $\tilde{S}_{1}^{8/3}$ & $q_{L}$ & $u_{R}$ & $d_{R}$ &
$\ell_{L}$ & $e_{R}$ & $\nu_{R}$\\\hline
$U(1)_{PQ}$ & $\dfrac{1}{x}-x\rule[-0.14in]{0in}{0.36in}$ & $-2x$ & $0$ & $x$
& $\dfrac{1}{x}$ & $-\dfrac{1}{x}-x$ & $-x$ & $-\dfrac{1}{x}$\\
$U(1)_{\mathcal{B}-\mathcal{L}}$ & $-2/3$ & $-2/3$ & $1/3$ & $1/3$ & $1/3$ &
$-1$ & $-1$ & $-1$\\\hline
\end{tabular}
\end{equation}
with $\phi$, $H_{u}$, and $H_{d}$ neutral under $U(1)_{\mathcal{B}-\mathcal{L}}$. This shows that $U(1)_{\mathcal{B}+\mathcal{L}}\subset U(1)_{PQ}\subset U(1)_{Hu}\otimes U(1)_{Hd}\otimes U(1)_{\mathcal{B}}\otimes U(1)_{\mathcal{L}}$ is spontaneously broken. Note well that the quoted $U(1)_{PQ}$ charges are just one possible choice, since $U(1)_{\mathcal{B}-\mathcal{L}}$ remains as an ambiguity. We could also have written the charges as
\begin{equation}%
\begin{tabular}[c]{ccccccccc}\hline
& $S_{1}^{8/3}$ & $\tilde{S}_{1}^{8/3}$ & $q_{L}$ & $u_{R}$ & $d_{R}$ &
$\ell_{L}$ & $e_{R}$ & $\nu_{R}$\\\hline
$U(1)_{PQ}$ & $\dfrac{1}{x}-x-2\xi\rule[-0.14in]{0in}{0.36in}$ & $-2x-2\xi$ &
$\xi$ & $x+\xi$ & $\dfrac{1}{x}+\xi$ & $-\dfrac{1}{x}-x-3\xi$ & $-x-3\xi$ &
$-\dfrac{1}{x}-3\xi$\\\hline
\end{tabular}
\end{equation}
with $\xi$ the free parameter corresponding to $U(1)_{\mathcal{B}-\mathcal{L}}$. We can also see that this corresponds to Eq.~(\ref{DFSZfermions}) with $\beta=\xi$ and $\gamma=-\dfrac{1}{x}-x-3\xi$. This shows that the dimension-five axion to gauge boson couplings are unaffected by the LQ/DQ, since they are independent of $\beta$ and $\gamma$. Also, one should not conclude that the axion does not couple to $q_{L}$, even though that coupling is absent from the axion effective Lagrangian since $PQ(q_{L})$ is set to zero.

Concerning the axion $\mathcal{B}+\mathcal{L}$ violating operator, the same effective interactions arises as in the KSVZ scenario, see Eq.~(\ref{EffHKSVZ1}). This is evident from Fig.~\ref{Fig2}, which is independent of how the axion emerges. The only difference is that the pseudoscalar component of $\phi$ is not purely the axion, but this is only a totally negligible $\mathcal{O}(v_{EW}/vs)$ effect, see Eq.~(\ref{DFSZA0}).

Finally, exactly as in the KSVZ scenario, the remaining $U(1)_{\mathcal{B}-\mathcal{L}}$ freedom permits to set up a PQ-induced seesaw mechanism by adding $\phi\bar{\nu}_{R}^{\mathrm{C}}\nu_{R}$ or $\phi^{\dagger}\bar{\nu}_{R}^{\mathrm{C}}\nu_{R}$. In both cases, this simply fixes the parameter $\xi$ and removes the remaining $U(1)_{\mathcal{B}-\mathcal{L}}$ ambiguity. Yet, the final PQ charges do not reflect at all the peculiar symmetry breaking pattern, with $U(1)_{\mathcal{B}+\mathcal{L}}$ and $U(1)_{\mathcal{L}}$ being separately, but concurrently, spontaneously broken at the PQ scale. By the way, exactly the same PQ charges arise if $\phi^{2}S_{1}^{8/3\dagger}\tilde{S}_{1}^{8/3}$ is replaced by $H_{d}^{\dagger}H_{u}S_{1}^{8/3\dagger}\tilde{S}_{1}^{8/3}$, though as discussed before, the symmetry breaking chain is very different, as are the axion couplings.

\subsubsection{Spontaneous breaking of $\mathcal{B}-\mathcal{L}$}

Pursuing our adaptation of the KSVZ scenario, let us consider now%
\begin{equation}
\mathcal{L}_{\mathrm{DFSZ+LQ}}=\mathcal{L}_{\mathrm{DFSZ}}+S_{1}^{4/3}\bar
{d}_{R}^{\mathrm{C}}d_{R}+S_{2}^{7/3}\bar{u}_{R}\ell_{L}+\phi H_{u}%
S_{2}^{7/3\dagger}S_{1}^{4/3}+h.c.\ . \label{LagrDFSZ2}%
\end{equation}
Both fermionic couplings of $S_{2}^{7/3}$ cannot be present at the same time for the PQ symmetry to exist, so we choose to keep $S_{2}^{7/3}\bar{u}_{R}\ell_{L}$ and discard $S_{2}^{7/3}\bar{q}_{L}e_{R}$. Also, we introduced $H_{u}$ in the quartic scalar coupling, but could equally well have used $H_{d}$. From this Lagrangian, the PQ charges are found to be
\begin{equation}%
\begin{tabular}[c]{ccccccccc}\hline
& $S_{1}^{4/3}$ & $S_{2}^{7/3}$ & $q_{L}$ & $u_{R}$ & $d_{R}$ & $\ell_{L}$ &
$e_{R}$ & $\nu_{R}$\\\hline
$U(1)_{PQ}$ & $-\dfrac{2}{x}\rule[-0.14in]{0in}{0.36in}$ & $\dfrac{3x}%
{2}-\dfrac{3}{2x}$ & $0$ & $x$ & $\dfrac{1}{x}$ & $\dfrac{3}{2x}-\dfrac{x}{2}$
& $\dfrac{5}{2x}-\dfrac{x}{2}$ & $\dfrac{3}{2x}+\dfrac{x}{2}$\\
$U(1)_{\mathcal{B}+\mathcal{L}}$ & $-2/3$ & $-2/3$ & $1/3$ & $1/3$ & $1/3$ &
$1$ & $1$ & $1$\\\hline
\end{tabular}
\end{equation}
So, $U(1)_{\mathcal{B}-\mathcal{L}}$ is spontaneously broken at the PQ scale, but $U(1)_{\mathcal{B}+\mathcal{L}}$ remains. As before, we could rewrite the PQ charge introducing a free parameter to reflect the exact $U(1)_{\mathcal{B}+\mathcal{L}}$ symmetry, hence one should not interpret $PQ(q_{L})=0$ as meaning it has no coupling to the axion. The $U(1)_{\mathcal{B}+\mathcal{L}}$ ambiguity can then be used to allow for a $\phi\bar{\nu}_{R}^{\mathrm{C}}\nu_{R}$ or $\phi^{\dagger}\bar{\nu}_{R}^{\mathrm{C}}\nu_{R}$ coupling, and set up the seesaw mechanism.

The final $(\Delta\mathcal{B},\Delta\mathcal{L})=(1,-1)$ operator is again one of the dimension-seven operators listed in Table~\ref{TableLQBL}. Note, though, that because the PQ symmetry restricts the LQ couplings to SM fermions, only a single operator is induced. This is a generic characteristic of the DFSZ implementation: compared to the KSVZ case, it is more restrictive. Phenomenologically, this could show up as definite decay patterns for the proton (if ever observed). Starting from Eq.~(\ref{LagrDFSZ2}), the operator arising at tree level is
\begin{equation}
\mathcal{H}_{(\Delta\mathcal{B},\Delta\mathcal{L})=(1,-1)}^{eff}=\frac{1}{m_{S}^{2}m_{\tilde{S}}^{2}}\phi H_{u}\bar{d}_{R}d_{R}^{\mathrm{C}}\bar
{u}_{R}\ell_{L}+h.c.\ ,
\end{equation}
Note that some other gauge and PQ invariant operators may arise at higher loops via Yukawa insertions, but those are more suppressed. The leading proton decay operator is thus proportional to $v_{\phi}v_{u}/m_{S}^{4}$, and the constraints are similar as in the KSVZ scenario. Concerning the axion coupling, notice that
\begin{equation}
\phi H_{u}\ell_{L}\rightarrow\frac{1}{2}v_{u}v_{\phi}\exp i\left(  \frac{\eta
_{u}}{v_{u}}+\frac{v_{\phi}}{v_{\phi}}\right)  \nu_{L}\ ,
\end{equation}
so the combination that occurs in the effective operator is
\begin{equation}
\mathcal{H}_{(\Delta\mathcal{B},\Delta\mathcal{L})=(1,-1)}^{eff}=\frac
{v_{\phi}v_{u}}{m_{S}^{2}m_{\tilde{S}}^{2}}\left(  1+i\frac{G^{0}}{v}%
+i\frac{a^{0}}{v_{\phi}}\frac{3x^{2}+1}{x^{2}+1}+i\frac{\pi^{0}}{v}x\right)
\bar{d}_{R}d_{R}^{\mathrm{C}}\bar{u}_{R}\nu_{L}\ .
\end{equation}
For comparison, the $\mu H_{u}S_{2}^{7/3\dagger}S_{1}^{4/3}$ coupling would lead to
\begin{equation}
\mathcal{H}_{(\Delta\mathcal{B},\Delta\mathcal{L})=(1,-1)}^{eff}=\frac{\mu
v_{u}}{m_{S}^{2}m_{\tilde{S}}^{2}}\left(  1+i\frac{G^{0}}{v}+i\frac{a^{0}%
}{v_{\phi}}\frac{2x^{2}}{x^{2}+1}+i\frac{\pi^{0}}{v}x\right)  \bar{d}_{R}%
d_{R}^{\mathrm{C}}\bar{u}_{R}\nu_{L}\ ,
\end{equation}
with $\mu$ some mass scale. The $(\Delta\mathcal{B},\Delta\mathcal{L})=(1,-1)$ operator arises at the $v_{\phi}$ scale from $\phi H_{u}S_{2}^{7/3\dagger}S_{1}^{4/3}$, but at a lower scale from $H_{u}S_{2}^{7/3\dagger}S_{1}^{4/3}$ since we would expect $\mu$ to be at the LQ/DQ scale, $\mu\sim m_{S}$, or even at the electroweak scale, $\mu\sim v$. Note that in both cases, the $G^{0}$ enters as expected for a would-be Goldstone, and would disappear in the unitary gauge. The axion coupling is $\mathcal{O}(v_{EW}/v_{\phi})$ compared to the four-fermion operator, exactly like in the KSVZ scenario.

\subsubsection{Spontaneous breaking of $\mathcal{B}$}

Neutron-antineutron oscillations can be induced in the same way in the DFSZ and KSVZ models, see Fig.~\ref{Fig4a}. Starting with
\begin{equation}
\mathcal{L}_{\mathrm{DFSZ+LQ}}=\mathcal{L}_{\mathrm{DFSZ}}+S_{1}^{4/3}\bar
{d}_{R}^{\mathrm{C}}d_{R}+S_{1}^{8/3}\bar{u}_{R}^{\mathrm{C}}u_{R}+\phi
S_{1}^{4/3}S_{1}^{4/3}S_{1}^{8/3}+h.c.\ , \label{LagrDFSZ3}%
\end{equation}
where $S_{1}^{4/3}\sim(\mathbf{3},\mathbf{1},+4/3)$ and $S_{1}^{8/3}\sim(\mathbf{\bar{6}},\mathbf{1},-8/3)$, we get the PQ charges%
\begin{equation}%
\begin{tabular}[c]{ccccccccc}\hline
& $S_{1}^{4/3}$ & $S_{1}^{8/3}$ & $q_{L}$ & $u_{R}$ & $d_{R}$ & $\ell_{L}$ &
$e_{R}$ & $\nu_{R}$\\\hline
$U(1)_{PQ}$ & $\dfrac{x}{2}-\dfrac{5}{6x}\rule[-0.14in]{0in}{0.36in}$ &
$\dfrac{7}{6x}-\dfrac{3x}{2}$ & $-\dfrac{x}{4}-\dfrac{7}{12x}$ & $\dfrac
{3x}{4}-\dfrac{7}{12x}$ & $\dfrac{5}{12x}-\dfrac{x}{4}$ & $0$ & $\dfrac{1}{x}$
& $x$\\
$U(1)_{\mathcal{L}}$ & $0$ & $0$ & $0$ & $0$ & $0$ & $1$ & $1$ & $1$\\\hline
\end{tabular}
\end{equation}
Thus, $U(1)_{\mathcal{B}}$ is broken spontaneously, but $U(1)_{\mathcal{L}}$ stays exact. The phenomenology is the same as in the KSVZ model, see Fig.~\ref{Fig4a} and Eq.~(\ref{ScaleNN}). Majorana neutrino masses can be generated spontaneously with $\phi^{\dagger}\bar{\nu}_{R}^{\mathrm{C}}\nu_{R}$, but not with $\phi\bar{\nu}_{R}^{\mathrm{C}}\nu_{R}$ as the PQ charges of the leptons would then allow for the $S_{1}^{4/3}\bar{u}_{R}\nu_{R}^{\mathrm{C}}$ coupling, and thereby to tree-level proton decay.

\subsubsection{Spontaneous breaking of $\mathcal{B}\pm3\mathcal{L}$}

The last two scenarios are those producing exotic $(\Delta\mathcal{B},\Delta\mathcal{L})=(1,\pm3)$ proton decay operators. Let us start with the $(\Delta\mathcal{B},\Delta\mathcal{L})=(1,3)$ case, and the Lagrangian
\begin{equation}
\mathcal{L}_{\mathrm{DFSZ+LQ}}=\mathcal{L}_{\mathrm{DFSZ}}+S_{1}^{2/3}\bar
{q}_{L}\ell_{L}^{\mathrm{C}}+V_{2,\mu}^{1/3}(\bar{u}_{R}\gamma^{\mu}\ell
_{L}^{\mathrm{C}}+\bar{q}_{L}\gamma^{\mu}\nu_{R}^{\mathrm{C}})+\phi^{\dagger}
S_{1}^{2/3}V_{2,\mu}^{1/3}V_{2}^{1/3,\mu}+h.c.\ . \label{LagrDFSZ4}%
\end{equation}
The $S_{1}^{2/3}(\bar{d}_{R}\nu_{R}^{\mathrm{C}}+\bar{u}_{R}e_{R}^{\mathrm{C}})$ and $S_{1}^{2/3}\bar{q}_{L}\ell_{L}^{\mathrm{C}}$ couplings cannot both be present, and we take only the latter, while the $V_{2,\mu}^{1/3}(\bar{u}_{R}\gamma^{\mu}\ell_{L}^{\mathrm{C}}+\bar{q}_{L}\gamma^{\mu}\nu _{R}^{\mathrm{C}})$ couplings are compatible with each other. The $U(1)$ charges are then
\begin{equation}%
\begin{tabular}[c]{ccccccccc}\hline
& $S_{1}^{2/3}$ & $V_{2,\mu}^{1/3}$ & $q_{L}$ & $u_{R}$ & $d_{R}$ & $\ell_{L}$
& $e_{R}$ & $\nu_{R}$\\\hline
$U(1)_{PQ}$ & $\dfrac{1}{6x}-\dfrac{x}{2}\rule[-0.14in]{0in}{0.36in}$ &
$\dfrac{x}{2}+\dfrac{1}{6x}$ & $0$ & $x$ & $\dfrac{1}{x}$ & $\dfrac{1}%
{6x}-\dfrac{x}{2}$ & $\dfrac{7}{6x}-\dfrac{x}{2}$ & $\dfrac{1}{6x}+\dfrac
{x}{2}$\\
$U(1)_{3\mathcal{B}-\mathcal{L}}$ & $0$ & $0$ & $1$ & $1$ & $1$ & $-1$ & $-1$
& $-1$\\\hline
\end{tabular}
\end{equation}
The $U(1)_{3\mathcal{B}-\mathcal{L}}$ symmetry remains, and its orthogonal combination $U(1)_{\mathcal{B}+3\mathcal{L}}$ is spontaneously broken. Dimension-nine $(\Delta\mathcal{B},\Delta\mathcal{L})=(1,3)$ proton decay operators thus appear at the low scale (as well as semileptonic $(\Delta\mathcal{B},\Delta\mathcal{L})=(0,0)$ operators since these charges allow for the $D^{\mu}HS_{1}^{2/3}V_{2,\mu}^{1/3\dagger}$ couplings). Once more, there is enough room for a seesaw mechanism with $\phi\bar{\nu}_{R}^{\mathrm{C}}\nu_{R}$ and/or $\phi^{\dagger}\bar{\nu}_{R}^{\mathrm{C}}\nu_{R}$. Depending on the LQ couplings of $S_{1}^{2/3}$, it is always possible to choose the seesaw operator that sets PQ charges forbidding the DQ couplings of both $S_{1}^{2/3}$ and $V_{2,\mu}^{1/3}$, and thus proton decay via dimension-six operators.

Concerning the $(\Delta\mathcal{B},\Delta\mathcal{L})=(1,-3)$ operators, we start from
\begin{align}
\mathcal{L}_{\mathrm{DFSZ+LQ}}  &  =\mathcal{L}_{\mathrm{DFSZ}}+S_{2}%
^{1/3}\bar{d}_{R}\ell_{L}+V_{1,\mu}^{2/3}\bar{d}_{R}\gamma^{\mu}\nu
_{R}\nonumber\\
&  +\phi(H_{u}S_{2}^{1/3\dagger}S_{1}^{2/3}+H_{d}^{\dagger}V_{1}%
^{2/3\dagger,\mu}V_{2,\mu}^{1/3}+S_{1}^{2/3\dagger}V_{2,\mu}^{1/3\dagger}
V_{2}^{1/3,\mu\dagger})+h.c.\ , \label{LagrDFSZ5}%
\end{align}
with the $S_{2}^{1/3}\bar{q}_{L}\nu_{R}$ removed. Several choices are possible for introducing the doublets $H_{u}$ and $H_{d}$ in these couplings, and we opt for the one most symmetrical with the Yukawa couplings, see Eq.~(\ref{DFSZ0}). Only one of the fermionic couplings of $S_{2}^{1/3}$ can be turned on, and we choose $S_{2}^{1/3}\bar{d}_{R}\ell_{L}$. Then, the $U(1)$
charges are found to be
\begin{equation}%
\begin{tabular}
[c]{ccccccccccc}\hline
& $S_{1}^{2/3}$ & $V_{2,\mu}^{1/3}$ & $S_{2}^{1/3}$ & $V_{1,\mu}^{2/3}$ &
$q_{L}$ & $u_{R}$ & $d_{R}$ & $\ell_{L}$ & $e_{R}$ & $\nu_{R}$\\\hline
$U(1)_{PQ}\rule[-0.14in]{0in}{0.39in}$ & $\dfrac{x^{2}+5}{6x}$ & $\dfrac
{x^{2}-1}{6x}$ & $\dfrac{5x^{2}+4}{3x}$ & $\dfrac{2x^{2}+4}{3x}$ & $0$ & $x$ &
$\dfrac{1}{x}$ & $\dfrac{5x^{2}+1}{-3x}$ & $\dfrac{5x^{2}-2}{-3x}$ &
$\dfrac{2x^{2}+1}{-3x}$\\
$U(1)_{3\mathcal{B}+\mathcal{L}}$ & $0$ & $0$ & $0$ & $0$ & $1$ & $1$ & $1$ &
$1$ & $1$ & $1$\\\hline
\end{tabular}
\end{equation}
This time, $U(1)_{3\mathcal{B}+\mathcal{L}}$ remains and $U(1)_{\mathcal{B}-3\mathcal{L}}$ is spontaneously broken. The induced operator, from a process easily adapted from that of Fig.~\ref{Fig4b}, is
\begin{equation}
\mathcal{H}_{(\Delta\mathcal{B},\Delta\mathcal{L})=(1,-3)}^{eff}=\frac
{\phi^{4}(H_{d}^{\dagger}H_{d}^{\dagger}H_{u})}{m_{S1/3}^{2}m_{V2/3}^{4}m_{S2/3}^{2}%
m_{V1/3}^{4}}\bar{d}_{R}\ell_{L}\bar{d}_{R}\gamma_{\mu}\nu
_{R}\bar{d}_{R}\gamma^{\mu}\nu_{R}\ .
\end{equation}
Again, phenomenologically, there is not much difference between the DFSZ and KSVZ implementation.

\subsection{Axion-induced proton decay and neutron-antineutron oscillations\label{SecSpont}}

In both the KSVZ and DFSZ cases, we can induce spontaneously proton decay or neutron-antineutron oscillations. But, in all the scenarios discussed up to now, the processes involving the axion were $\mathcal{O}(v_{EW}/v_{\phi})$ with respect to that without it. The reason is of course that in all cases, some coupling of $\phi$ to the LQ/DQ states was introduced, and $\phi=(v_{\phi}+\rho)\exp(i\eta_{\phi}/v_{\phi})\approx v_{\phi}+\rho+ia^{0}$ (see Eq.~(\ref{ExampleSSS})). The purpose here is to kill off the leading term, leaving only axion-induced $\mathcal{B}$ and/or $\mathcal{L}$ violating processes. The only way to achieve this is to consider derivative couplings of $\phi$ to pairs of LQ/DQs, and there are only three renormalizable options
\begin{equation}
\partial_{\mu}\phi S_{2}^{1/3\dagger}V_{2}^{1/3,\mu}\ ,\ \partial_{\mu}\phi
S_{1}^{2/3\dagger}V_{1}^{2/3,\mu}\ ,\ \partial_{\mu}\phi S_{1}^{4/3\dagger
}V_{1}^{4/3,\mu}.
\label{DerScenars}
\end{equation}
In these cases, the axion enters as $\partial_{\mu}\phi\approx\partial_{\mu}\rho+i\partial_{\mu}a^{0}$, without a leading term tuned by $v_{\phi}$. Though we have not attempted at constructing UV complete models generating such interactions, their structure is evocative of that which could arise if both $\phi$ and scalar LQ/DQ were somehow related to the fields giving masses to the vector LQ/DQ. Such a situation can happen in simple GUT models: In Ref.~\cite{Wise:1981ry} for example, $\phi$ is identified with the phase of the complex $H_{24}$ field breaking $SU(5)$ down to the SM gauge group. Note, though, that the PQ breaking scale and the LQ/DQ mass scale would be related in such models. In the present section, the two will be kept independent, with the latter usually much lower than the former.

Let us see which symmetry breaking patterns can be achieved with these building blocks. We will use the KSVZ setting throughout as the alignments of the PQ with some combination of $\mathcal{B}$ and $\mathcal{L}$ are manifest, but the adaptation to the DFSZ scenario is immediate. Also, we will discard $\Psi_{L,R}$ from the discussion. As in Sec.~\ref{SecKSVZ}, their charge can always be set separately by introducing some couplings to the LQ/DQ.

\subsubsection{Spontaneous breaking of $\mathcal{B}-\mathcal{L}$}

The scenarios with $(\Delta\mathcal{B},\Delta\mathcal{L})=(1,-1)$ operators are immediately obtained using any one of the three couplings in Eq.~(\ref{DerScenars}). For example, we can take
\begin{equation}
\mathcal{L}_{\mathrm{KSVZ+LQ}}=\mathcal{L}_{\mathrm{KSVZ}}+S_{1}^{2/3}(\bar
{q}_{L}^{\mathrm{C}}q_{L}+\bar{d}_{R}^{\mathrm{C}}u_{R})+V_{1,\mu}^{2/3}%
\bar{d}_{R}\gamma^{\mu}\nu_{R}+\partial_{\mu}\phi S_{1}^{2/3\dagger}%
V_{1}^{2/3,\mu}+h.c.\ , \label{LagrSSB1}%
\end{equation}
and get two active $U(1)$s, with charges
\begin{equation}%
\begin{tabular}[c]{cccccccccc}\hline
& $\phi$ & $S_{1}^{2/3}$ & $V_{1,\mu}^{2/3}$ & $q_{L}$ & $u_{R}$ & $d_{R}$ &
$\ell_{L}$ & $e_{R}$ & $\nu_{R}$\\\hline
$U(1)_{\mathcal{B}}$ & $-1$ & $-2/3$ & $1/3$ & $1/3$ & $1/3$ & $1/3$ & $0$ &
$0$ & $0$\\
$U(1)_{\mathcal{L}}$ & $1$ & $0$ & $-1$ & $0$ & $0$ & $0$ & $1$ & $1$ &
$1$\\\hline
\end{tabular}
\end{equation}
Thus, $\phi$ spontaneously breaks $U(1)_{\mathcal{B}-\mathcal{L}}$, leaving $\mathcal{B}+\mathcal{L}$ as an exact global symmetry. As before, we can add a $\phi\bar{\nu}_{R}^{\mathrm{C}}\nu_{R}$ coupling to generate neutrino masses and forbid the LQ couplings of $S_{1}^{2/3}$.

\begin{figure}[t]
\begin{center}
\includegraphics[height=1.5766in,width=4.7547in]{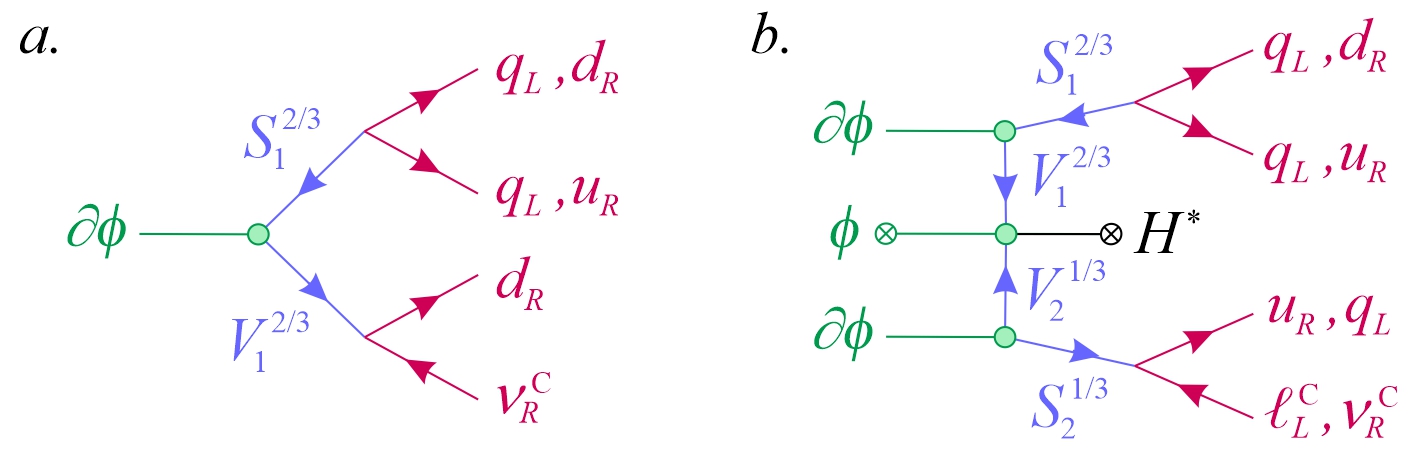}%
\caption{The $(\Delta\mathcal{B},\Delta\mathcal{L})=(1,-1)$ operators involving one ($a.$) and two axions ($b.$).}%
\label{Fig5}%
\end{center}
\end{figure}

The situation starting from the other derivative interaction is similar, hence we can generate:
\begin{align}
\partial_{\mu}\phi S_{1}^{2/3\dagger}V_{1}^{2/3,\mu}  &  \rightarrow
\mathcal{H}_{(\Delta\mathcal{B},\Delta\mathcal{L})=(1,-1)}^{eff}=\frac
{1}{m_{S}^{2}m_{V}^{2}}\partial_{\mu}\phi(\bar{q}_{L}^{\mathrm{C}}q_{L}%
+\bar{d}_{R}^{\mathrm{C}}u_{R})\bar{d}_{R}^{\mathrm{C}}\gamma^{\mu}\nu
_{R}^{\mathrm{C}}\ ,\\
\partial_{\mu}\phi S_{2}^{1/3\dagger}V_{2}^{1/3,\mu}  &  \rightarrow
\mathcal{H}_{(\Delta\mathcal{B},\Delta\mathcal{L})=(1,-1)}^{eff}=\frac
{1}{m_{S}^{2}m_{V}^{2}}\partial_{\mu}\phi\bar{d}_{R}^{\mathrm{C}}\gamma^{\mu
}q_{L}(\bar{d}_{R}^{\mathrm{C}}\ell_{L}^{\mathrm{C}}+\bar{q}_{L}^{\mathrm{C}%
}\nu_{R}^{\mathrm{C}})\ ,\\
\partial_{\mu}\phi S_{1}^{4/3\dagger}V_{1}^{4/3,\mu}  &  \rightarrow
\mathcal{H}_{(\Delta\mathcal{B},\Delta\mathcal{L})=(1,-1)}^{eff}=\frac
{1}{m_{S}^{2}m_{V}^{2}}\partial_{\mu}\phi\bar{d}_{R}^{\mathrm{C}}d_{R}(\bar
{u}_{R}^{\mathrm{C}}\gamma^{\mu}\nu_{R}^{\mathrm{C}}+\bar{d}_{R}^{\mathrm{C}%
}\gamma^{\mu}e_{R}^{\mathrm{C}}+\bar{q}_{L}^{\mathrm{C}}\gamma^{\mu}\ell
_{L}^{\mathrm{C}})\ .
\end{align}
All these situations induce $(\Delta\mathcal{B},\Delta\mathcal{L})=(1,-1)$ operators, see Fig.~\ref{Fig5}$a$, but necessarily involving the axion, with for example
\begin{align}
\frac{\partial_{\mu}\phi}{m_{S}^{2}m_{V}^{2}}\bar{d}_{R}\gamma^{\mu}%
q_{L}^{\mathrm{C}}\bar{d}_{R}\ell_{L}+h.c. & \rightarrow\frac{1}{m_{S}^{2}%
m_{V}^{2}}\partial_{\mu}a^{0}\bar{d}_{R}\gamma^{\mu}q_{L}^{\mathrm{C}}\bar
{d}_{R}\ell_{L}+h.c. \nonumber \\ 
 & \rightarrow\frac{\Lambda_{QCD}^{3}}{m_{S}^{2}m_{V}^{2}%
}(\partial_{\mu}a^{0}\bar{p}\gamma^{\mu}\left(  1-\gamma^{5}\right)
\ell + \partial_{\mu}a^{0}\bar{n}\gamma^{\mu}\left(  1-\gamma^{5}\right)
\nu + ... + h.c.)\ ,
\label{OpsAPL}
\end{align}
where (...) denotes operators with additional light mesons. Given the proton decay bounds, and taking $\Lambda_{QCD}\approx 300$~MeV, this imposes a quite high bound $m_{S}\approx m_{V}>10^{4}$~TeV, close to the PQ breaking scale and quite lower than the GUT scale. With those values, such $(\Delta\mathcal{B},\Delta\mathcal{L})=(1,-1)$ operators cannot affect significantly the phenomenology of the axion, as its coupling to photons or gluons remain much larger.

The situation can be different for an axion-like particle (ALP) with a mass above that of the proton. If the mass is just slightly above but below that of the neutron, $m_{p}-m_{e}<m_{a}<m_{n}$, it could explain the neutron lifetime anomaly by opening up $n\rightarrow a^{0}+\nu$, realizing the mechanism proposed in Ref.~\cite{NeutronTau}. The branching ratio for that latter process can reach $1\%$ for $m_{S}^{2}\approx m_{V}^{2}$ around the TeV, i.e., very close to the bounds from direct LHC searches~\cite{CMS:2020gru,ATLAS:2020dsk,ATLAS:2020xov,CMS:2020wzx,ATLAS:2021oiz}. Such a large rate may be surprising for TeV scale LQ/DQ, but can be understood from the fact that most of the $m_S$ powers in the operator of Eq.~(\ref{OpsAPL}) are compensated by $\Lambda_{QCD}$. For comparison, the electroweak neutron beta decay rate is proportional to $G_F m_e^5$, with $m_e$ the electron mass and $G_F \sim 1/v_{EW}^2$, bringing a much harsher scale suppression.

\begin{figure}[t]
\begin{center}
\includegraphics[width=11cm]{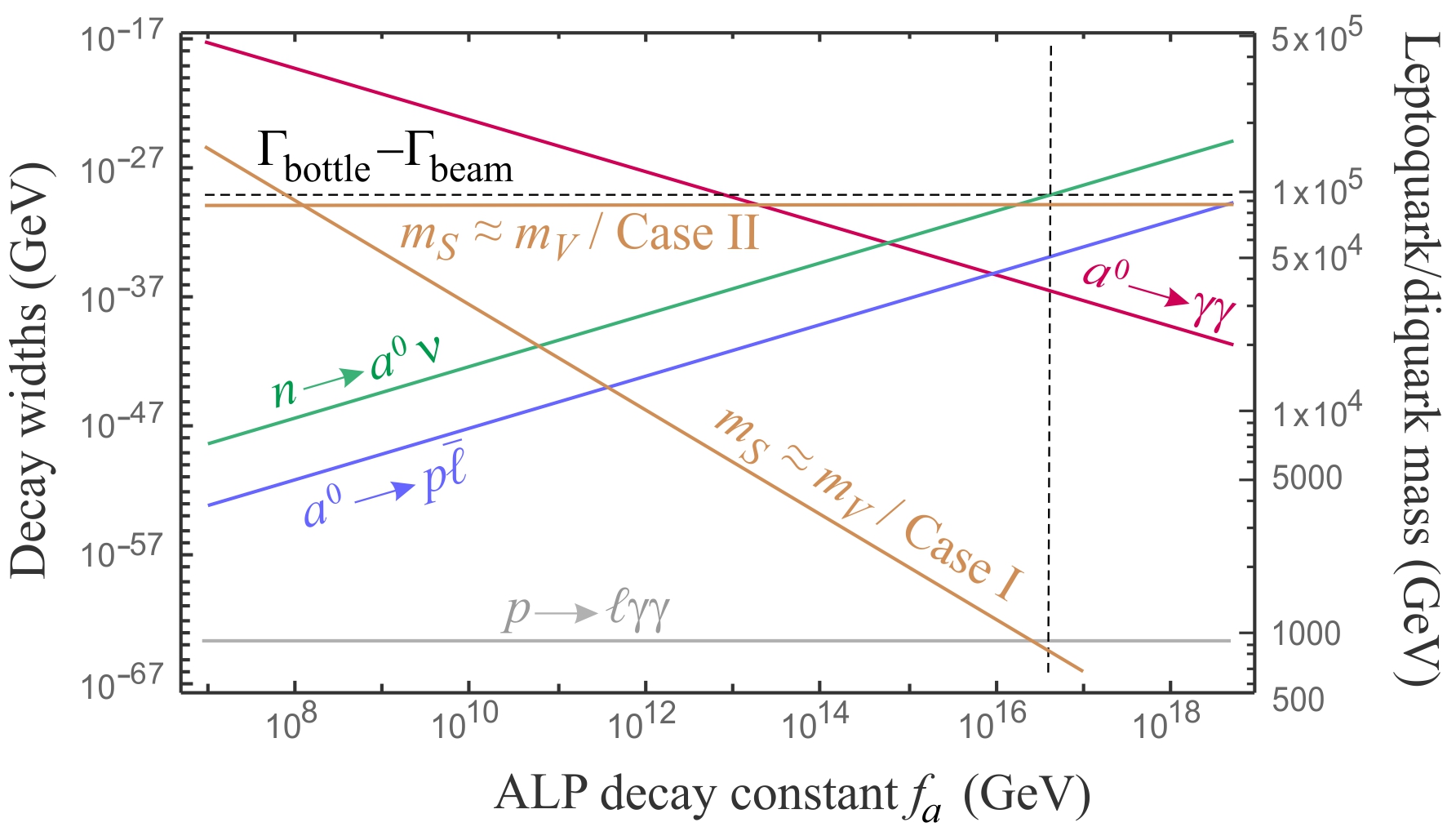}%
\caption{Evolution of the $a^{0}\rightarrow\gamma\gamma$, $a^{0}\rightarrow p\bar{\ell}$, $n\rightarrow a^{0}+\nu$ and $p\rightarrow\ell(a^{0\ast}\rightarrow\gamma\gamma)$ widths as functions of $f_a$. The dashed line indicates the observed neutron lifetime discrepancy, $\Gamma_{bottle}-\Gamma_{beam} \approx 7.1\times 10^{-30}$~GeV~\cite{NeutronTau}. The ALP mass is kept fixed at $m_a =0.9384$~GeV. The LQ/DQ mass is adjusted so that $\tau \left( p\rightarrow\ell(a^{0\ast}\rightarrow\gamma\gamma) \right) = 10^{32}$~yr. For the $(\Delta\mathcal{B},\Delta\mathcal{L})=(1,-1)$ operator, Eq.~(\ref{OpsAPL}), $m_{S}\approx m_{V}$ then follows the indicated line (Case I), and must be a bit below the TeV to reproduce the observed $\Gamma_{bottle}-\Gamma_{beam}$ discrepancy. Concerning the $(\Delta\mathcal{B},\Delta\mathcal{L})=(1,1)$ scenario, for which leptons and antileptons should be interchanged, the extra factor of $f_a$ in Eq.~(\ref{OpsAPantiL}) sets $m_{S}\approx m_{V}\approx 90$~TeV independently of $f_a$ (Case II).}
\label{FigNeut}%
\end{center}
\end{figure}

For this scenario to be viable, the ALP should not decay too quickly to light particles, since that would allow $p\rightarrow e(a^{0\ast}\rightarrow X)$ at an unacceptable rate. Usually, ALPs have two gluon and/or two photon couplings. To analyze this situation, let us turn on the usual ALP photon coupling tuned by some very high $f_{a}$ scale:
\begin{equation}
\frac{e^{2}}{f_{a}}a^{0}F_{\mu\nu}\tilde{F}^{\mu\nu}\rightarrow\Gamma
(a^{0}\rightarrow\gamma\gamma)=\frac{4\pi\alpha^{2}m_{a}^{3}}{f_{a}^{2}}\ .
\end{equation}
By comparison, the decay rate into proton plus antilepton (or antiproton plus lepton) is
\begin{equation}
\Gamma(a^{0}\rightarrow p\bar{\ell})=\frac{m_{a}}{4\pi}\left(  1-\frac{m_{p}^{2}%
}{m_{a}^{2}}\right)  ^{2}\left(  \frac{m_{p}\Lambda_{QCD}^{3}}{m_{S}^{2}%
m_{V}^{2}}\right)  ^{2}\ ,
\end{equation}
while the neutron decay rate into ALP is
\begin{equation}
\Gamma(n\rightarrow a^{0}\nu)=\frac{m_{n}}{4\pi}\left(  1-\frac{m_{a}^{2}%
}{m_{n}^{2}}\right)  ^{2}\left(  \frac{m_{n}\Lambda_{QCD}^{3}}{m_{S}^{2}%
m_{V}^{2}}\right)  ^{2}\ .
\label{neutdec}
\end{equation}
Now, kinematically, proton decay can occur via the process $p\rightarrow\ell(a^0\rightarrow\gamma\gamma)$, whose decay rate is ($r_{a}\equiv m_{a}/m_{p}$, and the lepton mass is set to zero)
\begin{equation}
\Gamma\left(  p\rightarrow\ell(a^{0\ast}\rightarrow\gamma\gamma)\right)
=\frac{2\alpha^{2}m_{p}^{3}}{\pi f_{a}^{2}}\left(  \frac{m_{p}\Lambda_{QCD}%
^{3}}{m_{S}^{2}m_{V}^{2}}\right)  ^{2}\int_{0}^{1}dz\frac{z^{2}\left(
1-z\right)  ^{2}}{(r_{a}^{2}-z)^{2}}\overset{r_{a}\rightarrow1}{\approx}%
\frac{2\alpha^{2}m_{p}^{3}}{3\pi f_{a}^{2}}\left(  \frac{m_{p}\Lambda_{QCD}%
^{3}}{m_{S}^{2}m_{V}^{2}}\right)  ^{2}\;.
\end{equation}
For $m_{p}<m_{a}<m_{n}$, if we impose that the $p\rightarrow\ell(a^{\ast}\rightarrow\gamma\gamma)$ lifetime is greater than $10^{32}$ years, 
\begin{equation}
m_S \approx m_V > 900\ \text{GeV} \left( \frac{10^{16}\ \text{GeV}}{f_a} \right)^{1/4}\;.
\end{equation}
Plugging this in Eq.~(\ref{neutdec}), the branching ratio for $n\rightarrow a^{0}+\nu$ is at around $1\%$ provided $f_{a}$ is pushed at the GUT scale, $f_a \approx 10^{16}\ \text{GeV}$, see Fig.~\ref{FigNeut}. Note that for that value of $f_{a}$, the ALP still decays mostly into $\gamma\gamma$ as $\Gamma(a^{0}\rightarrow p\bar{\ell})>\Gamma(a^{0}\rightarrow\gamma\gamma)$ requires $f_{a}$ about an order of magnitude larger. The $p\rightarrow\ell(a^{0\ast}\rightarrow\gamma\gamma)$ decay can happen only for $\ell=e,\mu$, but the underlying LQ couplings could actually exhibit non-trivial flavor hierarchies. If they couple preferentially to the $\tau$, then proton decay would be forced to occur via more suppressed channels, e.g. via $p\rightarrow\pi\nu_{\tau}(a^{0\ast}\rightarrow\gamma\gamma)$, and $f_{a}$ could be brought down by a few orders of magnitude. Thus, an ALP could realize the scenario proposed in Ref.~\cite{NeutronTau} to solve the neutron lifetime puzzle, though it does not alleviate its inherent fine tuning of the dark particle mass. 

For heavier ALPs, neutron decay is kinematically closed, and the $\Gamma(a^{0}\rightarrow p\bar{\ell},n\bar{\nu})>\Gamma(a^{0}\rightarrow\gamma\gamma)$ pattern can arise for lower $f_{a}$ values (though still very large from the axion point of view), with for example $\Gamma(a^{0}\rightarrow p\bar{\ell},n\bar{\nu})>\Gamma(a^{0}\rightarrow\gamma\gamma)$ for $f_{a}>10^{13}$ GeV if $m_{a}=100~$GeV. This, however, requires also to boost the $a\rightarrow p\bar{\ell},n\bar{\nu}$ rate by allowing light LQ/DQ at around the TeV scale. Even if $\Gamma(a^{0}\rightarrow p\bar{\ell},n\bar{\nu})$ does not dominate, such decay channels could have some cosmological implications. From a baryogenesis point of view, it is interesting to remark that the present scenario has all the necessary ingredients. Provided the LQ/DQ couple to more than one SM fermion states, several operators will simultaneously contribute to the $a^{0}\rightarrow p\bar{\ell},n\bar{\nu}$ and $a^{0}\rightarrow\bar{p}\ell,\bar{n}\nu$ decay processes, and since the LQ/DQ couplings to SM fermions are a priori complex, their rates would be different (slightly, as rescattering is required). In this picture, note that if there are several LQ/DQ states with a non-trivial mass spectrum, their decay chains may first generate an asymmetry when $m_{S,V}>m_{a}$~\cite{Arnold:2012sd}, but it would be washed out and regenerated at a lower scale by the ALP decays. Whether this mechanism is sufficient to generate the observed baryon asymmetry is left for a future study.

To close this section, let us mention another scenario in which $(\Delta\mathcal{B},\Delta\mathcal{L})=(1,-1)$ operators require two derivative couplings, and proton decay is associated to axion pair production. Specifically, if we start from
\begin{align}
\mathcal{L}_{\mathrm{KSVZ+LQ}}  &  =\mathcal{L}_{\mathrm{KSVZ}}+S_{1}%
^{2/3}(\bar{q}_{L}^{\mathrm{C}}q_{L}+\bar{d}_{R}^{\mathrm{C}}u_{R}%
)+S_{2}^{1/3}(\bar{d}_{R}\ell_{L}+\bar{q}_{L}\nu_{R})\nonumber\\
&  \ \ \ \ +\partial_{\mu}\phi S_{1}^{2/3\dagger}V_{1}^{2/3,\mu}+\partial
_{\mu}\phi V_{2}^{1/3,\mu\dagger}S_{2}^{1/3}+\phi H^{\dagger}V_{1,\mu
}^{2/3\dagger}V_{2}^{1/3,\mu}+h.c.\ ,
\end{align}
the mixing between $S_{1}^{2/3}$ and $S_{2}^{1/3}$ can only occur through that of $V_{1,\mu}^{2/3}$ and $V_{2,\mu}^{1/3}$. Specifically, with this specific choice of couplings,
\begin{equation}%
\begin{tabular}[c]{cccccccccccc}\hline
& $\phi$ & $S_{1}^{2/3}$ & $V_{1,\mu}^{2/3}$ & $S_{2}^{1/3}$ & $V_{2,\mu
}^{1/3}$ & $q_{L}$ & $u_{R}$ & $d_{R}$ & $\ell_{L}$ & $e_{R}$ & $\nu_{R}%
$\\\hline
$U(1)_{\mathcal{B}}$ & $-1/3$ & $-2/3$ & $-1/3$ & $1/3$ & $0$ & $1/3$ & $1/3$
& $1/3$ & $0$ & $0$ & $0$\\
$U(1)_{\mathcal{L}}$ & $1/3$ & $0$ & $-1/3$ & $-1$ & $-2/3$ & $0$ & $0$ & $0$
& $1$ & $1$ & $1$\\\hline
\end{tabular}
\end{equation}
Thanks to these charges, which crucially follow from whether $\phi$ or $\phi^{\dagger}$ are introduced in the couplings, no other SM fermion couplings of the LQ/DQ, nor any other renormalizable couplings among the LQ/DQ, is allowed. Turning on $\phi\bar{\nu}_{R}^{\mathrm{C}}\nu_{R}$ does not change this picture, except for the operator $\phi^{\dagger}S_{1}^{2/3}S_{2}^{1/3}S_{2}^{1/3}$. This is quite natural looking at the Lagrangian, since $\phi^{\dagger}S_{1}^{2/3}S_{2}^{1/3}S_{2}^{1/3}$ followed by $S_{1}^{2/3}\rightarrow\bar{q}_{L}^{\mathrm{C}}q_{L}$ and $S_{2}^{1/3}\rightarrow\bar{q}_{L}\nu_{R}$ permits to recover $\phi\bar{\nu}_{R}^{\mathrm{C}}\nu_{R}$ by closing the $q_{L}$ loops. As the $\phi^{\dagger}S_{1}^{2/3}S_{2}^{1/3}S_{2}^{1/3}$ and $\phi\bar{\nu}_{R}^{\mathrm{C}}\nu_{R}$ have the same quantum numbers, both carrying $(\Delta\mathcal{B},\Delta\mathcal{L})=(0,2)$, they are both able to generate neutrino masses only, and do not affect the $(\Delta\mathcal{B},\Delta\mathcal{L})=(1,-1)$ pattern.

The leading operator for proton decay is now (Fig.~\ref{Fig5}$b$):%
\begin{equation}
\mathcal{H}_{(\Delta\mathcal{B},\Delta\mathcal{L})=(1,-1)}^{eff}=\frac
{1}{m_{S}^{4}m_{V}^{4}}\phi\partial_{\mu}\phi\partial^{\mu}\phi H^{\dagger
}(\bar{q}_{L}^{\mathrm{C}}q_{L}+\bar{d}_{R}^{\mathrm{C}}u_{R})(\bar{d}%
_{R}^{\mathrm{C}}\ell_{L}^{\mathrm{C}}+\bar{q}_{L}^{\mathrm{C}}\nu
_{R}^{\mathrm{C}})\ ,
\end{equation}
and it contains in particular
\begin{equation}
\mathcal{H}_{(\Delta\mathcal{B},\Delta\mathcal{L})=(1,-1)}^{eff}=\frac
{v_{\phi}v\Lambda_{QCD}^{3}}{m_{S}^{4}m_{V}^{4}}\partial_{\mu}a^{0}%
\partial^{\mu}a^{0}\bar{p}\left(  1-\gamma^{5}\right)  \ell+...+h.c.\ .
\end{equation}
Phenomenologically, proton decay is suppressed, even for relatively low $m_{S}\approx m_{V}$ of $\mathcal{O}(10~$TeV$)$. On the other hand, if $a^{0}$ is an ALP with twice its mass above the proton but below the neutron mass, this setting is less interesting because the LQ/DQ masses need to be too low to reach $B(n\rightarrow a^{0}+a^{0}+\bar{\nu})\approx1\%$. Whether ALP or axions, the cosmological implications of this scenario would be worth further study though, as the consequences of opening up (possibly CP-violating) $a^{0}+p\leftrightarrow a^{0}+\ell$ and $a^{0}+\bar{p}\leftrightarrow a^{0}+\bar{\ell}$ scattering processes could provide a new baryogenesis mechanism.

\subsubsection{Spontaneous breaking of $\mathcal{B}+\mathcal{L}$}

To attain $(\Delta\mathcal{B},\Delta\mathcal{L})=(1,1)$ operators, the trick is to start from the previous scenario, but use some additional LQ couplings to switch $\mathcal{L}$ by two units. Specifically, we can consider%
\begin{align}
\mathcal{L}_{\mathrm{KSVZ+LQ}}  &  =\mathcal{L}_{\mathrm{KSVZ}}+S_{1}%
^{2/3}(\bar{q}_{L}^{\mathrm{C}}q_{L}+\bar{d}_{R}^{\mathrm{C}}u_{R})+V_{2,\mu
}^{1/3}(\bar{u}_{R}\gamma^{\mu}\ell_{L}^{\mathrm{C}}+\bar{q}_{L}\gamma^{\mu
}\nu_{R}^{\mathrm{C}})\nonumber\\
&  \ \ \ \ +\partial_{\mu}\phi S_{1}^{2/3\dagger}V_{1}^{2/3,\mu}+\phi
H^{\dagger}V_{1,\mu}^{2/3\dagger}V_{2}^{1/3,\mu}+h.c.\ . \label{LagrSSB2}%
\end{align}
Provided $V_{1,\mu}^{2/3}$ has no couplings to SM fermions, and only those two interactions among $\phi$ and the LQ/DQ are present, two $U(1)$s are present in the Lagrangian, with charges
\begin{equation}%
\begin{tabular}[c]{ccccccccccc}\hline
& $\phi$ & $S_{1}^{2/3}$ & $V_{1,\mu}^{2/3}$ & $V_{2}^{1/3,\mu}$ & $q_{L}$ &
$u_{R}$ & $d_{R}$ & $\ell_{L}$ & $e_{R}$ & $\nu_{R}$\\\hline
$U(1)_{\mathcal{B}}$ & $-1/2$ & $-2/3$ & $-1/6$ & $1/3$ & $1/3$ & $1/3$ &
$1/3$ & $0$ & $0$ & $0$\\
$U(1)_{\mathcal{L}}$ & $-1/2$ & $0$ & $1/2$ & $1$ & $0$ & $0$ & $0$ & $1$ &
$1$ & $1$\\\hline
\end{tabular}
\end{equation}
So, the $U(1)_{\mathcal{B}+\mathcal{L}}$ symmetry is spontaneously broken, while $\mathcal{B}-\mathcal{L}$ remains. If neutrino masses are generated by adding the $\phi\bar{\nu}_{R}\nu_{R}^{\mathrm{C}}$ coupling, the remaining exact $U(1)$ symmetry suffices to keep off all other interactions among $\phi$ and the LQ/DQ, as well as the LQ/DQ couplings to SM fermions not already present in the Lagrangian, except for a $\phi^{\dagger}S_{1}^{2/3}V_{2}^{1/3,\mu}V_{2,\mu}^{1/3}$ which carries the same quantum number as $\phi^{\dagger}{\bar{\nu}_{R}}^{\mathrm{C}}\nu_{R}$. Neither is able to open new $(\Delta\mathcal{B},\Delta\mathcal{L})$ patterns for proton decay.

\begin{figure}[t]
\begin{center}
\includegraphics[height=1.3517in,width=2.0254in]{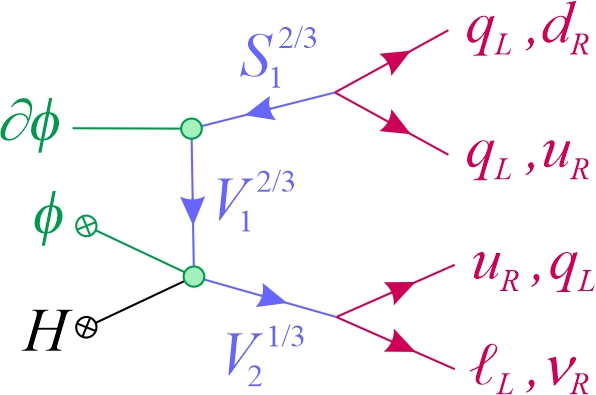}
\caption{Axion-induced $(\Delta\mathcal{B},\Delta\mathcal{L})=(1,1)$ operators.}
\label{Fig5b}
\end{center}
\end{figure}

The leading $\mathcal{B}+\mathcal{L}$ violating operator is (see Fig.~\ref{Fig5b})
\begin{align}
\mathcal{H}_{(\Delta\mathcal{B},\Delta\mathcal{L})=(1,1)}^{eff}  &  =\frac
{1}{m_{S}^{2}m_{V}^{4}}\phi\partial_{\mu}\phi H^{\dagger}(\bar{q}%
_{L}^{\mathrm{C}}q_{L}+\bar{d}_{R}^{\mathrm{C}}u_{R})(\bar{u}_{R}^{\mathrm{C}%
}\gamma^{\mu}\ell_{L}+\bar{q}_{L}^{\mathrm{C}}\gamma^{\mu}\nu_{R})\nonumber\\
&  \rightarrow\frac{v_{EW}v_{\phi}}{m_{S}^{2}m_{V}^{4}}\partial_{\mu}a^{0}(\bar
{q}_{L}^{\mathrm{C}}q_{L}+\bar{d}_{R}^{\mathrm{C}}u_{R})(\bar{u}%
_{R}^{\mathrm{C}}\gamma^{\mu}e_{L}+\bar{d}_{L}^{\mathrm{C}}\gamma^{\mu}\nu
_{R})\ .
\label{OpsAPantiL}
\end{align}
Phenomenologically, thanks to the $v_{EW}v_{\phi}$ from the $\phi HV_{2,\mu}^{1/3\dagger}V_{1}^{2/3,\mu}$ coupling, the LQ/DQ scale can be lower by about an order of magnitude without violating proton decay bounds. For ALPs, the main difference with the $(\Delta\mathcal{B},\Delta\mathcal{L})=(1,-1)$ scenario is that $f_{a}=v_{\phi}$ occurs in the $n\rightarrow a^{0}\bar{\nu}$ and $a^0\rightarrow p\ell$ amplitudes, but cancels out from the $p\rightarrow\bar{\ell}(a^0\rightarrow\gamma\gamma)$ rate. This means that $m_{S}\approx m_{V}$ cannot be as low as before, but must above $90$~TeV. Yet, this high scale is compensated in the $n\rightarrow a^{0}\bar{\nu}$ rate by the $v_{\phi}$ factor, so its branching ratio can still reach $\mathcal{O}(1\%)$. Actually, the dependencies of the various rates on $f_a$ stays exactly as depicted in Fig.~\ref{FigNeut}, but for $m_{S}\approx m_{V}\approx 90$~TeV.

Note, finally, that $\mathcal{B}+\mathcal{L}$ violating operators are not easily forced to involve pairs of axions. The pattern of LQ/DQ couplings to SM fermions, and their hypercharge, does not leave many options if only renormalizable operators are allowed. The simplest we could find would require two different Higgs doublets, so would be suitable for the DFSZ model
\begin{align}
\mathcal{L}_{\mathrm{DFSZ+LQ}}  &  =\mathcal{L}_{\mathrm{DFSZ}}+S_{1}%
^{4/3}\bar{d}_{R}^{\mathrm{C}}d_{R}+S_{1}^{2/3}(\bar{d}_{R}\nu_{R}%
^{\mathrm{C}}+\bar{u}_{R}e_{R}^{\mathrm{C}}+\bar{q}_{L}\ell_{L}^{\mathrm{C}%
})\nonumber\\
&  \ \ \ \ +\partial_{\mu}\phi S_{1}^{2/3\dagger}V_{1}^{2/3,\mu}+\partial
_{\mu}\phi S_{1}^{4/3\dagger}V_{1}^{4/3,\mu}+H_{u}^{\dagger}H_{d}^{\dagger
}V_{1,\mu}^{2/3\dagger}V_{1}^{4/3,\mu}+h.c.\ .
\end{align}
As the phenomenology is similar as that for $\mathcal{B}-\mathcal{L}$ violating operators, we do not detail this situation further.

\subsubsection{Spontaneous breaking of $\mathcal{B}$}

Given that we want to start from the derivative couplings, which are all quadratic in the LQ/DQ, we will need to add at least some cubic interactions. This quickly increases the number of new state needed, and phenomenologically, the longest the chain, the smallest the predicted rate given that LQ/DQ masses are at least of a few TeV.

The simplest processes correspond to the skeleton graph $\partial_{\mu}\phi\rightarrow X_{i}(X_{l}\rightarrow X_{j}X_{k})$, with the final
$X_{i}X_{j}X_{k}$ set allowing for $(\Delta\mathcal{B},\Delta\mathcal{L})=(2,0)$ transitions, so $X_{i,j,k}=S_{1}^{y}$ or $V_{2}^{y}$ for some $y$. If $X_{l}$ is integrated out, the effective operator involves $\partial_{\mu}\phi X_{i}X_{j}X_{k}$ plus some Higgs fields. The simplest such operators are of dimension six, and only seven of them are compatible with the gauge symmetries,
\begin{align}
&  \partial_{\mu}\phi~H^{\dagger}~(S_{1}^{2/3}S_{1}^{4/3}V_{2}^{1/3,\mu
},\ S_{1}^{4/3}S_{1}^{4/3}V_{2}^{5/3,\mu},\ V_{2}^{1/3,\mu}V_{2,\nu}%
^{1/3}V_{2}^{1/3,\nu})\ ,\\
&  \partial_{\mu}\phi~H~(V_{2}^{5/3,\mu}V_{2,\nu}^{1/3}V_{2}^{1/3,\nu}%
,\ S_{1}^{2/3}S_{1}^{2/3}V_{2}^{1/3,\mu},\ S_{1}^{2/3}S_{1}^{4/3}%
V_{2}^{5/3,\mu},\ S_{1}^{4/3}S_{1}^{8/3}V_{2}^{1/3,\mu})\ ,
\end{align}
where $\partial_{\mu}\phi$ could be replaced by $\partial_{\mu}\phi^{\dagger}$ wherever required. Starting from the three derivative interactions of Eq.~(\ref{DerScenars}), there are several ways to reach these operators using a $HX_{l}X_{j}X_{k}$ or $H^{\dagger}X_{l}X_{j}X_{k}$ coupling. Since $X_{l}=S_{2}^{y}$ or $V_{1}^{y}$, these operators alone cannot induce $(\Delta\mathcal{B},\Delta\mathcal{L})=(2,0)$ processes. Further, if $X_{l}$ transforms as a $6$, it does not couple to SM fermions hence these operators cannot lead to proton decay either. If $X_{l}$ transforms as a $3$, one must make sure the PQ charges forbid $X_{l}\rightarrow\ell q$. All this nevertheless leaves many possible mechanisms, though many of them turn out to be essentially equivalent phenomenologically, so let us take an example.

\begin{figure}[t]
\begin{center}
\includegraphics[height=1.9787in,width=4.7158in]{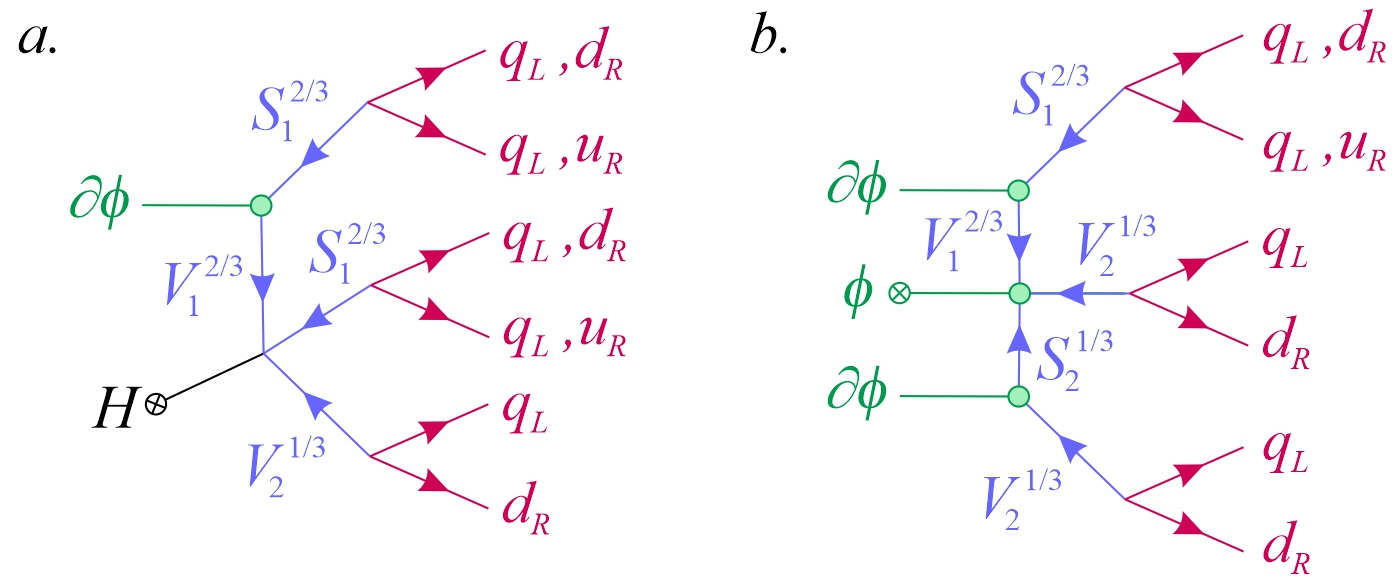}
\caption{One and two axion induced neutron-antineutron oscillation $(\Delta\mathcal{B},\Delta\mathcal{L})=(2,0)$ operators.}
\label{Fig6}
\end{center}
\end{figure}

Consider the Lagrangian
\begin{align}
\mathcal{L}_{\mathrm{KSVZ+LQ}}  &  =\mathcal{L}_{\mathrm{KSVZ}}+S_{1}%
^{2/3}(\bar{q}_{L}^{\mathrm{C}}q_{L}+\bar{d}_{R}^{\mathrm{C}}u_{R})+V_{2,\mu
}^{1/3}\bar{d}_{R}^{\mathrm{C}}\gamma^{\mu}q_{L}\nonumber\\
&  +\partial_{\mu}\phi V_{1}^{2/3,\mu\dagger}S_{1}^{2/3}+HV_{1,\mu}^{2/3}%
S_{1}^{2/3}V_{2}^{1/3,\mu}+h.c.\ , \label{LagrSSB3a}%
\end{align}
where $S_{1}^{2/3}$ and $V_{1}^{2/3,\mu}$ transform as $\mathbf{3}$, but $V_{2}^{1/3,\mu}\sim\mathbf{\bar{6}}$ since the final operator $\partial_{\mu}\phi HS_{1}^{2/3}S_{1}^{2/3}V_{2}^{1/3,\mu}$ would cancel for $V_{2}^{1/3,\mu}\sim\mathbf{3}$. Dropping the $\Psi_{L,R}$, as their charge can independently be fixed by turning on some couplings to the LQ, the active $U(1)$s are then
\begin{equation}%
\begin{tabular}[c]{ccccccccccc}\hline
& $\phi$ & $S_{1}^{2/3}$ & $V_{2,\mu}^{1/3}$ & $V_{1,\mu}^{2/3}$ & $q_{L}$ &
$u_{R}$ & $d_{R}$ & $\ell_{L}$ & $e_{R}$ & $\nu_{R}$\\\hline
$U(1)_{\mathcal{B}}$ & $2$ & $-2/3$ & $-2/3$ & $4/3$ & $1/3$ & $1/3$ & $1/3$
& $0$ & $0$ & $0$\\
$U(1)_{\mathcal{L}}$ & $0$ & $0$ & $0$ & $0$ & $0$ & $0$ & $0$ & $1$ & $1$ &
$1$\\\hline
\end{tabular}
\ \label{SponB}%
\end{equation}
Turning on any of the LQ couplings of $S_{1}^{2/3}$ or $V_{1,\mu}^{2/3}$ would break $U(1)_{\mathcal{B}+\mathcal{L}}$, and induce proton decay (compare Eq.~(\ref{LagrSSB3a}) with Eq.~(\ref{LagrSSB2})). At this level, their presence is thus forbidden by the still active $U(1)_{\mathcal{B}}$ and $U(1)_{\mathcal{L}}$ symmetries. For an even stricter protection, the PQ symmetry can be extended to prevent these couplings. It suffices to add a seesaw mechanism with the $\phi^{\dagger}\bar{\nu}_{R}^{\mathrm{C}}\nu_{R}$ coupling, something we should do anyway (the $\phi\bar{\nu}_{R}^{\mathrm{C}}\nu_{R}$ would instead allow all the LQ couplings). Note that $S_{1}^{2/3}$ and $V_{2,\mu}^{1/3}$ can mix via a $D^{\mu}HS_{1}^{2/3\dagger}V_{2,\mu}^{1/3}$ term, but this is inessential since they have the same $\mathcal{B}$ and $\mathcal{L}$ quantum numbers. This scenario lead to neutron-antineutron oscillation operators, with the diagram of Fig.~\ref{Fig6}$a$, corresponding to
\begin{align}
\mathcal{H}_{(\Delta\mathcal{B},\Delta\mathcal{L})=(2,0)}^{eff}  &  =\frac
{1}{m_{S}^{8}}\partial_{\mu}\phi H(\bar{q}_{L}^{\mathrm{C}}q_{L}+\bar{d}%
_{R}^{\mathrm{C}}u_{R})(\bar{q}_{L}^{\mathrm{C}}q_{L}+\bar{d}_{R}^{\mathrm{C}%
}u_{R})\bar{d}_{R}^{\mathrm{C}}\gamma^{\mu}q_{L}+h.c.\ \nonumber\\
&  \rightarrow\frac{v\Lambda_{QCD}^{6}}{m_{S}^{8}}\partial_{\mu}a^{0}\bar
{n}^{\mathrm{C}}\gamma^{\mu}\gamma_{5}n + ... + h.c.\ , \label{SponB2}%
\end{align}
where we have set all the DQ masses to a common $m_{S}$ value. Since there is no associated $n-\bar{n}$ operator, this scale can in principle be quite low. The best low energy limits come from nuclear transitions, as this operator also contributes to $nn\rightarrow a$, but those do not push $m_{S}$ well above the TeV scale~\cite{Heeck:2020nbq}. The main constraint thus come from LHC searches~\cite{CMS:2020gru,ATLAS:2020dsk,ATLAS:2020xov,CMS:2020wzx,ATLAS:2021oiz}. Note, though, that the generic leptoquark searches may not apply to this case: all these states decay to diquark pairs and, furthermore, $V_{1}^{2/3,\mu}$ could end up quite long lived if it is lighter than $V_{2}^{1/3,\mu}$ and $S_{1}^{2/3,\mu}$, and would show up in channels with at least four jets.

Even if $m_{S}$ can be quite low, at around the TeV say, the $a^{0}nn$ coupling is significantly smaller than the other couplings, including to $n\bar{n}$, as can be estimated setting $f_{a}\equiv v_{\phi}$:
\begin{equation}
\frac{1}{f_{a}}\approx\frac{v\Lambda_{QCD}^{6}}{m_{S}^{8}}\Leftrightarrow
m_{S}\approx10~\text{GeV}\times\left(  \frac{f_{a}}{10^{9}\ \text{GeV}%
}\right)  ^{1/8}\ ,
\end{equation}
for $\Lambda_{QCD}\approx300$ MeV. Even with $f_{a}$ close to the Planck scale, the LQ mass would need to be well below the TeV scale, which would again be ruled out by direct searches. For $m_{S}$ around the TeV, the $a^{0}nn$ coupling is at best $10^{-16}$ smaller than that to $a^{0}n\bar{n}$. Thus, $a^{0}nn$ does not represent a competitive signature for direct axion searches.

Indirectly, the $a^{0}nn$ coupling may nevertheless open new routes by relying instead on neutron-antineutron oscillation phenomena. Indeed, while $a^{0}nn$ cannot generate $n\rightarrow\bar{n}$ in vacuum, oscillations could now be catalyzed by an axion dark matter background. While the typical high frequency of the coherent axion field precludes any observation using standard beam searches for $n-\bar{n}$ oscillations (the induced $\delta m_{n-\bar{n}}$ would average to zero), transient variations of the axion field may be observable in this way. Another possibility would be to exploit the magnetic splitting between $n$ and $\bar{n}$ states, which in a $1~$T magnetic field would be of about $10^{-7}$~eV~\cite{Phillips:2014fgb}, larger than the axion mass if $f_{a}>10^{14}$~GeV. Note that the neutron beam go through a 4.6~T magnet in neutron lifetime experiments, Ref.~\cite{Nico:2004ie,Yue:2013qrc}, and that axion-induced mixing effects, if they occur, would not have been excluded by the recent mirror neutron search of Ref.~\cite{Broussard:2021eyr}, which relies on hypothesized mirror neutrons capabilities to pass through normal matter.

Two other features compared to the usual neutron oscillations are worth mentioning: the coupling is axial, $\bar{n}^{\mathrm{C}}\gamma^{\mu}\gamma_{5}n$, instead of the usual scalar $\bar{n}^{\mathrm{C}}n$ oscillation operator, so the spin dependencies are different~\cite{Gardner:2014cma}, and the $\partial_{\mu}a^{0}\bar{n}^{\mathrm{C}}\gamma^{\mu}\gamma_{5}n$ coupling can be CP violating~\cite{Berezhiani:2015uya,McKeen:2015cuz,Berezhiani:2018xsx} since the DQ couplings are a priori complex, so $n$ and $\bar{n}$ may react differently to an axionic background. Also, compared to neutron-mirror neutron oscillations, like those invoked to explain the neutron lifetime anomaly~\cite{Berezhiani:2018eds}, the antineutron would not be invisible but would either decay to antiproton, or annihilate with the surrounding matter. A quantitative analysis of these signatures is clearly called for but would require a detailed study, which go beyond our scope. Also, other manifestations of the $a^{0}nn$ coupling in an astrophysical and cosmological context are left for a future study.

With only three LQ, another rather simple scenario can lead to the $\partial_{\mu}\phi H^{\dagger}V_{2}^{1/3,\mu}V_{2,\nu}^{1/3}V_{2}^{1/3,\nu}$ operator by virtual $S_{2}^{1/3}$ exchanges, and involves only states transforming as $\mathbf{3}$:
\begin{align}
\mathcal{L}_{\mathrm{KSVZ+LQ}}  &  =\mathcal{L}_{\mathrm{KSVZ}}+S_{1}%
^{2/3}(\bar{q}_{L}^{\mathrm{C}}q_{L}+\bar{d}_{R}^{\mathrm{C}}u_{R})+V_{2,\mu
}^{1/3}\bar{d}_{R}^{\mathrm{C}}\gamma^{\mu}q_{L}\nonumber\\
&  +\partial_{\mu}\phi S_{2}^{1/3\dagger}V_{2}^{1/3,\mu}+H^{\dagger}%
S_{2}^{1/3}V_{2,\nu}^{1/3}V_{2}^{1/3,\nu}+h.c.\ . \label{LagrSSB3b}%
\end{align}
The same $U(1)_{\mathcal{B}}$ charges are found as in Eq.~(\ref{SponB}), with $V_{1,\mu}^{2/3}\rightarrow S_{2}^{1/3}$. Also, as before, adding the $\phi^{\dagger}\bar{\nu}_{R}^{\mathrm{C}}\nu_{R}$ coupling prevents all the LQ couplings of $V_{2,\mu}^{1/3}$, $S_{1}^{2/3}$, and $S_{2}^{1/3}$. Proton decay is now forbidden by the existence of the PQ symmetry at the high scale, and does not arise at the low scale thanks to the specific $(\Delta\mathcal{B},\Delta\mathcal{L})=(2,0)$ and $(\Delta\mathcal{B},\Delta\mathcal{L})=(0,2)$ symmetry breaking pattern. The final operator is phenomenologically similar to that in Eq.~(\ref{SponB2}).

Many other choices of DQ states are possible, but they lead to similar patterns. We will not investigate more complicated processes, except for the following that leads to a different phenomenology:
\begin{align}
\mathcal{L}_{\mathrm{KSVZ+LQ}} &  =\mathcal{L}_{\mathrm{KSVZ}}+S_{1}%
^{2/3}(\bar{q}_{L}^{\mathrm{C}}q_{L}+\bar{d}_{R}^{\mathrm{C}}u_{R})+V_{2,\mu
}^{1/3}\bar{d}_{R}^{\mathrm{C}}\gamma^{\mu}q_{L}\nonumber\\
&  \ \ \ \ +\partial_{\mu}\phi S_{2}^{1/3\dagger}V_{2}^{1/3,\mu}+\partial
_{\mu}\phi V_{1}^{2/3,\mu\dagger}S_{1}^{2/3}+\phi^{\dagger}S_{2}^{1/3}V_{2,\mu}%
^{1/3}V_{1}^{2/3,\mu}+h.c.\ .\label{LagrSSB3c}%
\end{align}
In some senses, it combines the previous two scenarios, and gives the same charges as in Eq.~(\ref{SponB}), with $V_{1,\mu}^{2/3}$ and $S_{2}^{1/3}$ having $\mathcal{B}=4/3$. Also, the $\phi^{\dagger}\bar{\nu}_{R}^{\mathrm{C}}\nu_{R}$ coupling now suffices to prevent the LQ couplings of the four states, $V_{2,\mu}^{1/3}$, $V_{1,\mu}^{2/3}$, $S_{1}^{2/3}$, $S_{2}^{1/3}$. What differs however is how the $(\Delta\mathcal{B},\Delta\mathcal{L})=(2,0)$ effects are induced at the low-energy scale. The two derivative couplings are needed, and $\phi$ further occurs in the cubic DQ coupling, so the leading operator is (see Fig.~\ref{Fig6}$b$)
\begin{align}
\mathcal{H}_{(\Delta\mathcal{B},\Delta\mathcal{L})=(2,0)}^{eff} &  =\frac
{1}{m_{S}^{10}}\phi\partial_{\mu}\phi^\dagger\partial_{\nu}\phi^\dagger(\bar{q}_{L}%
^{\mathrm{C}}q_{L}+\bar{d}_{R}^{\mathrm{C}}u_{R})\bar{d}_{R}^{\mathrm{C}%
}\gamma^{\mu}q_{L}\bar{d}_{R}^{\mathrm{C}}\gamma^{\nu}q_{L}+h.c.\ \nonumber\\
&  \rightarrow\frac{v_{\phi}\Lambda_{QCD}^{6}}{m_{S}^{10}}\partial_{\mu}%
a^{0}\partial^{\mu}a^{0}\bar{n}^{\mathrm{C}}\gamma_{5}n+...+h.c.\ .
\end{align}
Though this operator is now of dimension 14 instead of that of dimension 12 in Eq.~(\ref{SponB2}), the extra suppression is compensated by the $v_{\phi}$ factor since $v_{\phi}\Lambda_{QCD}/m_{S}^{2}$ is of $\mathcal{O}(1)$ for $m_{S}$ around the tens of TeV scale and $v_{\phi}$ at around $10^{6}$~TeV. The nuclear transition bounds are thus similar as in the single axion case, and in any case not competitive with direct collider searches for new colored states. Phenomenologically, neutron-antineutron conversion now requires pairs of axions, and would occur through scattering processes like $a^{0}+n\leftrightarrow a^{0}+\bar{n}$ or $n+n\leftrightarrow a^0+a^0$ and $\bar{n}+\bar{n}\leftrightarrow a^0+a^0$. Though unlikely to be ever observed, these processes could play a cosmological role.

\subsubsection{Spontaneous breaking of $\mathcal{B}\pm3\mathcal{L}$}

The $(\Delta\mathcal{B},\Delta\mathcal{L})=(1,3)$ scenarios are trivially obtained from any of the $(\Delta\mathcal{B},\Delta\mathcal{L})=(2,0)$ Lagrangians of the previous section by switching all DQ couplings to LQ couplings. For example, starting from Eq.~(\ref{LagrSSB3b}),
\begin{align}
\mathcal{L}_{\mathrm{KSVZ+LQ}} &  =\mathcal{L}_{\mathrm{KSVZ}}+S_{1}%
^{2/3}(\bar{d}_{R}\nu_{R}^{\mathrm{C}}+\bar{u}_{R}e_{R}^{\mathrm{C}}+\bar
{q}_{L}\ell_{L}^{\mathrm{C}})+V_{2,\mu}^{1/3}(\bar{u}_{R}\gamma^{\mu}\ell
_{L}^{\mathrm{C}}+\bar{q}_{L}\gamma^{\mu}\nu_{R}^{\mathrm{C}})\ \nonumber\\
&  +\partial_{\mu}\phi S_{2}^{1/3\dagger}V_{2}^{1/3,\mu}+H^{\dagger}%
S_{2}^{1/3}V_{2,\nu}^{1/3}V_{2}^{1/3,\nu}+h.c.\ ,
\end{align}
leads to the charges
\begin{equation}%
\begin{tabular}[c]{ccccccccccc}\hline
& $\phi$ & $S_{1}^{2/3}$ & $V_{2,\mu}^{1/3}$ & $S_{1}^{2/3}$ & $q_{L}$ &
$u_{R}$ & $d_{R}$ & $\ell_{L}$ & $e_{R}$ & $\nu_{R}$\\\hline
$U(1)_{\mathcal{B}}$ & $-1$ & $-2/3$ & $1/3$ & $1/3$ & $1/3$ & $1/3$ & $1/3$ &
$0$ & $0$ & $0$\\
$U(1)_{\mathcal{L}}$ & $-3$ & $-2$ & $1$ & $1$ & $0$ & $0$ & $0$ & $1$ & $1$ &
$1$\\\hline
\end{tabular}
\end{equation}
By analogy, $(\Delta\mathcal{B},\Delta\mathcal{L})=(1,-3)$ transitions can be induced by taking the Lagrangian
\begin{equation}
\mathcal{L}_{\mathrm{KSVZ+LQ}}=\mathcal{L}_{\mathrm{KSVZ}}+S_{2}^{1/3}(\bar
{d}_{R}\ell_{L}+\bar{q}_{L}\nu_{R})+V_{1,\mu}^{2/3}\bar{d}_{R}\gamma^{\mu}%
\nu_{R}+\partial_{\mu}\phi V_{1}^{2/3,\mu\dagger}S_{1}^{2/3}+\phi S_{1}%
^{2/3}S_{2}^{1/3}S_{2}^{1/3}+h.c.\ ,
\end{equation}
with the charges are
\begin{equation}%
\begin{tabular}[c]{ccccccccccc}\hline
& $\phi$ & $S_{1}^{2/3}$ & $S_{2}^{1/3}$ & $V_{1,\mu}^{2/3}$ & $q_{L}$ &
$u_{R}$ & $d_{R}$ & $\ell_{L}$ & $e_{R}$ & $\nu_{R}$\\\hline
$U(1)_{\mathcal{B}}$ & $1/2$ & $-1/6$ & $1/3$ & $1/3$ & $1/3$ & $1/3$ & $1/3$
& $0$ & $0$ & $0$\\
$U(1)_{\mathcal{L}}$ & $-3/2$ & $1/2$ & $-1$ & $-1$ & $0$ & $0$ & $0$ & $1$ &
$1$ & $1$\\\hline
\end{tabular}
\end{equation}
Note that for each case, additional $(\Delta\mathcal{B},\Delta\mathcal{L})=(0,0)$ couplings involving pairs of LQs are possible, like $\phi HS_{2}^{1/3\dagger}S_{1}^{2/3}$ or $D^{\mu}HS_{2}^{1/3\dagger}V_{1,\mu}^{2/3}$ for the $(\Delta\mathcal{B},\Delta\mathcal{L})=(1,-3)$ scenario. Those can neither affect the symmetry pattern, nor open new routes for proton decay.

Phenomenologically, these scenarios are very similar to the $(\Delta\mathcal{B},\Delta\mathcal{L})=(1,\pm1)$ ones described before, so we will not detail them further. The main difference is the extra suppression of proton and neutron decays due to the higher dimensionality of the operators, and of the many particles in the final states. These scenarios thus have essentially the same phenomenology whenever these suppressions can be compensated by lowering the LQ mass scale without violating LHC bounds.

\section{Conclusions\label{Ccl}}

In this paper, the opportunities arising from combining leptoquarks and diquarks with axions have been systematically analyzed. From a phenomenological standpoint, our main results are:

\begin{enumerate}
\item The PQ symmetry of which the axion is the Goldstone boson can be identified with any combination of baryon $\mathcal{B}$ and lepton $\mathcal{L}$ numbers. In this way, $\mathcal{B}$ and $\mathcal{L}$ appear partly protected by the PQ symmetry, which has to be exact above the PQ breaking scale. Reminiscent of the possible $\Delta\mathcal{B}$ and/or $\Delta\mathcal{L}$ operators made of SM fields (see Table~\ref{TableLQBL}), the simplest scenarios identify $U(1)_{PQ}$ with $U(1)_{\mathcal{B}\pm\mathcal{L}}$, $U(1)_{\mathcal{B}\pm3\mathcal{L}}$, $U(1)_{\mathcal{B}}$, or $U(1)_{\mathcal{L}}$, and induce spontaneously either proton decay, neutron-antineutron oscillations, or a Majorana mass terms for $\nu_{R}$ (or more generally, neutrinoless double beta decays).

\item All scenarios can be supplemented with a seesaw mechanism. The axion is then not only the Goldstone boson associated to $U(1)_{\mathcal{B}\pm\mathcal{L}}$, $U(1)_{\mathcal{B}\pm3\mathcal{L}}$, or $U(1)_{\mathcal{B}}$ breaking, but becomes also the Majoron associated to the $U(1)_{\mathcal{L}}$ breaking. Though no global symmetry (besides of course $U(1)_{PQ}$ itself) remains, each scenario retains a specific phenomenology. For example, when $U(1)_{PQ}$ is identified both with $U(1)_{\mathcal{B}}$ and $U(1)_{\mathcal{L}}$, $(\Delta\mathcal{B},\Delta\mathcal{L})=(2n,0)$ and $(\Delta\mathcal{B},\Delta\mathcal{L})=(0,2n)$ transitions are possible, but proton decay cannot occur.

\item For each pattern of symmetry breaking, it is also possible to prevent axion-free proton decay, neutron-antineutron oscillations, or neutrinoless double beta decays. In other words, one can make sure $(\Delta\mathcal{B},\Delta\mathcal{L})$ effects always involve at least one axion field. Phenomenologically, $(\Delta\mathcal{B},\Delta\mathcal{L})=(1,1)$ scenarios open the door to $p\rightarrow a^{0}+\ell,$ $n\rightarrow a^{0}+\nu$, $p\rightarrow2a^{0}+\ell,$ $n\rightarrow2a^{0}+\nu$, and scattering processes like $a^{0}+(p,n)\leftrightarrow a^{0}+(\ell,\nu)$. Scenarios with $(\Delta\mathcal{B},\Delta\mathcal{L})=(1,-1)$ or $(1,\pm3)$ are similar. Following the strategy proposed in Ref.~\cite{NeutronTau}, if $a^{0}$ is an ALP of just the right mass, such that proton decay is forbidden but neutron decay is not, these scenarios are able to solve the neutron lifetime puzzle, see Fig.~\ref{FigNeut}.

\item When applied to $(\Delta\mathcal{B},\Delta\mathcal{L})=(2,0)$ operators, being forced to include an axion field could lead to very peculiar effects. The phenomenology of the $\partial_{\mu}a^{0}\bar{n}^{\mathrm{C}}\gamma^{\mu}\gamma_{5}n$ and $\partial_{\mu}a^{0}\partial^{\mu}a^{0}\bar{n}^{\mathrm{C}}\gamma_{5}n$ interactions have, to our knowledge, not been investigated in detail yet. Though a dedicated analysis is called for, we do not expect these interactions to be phenomenologically relevant in vacuum, but they could open interesting channels in an axionic dark matter background, or transitions like $n\rightarrow\bar{n}+a^{0}$ or $n\rightarrow\bar{n}+a^{0}+a^0$ in an intense magnetic field.
\end{enumerate}

Besides these phenomenological aspects, we have also analyzed the consequences on the foundations of axion effective Lagrangians. Whenever the axion is associated to some patterns of $U(1)_{\mathcal{B}}$ and/or $U(1)_{\mathcal{L}}$ breaking, the SM fermions become charged under the PQ symmetry. Typically, they thus occur in the usual $f_{a}$-suppressed derivative interactions, but through vector current interactions, $\partial_{\mu}a^{0}\bar{\psi}\gamma^{\mu}\psi$ (since $\mathcal{B}$ and $\mathcal{L}$ are vectorial). Often, these interactions are discarded owing to the naive vector Ward identity, but this is incorrect for two reasons:

\begin{enumerate}
\item[5.] Axion-gauge field interaction are usually expected to be $(g_{X}^{2}/f_{a})\mathcal{N}_{X}a^{0}X_{\mu\nu}\tilde{X}^{\mu\nu}$, $X=G^{a}$, $W^{i}$, $B$, with $\mathcal{N}_{X}$ summing up the contribution of all the fields charged under both the PQ symmetry and the $X$ gauge interactions of strength $g_{X}$.\ Thus, $\mathcal{N}_{X}$ depend on the SM fermion charges, with in particular $\mathcal{N}_{W}$ and $\mathcal{N}_{B}$ depending on how $U(1)_{\mathcal{B}}$ and/or $U(1)_{\mathcal{L}}$ are embedded in $U(1)_{PQ}$. Yet, as shown in Ref.~\cite{Quevillon:2019zrd}, the SM fermion contributions to $\mathcal{N}_{W}$ and $\mathcal{N}_{B}$ arising from $U(1)_{\mathcal{B}}$ and/or $U(1)_{\mathcal{L}}$ systematically cancel with that coming from triangle graphs built on the corresponding $\partial_{\mu}a^{0}\bar{\psi}\gamma^{\mu}\psi$ interactions.\ At the end of the day, the $U(1)_{\mathcal{B}}$ and/or $U(1)_{\mathcal{L}}$ components of $U(1)_{PQ}$ do not alter the axion to gauge boson couplings, even though this is not apparent at the level of the effective Lagrangian.

\item[6.] The counting rule in powers of $1/f_{a}$, central in constructing the axion effective Lagrangian (see e.g. Ref.~\cite{Georgi:1986df}), is invalid when $\mathcal{B}$ and/or $\mathcal{L}$ are broken spontaneously along with the PQ symmetry. Indeed, the equations of motion of the SM fermions (or that of the leptoquarks if they have not been integrated out) inherit $\mathcal{O}((f^{\alpha})^{n}),$ $n\geq1$ terms, so that $\mathcal{O}(f_{a}^{n-1})$ interactions are hidden inside $f_{a}^{-1}\partial_{\mu}a^{0}\bar{\psi}\gamma^{\mu}\psi$. In practice, in the present paper, all these interactions were suppressed by some relatively high power of the leptoquark masses, which are pushed above the TeV by direct collider searches. Thus, in all the scenarios considered here, the $\mathcal{B}$ and/or $\mathcal{L}$ violating interactions are not expected to be dominant compared to e.g. the two photon or two gluon modes for $f_{a}$ below the Planck scale. Still, as this relative suppression has nothing to do with $f_{a}$, there is no guarantee it always happens.
\end{enumerate}

In conclusion, even if entangling the PQ symmetry with the accidental symmetries of the SM requires new leptoquarks states, and often several of them, these scenarios end up being more economical from a $U(1)$ global symmetry point of view. The axion becomes a central piece, not only solving the strong CP puzzle, and maybe making up for the observed dark matter, but also setting off the seesaw mechanism and introducing potentially CP violating baryon number violation. With all its capabilities, the axion could hold the keys to many of the standing cosmological enigmas.

\subsubsection*{Acknowledgements} This work is supported by the labex \textit{Enigmass}, and by the
CNRS/IN2P3 Master project \textit{Axions from Particle Physics to Cosmology}.

\end{document}